\documentclass[aps,reprint,showpacs,superscriptaddress]{revtex4-1}

\usepackage{graphicx}
\usepackage[version=3]{mhchem}
\usepackage{dcolumn}
\usepackage{bm}
\usepackage[usenames, dvipsnames]{xcolor}
\usepackage{threeparttable}
\usepackage{longtable}
\usepackage{afterpage}
\usepackage{multirow}
\AtBeginDocument{\usepackage{booktabs}}
\usepackage{url}
\usepackage[colorlinks=true,linkcolor=blue,citecolor=magenta,urlcolor=red]{hyperref}

\definecolor{greenGM}{rgb}{0.4353, 1, 0}
\definecolor{greenmax1}{rgb}{0.7843,0,0.7843}
\definecolor{greenmax2}{rgb}{0,0.6627,0.4118}
\definecolor{greenenv}{rgb}{0,0.8,0}
\definecolor{lightgray}{rgb}{0.65,0.65,0.65}
\definecolor{lightred}{rgb}{1,0.65,0.65}
\definecolor{red2000}{rgb}{1,0,0.2}
\definecolor{red200}{rgb}{1,0,0.4}
\definecolor{red20}{rgb}{1,0,0.6}
\definecolor{lightblue}{rgb}{0., 0.46, 0.8}
\definecolor{grayL}{rgb}{0.3,0.3,0.3}
\definecolor{grayLmk}{rgb}{0.6,0.6,0.6}
\definecolor{redL}{rgb}{1,0,0.3}
\definecolor{redLmk}{rgb}{1,0,0.6}
\definecolor{boxa}{rgb}{1,0.5,0}
\definecolor{boxs}{rgb}{0.5,0.5,0.5}
\definecolor{boxm}{rgb}{0,0.75,0.2}

\newcommand{\blue}[1]{\textcolor{lightblue}{#1}}

\newcommand{\orange}[1]{\textcolor{OrangeRed}{#1}}
\newcommand{\pblue}[1]{\textcolor{ProcessBlue}{#1}}

\newcommand{\green}[1]{\textcolor{ForestGreen}{#1}}

\begin{document}

\preprint{APS/123-QED}

\title{Computational Investigation of Half-Heusler Compounds for Spintronics Applications}

\author{Jianhua Ma}
\email{jm9yq@virginia.edu}
\affiliation{Department of Electrical and Computer Engineering, University of Virginia, Charlottesville,VA-22904, USA}%

\author{Vinay I. Hegde}
\affiliation{Department of Materials Science and Engineering, Northwestern University, Evanston, IL 60208, USA}

\author{Kamaram Munira}
\affiliation{Center for Materials for Information Technology, University of Alabama, Tuscaloosa, Alabama 35401, USA}%

\author{Yunkun Xie}
\affiliation{Department of Electrical and Computer Engineering, University of Virginia, Charlottesville,VA-22904, USA}%

\author{Sahar Keshavarz}
\affiliation{Center for Materials for Information Technology, University of Alabama, Tuscaloosa, Alabama 35401, USA}%
\affiliation{Department of Physics and Astronomy, University of Alabama, Tuscaloosa, Alabama 35401, USA}%

\author{David T. Mildebrath}
\affiliation{Center for Materials for Information Technology, University of Alabama, Tuscaloosa, Alabama 35401, USA}
\affiliation{Department of Physics and Astronomy, University of Alabama, Tuscaloosa, Alabama 35401, USA}

\author{C. Wolverton}
\affiliation{Department of Materials Science and Engineering, Northwestern University, Evanston, IL 60208, USA}

\author{Avik W. Ghosh}
\affiliation{Department of Electrical and Computer Engineering, University of Virginia, Charlottesville,VA-22904, USA}%

\author{W. H. Butler}
\email{wbutler@mint.ua.edu}
\affiliation{Center for Materials for Information Technology, University of Alabama, Tuscaloosa, Alabama 35401, USA}%
\affiliation{Department of Physics and Astronomy, University of Alabama, Tuscaloosa, Alabama 35401, USA}%

\date{\today}

\begin{abstract}
We present first-principles density functional calculations of the electronic structure, magnetism, and structural stability of 378 \textit{XYZ} half-Heusler compounds (with $X=$ Cr, Mn, Fe, Co, Ni, Ru, Rh; $Y=$ Ti, V, Cr, Mn, Fe, Ni; $Z=$ Al, Ga, In, Si, Ge, Sn, P, As, Sb). We find that a ``Slater-Pauling gap" in the  density of states, (i.e. a gap or pseudogap after nine states in  the three atom primitive cell)\  in at least one spin channel is a common feature in half-Heusler compounds. We find that the presence of such a gap at the Fermi energy in one or both spin channels contributes significantly to the stability of a half-Heusler compound. We calculate the formation energy of each compound and systematically investigate its stability against all other phases in the Open Quantum Materials Database (OQMD). We represent the thermodynamic phase stability of each compound as its distance from the convex hull of stable phases in the respective chemical space and show that the hull distance of a compound is a good measure of the likelihood of its experimental synthesis. We find low formation energies and mostly correspondingly low hull distances for compounds with $X=$ Co, Rh or Ni, $Y=$ Ti or V, and $Z=$ P, As, Sb or Si. We identify 26 18-electron semiconductors, 45 half-metals, and 34 near half-metals with negative formation energy, that follow the Slater-Pauling rule of three electrons per atom. Our calculations predict several new, as-yet unknown, thermodynamically stable phases which merit further experimental exploration --- RuVAs, CoVGe, FeVAs in the half-Heusler structure, and NiScAs, RuVP, RhTiP in the orthorhombic MgSrSi-type structure. Further, two interesting zero-moment half-metals, CrMnAs and MnCrAs, are calculated to have negative formation energy. In addition, our calculations predict a number of hitherto unreported semiconducting (e.g., CoVSn, RhVGe), half-metallic (e.g., RhVSb), and near half-metallic (e.g., CoFeSb, CoVP) half-Heusler compounds to lie close to the respective convex hull of stable phases, and thus may be experimentally realized under suitable synthesis conditions, resulting in potential candidates for various semiconducting and spintronics applications.
\end{abstract}

\pacs{63.22.-m, 66.70.-f, 44.10+i}

\maketitle

\section{INTRODUCTION}
\label{sec:introduction}

Half-Heusler, or semi-Heusler, compounds (space group $F\bar{4}3m$, Structurbericht designation $C1_b$) comprise a relatively large family of materials with diverse physical properties and applications. Functional materials based on these compounds include thermoelectric semiconductors~\cite{Graf20101216,sakurada2005effect,sootsman2009new}, piezoelectric semiconductors~\cite{PhysRevLett.109.037602}, optoelectronic semiconductors~\cite{PhysRevB.81.075208}, and topological insulators~\cite{PhysRevB.82.235121,chadov2010tunable}. A half-Heusler inspired the term ``half-metal'' when in 1983, de Groot and collaborators calculated the band structure of NiMnSb and observed that there was a gap at the Fermi energy for the minority spin channel, but not for the majority spin channel~\cite{PhysRevLett.50.2024,Helmholdt1984249}. Since then, the calculated electronic structures of many half-Heusler compounds show them to be half-metals or nearly half-metals, often with large band gaps. 

Because they have 100\% spin polarization at the Fermi level and can have relatively high Curie temperatures~\cite{PhysRevB.67.220403,an2008new}, Heusler-based half-metals have attracted significant interest for spintronics applications~\cite{hirohata2013heusler,hirohata2014future,felser2015basics}. Half-metals are considered ideal electrode materials for magnetic tunnel junctions (MTJs)~\cite{tanaka1999spin}, giant magnetoresistance devices (GMRs)~\cite{Hordequin1998225}, and for injecting spin-polarized currents into semiconductors~\cite{van2000epitaxial}. The huge number of possible half-Heusler compounds, their diverse properties and the recent realization that half-metallic Heuslers tend to remain half-metallic when layered with other Heuslers (including full-Heuslers)~\cite{Culbert,LayeredHeusler} raises the possibility of finding, tailoring or even designing  materials optimized for particular applications. 

Although numerous half-Heusler compounds have been predicted to be half-metallic by first-principles calculations~\cite{PhysRevB.66.134428,PhysRevB.51.10436,tobola2000electronic, kandpal2006covalent,galanakis2006electronic}, a comprehensive study of the structural, electronic and magnetic properties of the half-Heusler family is useful, because it is not clear which of the many half-metallic half-Heuslers that can be imagined, are stable. Thus, a systematic study of the structural stability of the half-Heusler ($C1_{b}$) family should provide guidance for future experiments. 

It is observed empirically that the calculated electronic structures of many half-Heusler compounds show a band gap at a band filling of  three electrons per atom in at least one of the spin channels. This feature is  known as the ``Slater-Pauling gap" \cite{galanakis2006electronic} and is a generalization of the ``Slater-Pauling rule" \cite{Slater1937,PhysRev.54.899}. The Slater-Pauling rule is based on the observation that the average magnetic moment in Bohr magnetons per atom, $M$, of many \textit{bcc}-based compounds is approximately, but closely, related to the average number of valence electrons per atom $N$ through $M=N-6$. Since the spin moment per atom is just the difference in the number of up and down electrons per atom ($M=N^\uparrow -N^\downarrow$), and since $N=N^\uparrow + N^\downarrow$, the Slater-Pauling rule implies $N^\downarrow=3$. The calculated electronic structure of these bcc compounds does not show gaps, but does often show a pseudogap (an energy range with a very low density of states) at a band-filling of approximately 3 in the minority channel. (There is a second part to the Slater-Pauling rule, not relevant to the Heuslers, that states that the magnetic moment per atom of many \textit{fcc}-based compounds is given by $M=10.6-N$.)     

The calculated electronic structure of many of the Heusler compounds show actual gaps at 3 electrons per atom.  We call these Slater-Pauling gaps.  When the Fermi energy falls in a Slater-Pauling gap, we will describe the system as a Slater-Pauling half-metal. We anticipate that a large, consistent database of calculated properties of half-Heuslers (both stable and unstable)  will allow the testing of hypotheses that may explain the occurrence and size of these Slater-Pauling band gaps in the Heusler compounds.

In this paper, we describe a computational investigation covering 378 half-Heusler compounds using first-principles methods. We have constructed a database of their electronic, magnetic and structural properties~\cite{Heusl89:online}, which enables us to identify potentially useful electrode/spacer materials for future spintronics applications. In Sec.~\ref{sec:computational_method} we present the details of our computational method. The techniques, codes and parameters used in our DFT calculations are described in Sec.~\ref{ssec:dft_calculations}. In Sec.~\ref{ssec:lowest_energy_structure} we discuss how we determine the structure with the lowest energy for each compound, and deal with the multiple solutions in energy and magnetic configuration that occur for some of these compounds. Sec.~\ref{ssec:energetic_quantities} describes our approach to investigating the stability of these  compounds through the calculation of their formation energies and the comparison of these calculated energies to the calculated energies of other possible phases and combinations of phases.  

In Sec.~\ref{ssec:formation_energy_hull_distance} we discuss the trends in (and the factors influencing) the formation energy and thermodynamic stability of these compounds across the periodic table. In Sec.~\ref{ssec:SP_semiconductors} we list some 18-electron Slater-Pauling half-Heusler semiconductors and analyze their electronic structures and chemical bonding characteristics. In Sec.~\ref{ssec:zero_moment_half_metals}, we discuss the possibility and implications of zero-moment half-metallic Slater-Pauling half-Heusler compounds. In Sec.~\ref{ssec:half_metallic_ferromagnets}, we present some half-metallic and near half-metallic ferromagnets that result from our calculations in terms of the Slater-Pauling rule. For all the half-Heusler compounds presented in Sec.~\ref{sec:results_discussion}, we systematically discuss their thermodynamic stability relative to other competing phases in the respective chemical space. Finally, we summarize our results and conclusions in Sec.~\ref{sec:summary_conclusion}.

\section{COMPUTATIONAL METHODS}
\label{sec:computational_method}

\begin{figure}[t]
\centering
\includegraphics[width=\columnwidth]{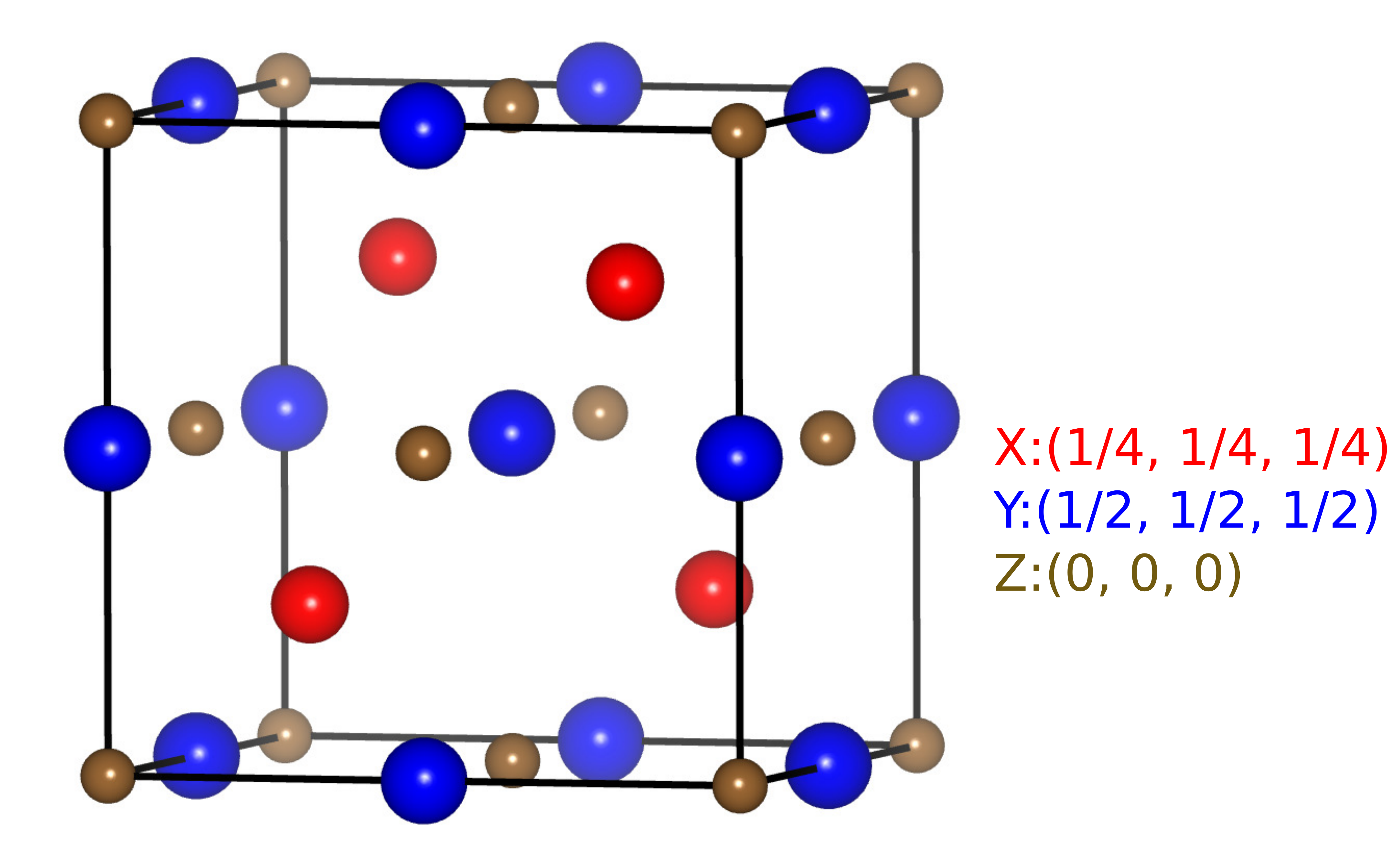} 
\caption{Schematic of the \textit{XYZ} half-Heusler $C1_{b}$ structure. It consists of three interpenetrating fcc sublattices with atomic sites $X$ $\left(\frac{1}{4},\, \frac{1}{4},\, \frac{1}{4}\right)$, $Y$ $\left(\frac{1}{2},\, \frac{1}{2},\, \frac{1}{2}\right)$, and $Z$ $(0,\, 0,\, 0)$. The $\left(\frac{3}{4},\, \frac{3}{4},\, \frac{3}{4}\right)$ site is vacant.}
\label{fig:c1b_crystal_structure}
\end{figure}

A half-Heusler compound, \textit{XYZ}, has a face-centered cubic structure with one formula unit per \emph{primitive} fcc unit cell. Its space group is $F\bar{4}3m$ (International Tables of Crystallography No.~216), and its Structurbericht designation is $C1_{b}$. The half-Heusler structure can be viewed as three interpenetrating fcc sublattices (Fig.~\ref{fig:c1b_crystal_structure}), occupied by $X$, $Y$, and $Z$ atoms, respectively. The $Z$ and $Y$ atoms are located at $(0,\, 0,\, 0)$ and $\left(\frac{1}{2},\, \frac{1}{2},\, \frac{1}{2}\right)$, and together form a rock salt sublattice. The $X$ atoms occupy $\left(\frac{1}{4},\, \frac{1}{4},\, \frac{1}{4}\right)$, and the site $\left(\frac{3}{4},\, \frac{3}{4},\, \frac{3}{4}\right)$ (which is occupied by an $X$ atom in an $X_2YZ$ full-Heusler compound) is vacant in the half-Heusler. The $X$ and $Y$ atoms considered without the $Z$ atoms would form a zincblende structure.  Similarly, the $X$ and $Z$ atoms considered without the $Y$ atoms would also form a zincblende structure. In the previous section it was implied that half-Heusler compounds are ``\textit{bcc}-based". The sense in which this assertion is valid follows from imagining that the $X$, $Y$ and $Z$ atoms as well as the vacancy site at $\left(\frac{3}{4},\, \frac{3}{4},\, \frac{3}{4}\right)$ are all replaced by atoms belonging to a single species. This would generate a bcc lattice.

In this study, (a) $X$ is one of 7 elements -- Cr, Mn, Fe, Co, Ni, Ru, or Rh, (b) $Y$ is one of 6 elements -- Ti, V, Cr, Mn, Fe, or Ni, and (c) $Z$ is one of 9 elements -- Al, Ga, In, Si, Ge, Sn, P, As, or Sb. In addition to these 378 ($7\times 6\times 9$) \textit{XYZ} systems, we also perform some calculations with $Y=$ Sc in order to study additional examples of half-metallic and semiconducting half-Heusler compounds. For each of the 378 potential half-Heusler compounds, we calculate its electronic and magnetic structure, stability against structural distortion, formation energy, and thermodynamic phase stability.

\subsection{Density Functional Theory Calculations}
\label{ssec:dft_calculations}
We perform all calculations using density-functional theory (DFT) as implemented in the Vienna Ab-initio Simulation Package (VASP)~\cite{Kresse199615} with a plane wave basis set and projector-augmented wave (PAW) potentials~\cite{PhysRevB.50.17953}. The set of PAW potentials for all elements and the plane wave energy cutoff of 520 eV for all calculations were both chosen for consistency with the Open Quantum Materials Database (OQMD)~\cite{raey,oqmd_npj_2015}. The Perdew-Burke-Ernzerhof (PBE) version of the generalized gradient approximation (GGA) to the exchange-correlation functional was adopted~\cite{PhysRevLett.77.3865}. The integrations over the irreducible Brillouin zone (IBZ) used the automatic mesh generation scheme within VASP with the mesh parameter (the number of $k$-points per {\AA}$^{-1}$ along each reciprocal lattice vector) set to 50, which usually generated a $15\times15\times15$ $\Gamma$-centered Monkhorst-Pack grid~\cite{PhysRevB.13.5188}, resulting in 288 $k$-points in the IBZ. The integrations employed the linear tetrahedron method with Bl\"ochl corrections~\cite{PhysRevB.49.16223}. To achieve a higher accuracy with respect to the magnetic moment, the interpolation formula of Vosko, Wilk, and Nusair~\cite{vosko1980accurate} was used in all calculations. Finally, during ionic relaxations, the convergence criterion for structural optimization was an energy change of less than $1\times10^{-5}$ eV between successive ionic steps.

\subsection{Determination of the Relaxed Structure}
\label{ssec:lowest_energy_structure}

We explain our procedure for obtaining the relaxed structures in some detail in order to make clear that the $C1_b$ structure is not guaranteed to minimize the energy of a particular equiatomic \textit{XYZ} system and that the possibility of multiple solutions to the DFT equations must be be considered  when there is more than one magnetic species in the unit cell. We calculate the formation energy using the pseudopotentials and convergence parameters  consistent with the OQMD~\cite{oqmd_npj_2015} so that the calculated formation energies of the half-Heusler compounds can be directly compared to those of many other phases in the OQMD. 

We performed full ionic relaxations within a 6-atom tetragonal cell for all of the 378 potential half-Heusler compounds. All relaxations started from the $C1_{b}$ structure with small displacements to avoid vanishing of the net force on each atom due to symmetry. 300 of these compounds were found to remain in the $C1_{b}$ structure, 6 relaxed to a tetragonal structure ($\left|c/a-1\right|>0.01$), while 72 compounds relaxed to a distorted structure that was neither cubic nor tetragonal.

For all systems, we performed DFT calculations using multiple initial magnetic configurations to start the iterative process that (usually)\ leads to a fixed point that minimizes the energy for a given  set of atoms and atomic coordinates.  As we shall see, a fixed point may be a local rather than a global energy minimum.  Use of multiple initial magnetic configurations including moment configurations in which the $X$ and $Y$ moments were parallel and anti-parallel  increased our chances of finding the global minimum.      

\begin{figure}
\includegraphics[width=\columnwidth]{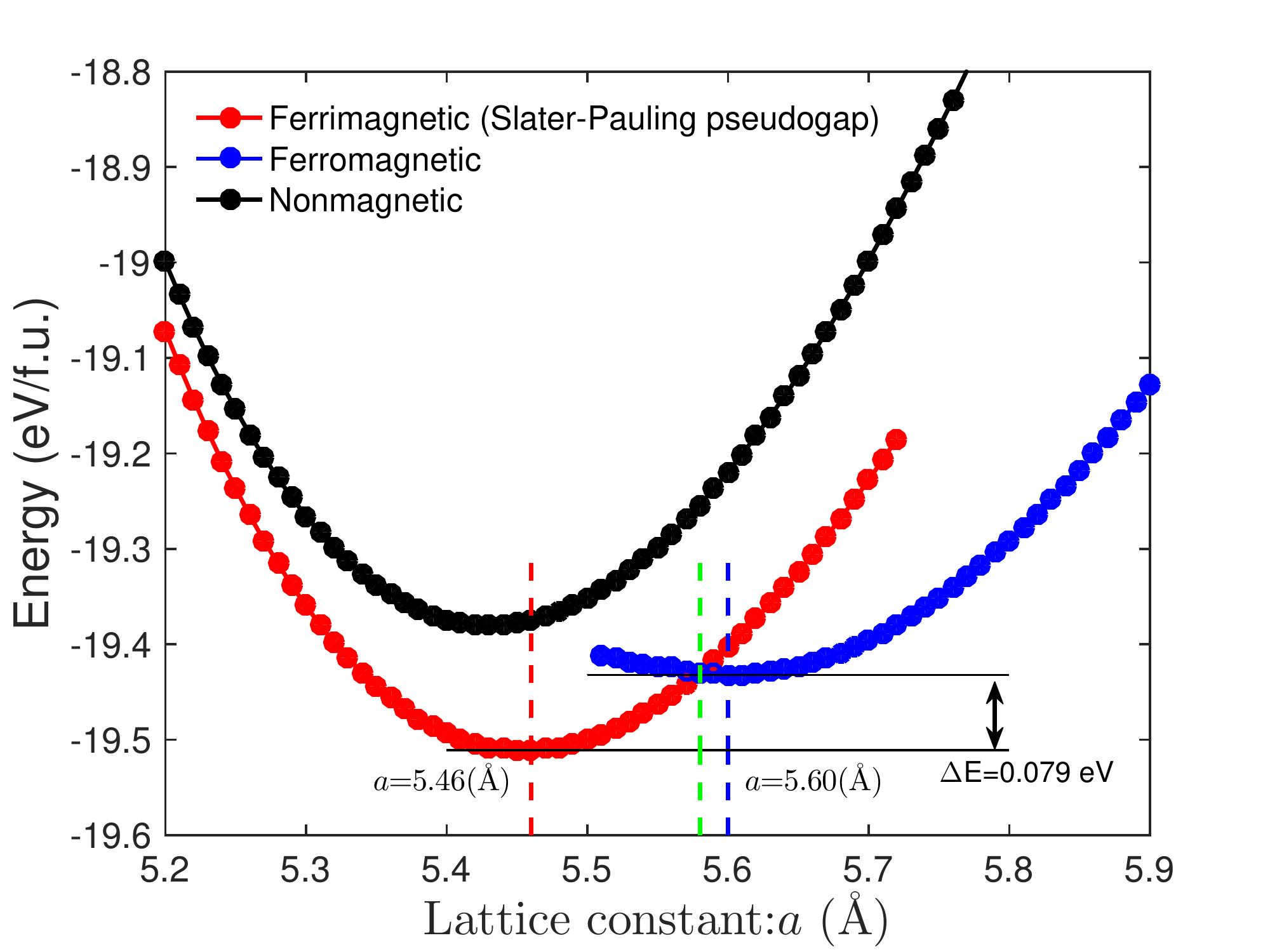}
\caption{Calculated total energies of CoMnAl in the half-Heusler $C1_{b}$ structure as a function of the lattice constant $a$ in ferrimagnetic, ferromagnetic, and nonmagnetic states.}
\label{fig:comnal_energy_vs_lattice_constant}
\end{figure}

To clarify this point, we provide a few examples of  \textit{XYZ} systems, for which we found multiple DFT solutions with different magnetic configurations at the same or similar lattice constants. An example of competition between a ferrimagnetic phase and a ferromagnetic phase is shown in Fig.~\ref{fig:comnal_energy_vs_lattice_constant} which displays the total energy as a function of the lattice parameter for CoMnAl in the $C1_{b}$ structure. Two energy minima occur at $a=5.46$ and $a=5.60$ {\AA}. For $a=5.46$~{\AA}, the moments within spheres of radius $1.45$~{\AA} surrounding each atom are $1.38$ for Mn, $-0.25$ for Co and $-0.10~\mu_{B}$ for Al, which indicates a ferrimagnetic state. For $a=5.60$~{\AA}, the compound has a total magnetic moment of 3.60~$\mu_{B}$ per formula unit (f.u.) and the magnetic configuration is ferromagnetic in the sense that Mn and Co have parallel moments. The moments within the 1.45~{\AA} spheres in this case are 3.20, 0.46 and $-0.09$~$\mu_B$ for Mn, Co, and Al, respectively. The ferromagnetic solution has an energy $0.079$~eV/f.u. higher than the ferrimagnetic solution. The lower-energy ferrimagnetic solution is a ``Slater-Pauling solution'', and the electronic density of states (DOS) shows a pseudogap near the Fermi energy that becomes a gap for slightly larger lattice constants.     

\begin{figure}
\includegraphics[width=\columnwidth]{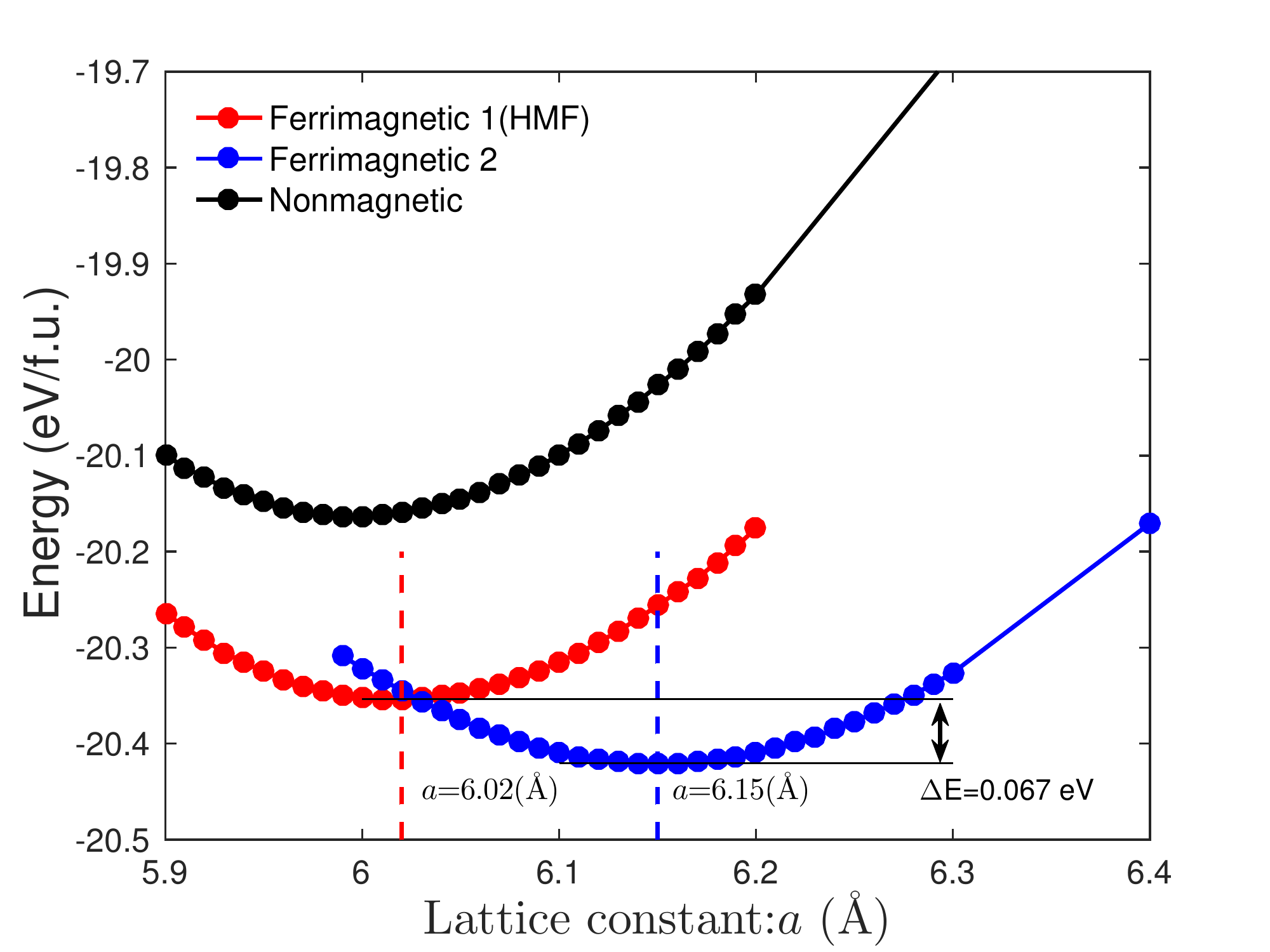}
\caption{Calculated total energies of RhCrSn in the half-Heusler $C1_{b}$ structure as a function of the lattice constant $a$ in two different ferrimagnetic and the nonmagnetic states.}
\label{fig:rhcrsn_energy_vs_lattice_constant}
\end{figure}

RhCrSn provides an example of competition between two different ferrimagnetic states as shown in Fig.~\ref{fig:rhcrsn_energy_vs_lattice_constant}. In this case, the energy minima at $a=6.02$ and $6.15$~{\AA} correspond to ferrimagnetic states with different atomic magnetic moments (see Table \ref{tab:magnetic_moments}). The $a=6.02$~{\AA} solution is half-metallic while the $a=6.15$~{\AA} solution is metallic. The metallic solution has a lower energy than the half-metallic solution by $0.067$~eV/f.u.

\begin{figure}
\includegraphics[width=\columnwidth]{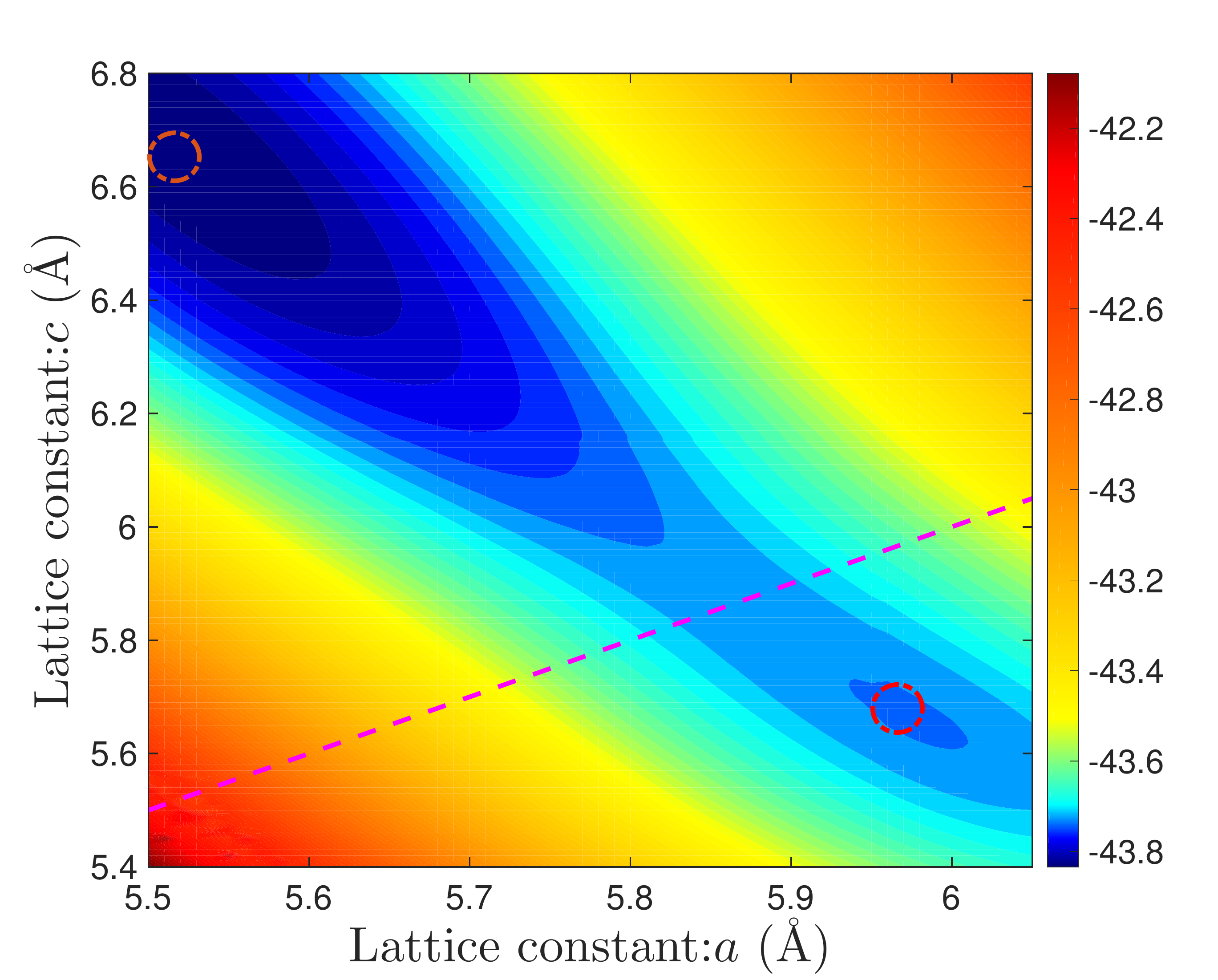}
\caption{Calculated total energies of tetragonal distorted CrTiAs as a function of two lattice constants $a$ and $c$. There are two half-metallic energy minima corresponding to two different tetragonal phases, labelled by dashed circles, at $(a, c)=(5.52, 6.66)$~{\AA} and $(a, c)=(5.97, 5.67)$~{\AA}. They differ in energy by $0.051$~eV/f.u. The pink dashed line corresponds to cubic structures.}
\centering
\label{fig:crtias_energy_vs_c_a}
\end{figure}

A few systems showed multiple local energy minima in a tetragonal structure, e.g., CrTiAs. Fig.~\ref{fig:crtias_energy_vs_c_a} presents the total energy as a function of the two lattice constants $a$ and $c$ for ordered CrTiAs. There are two energy minima (labeled by the dashed circles) with tetragonality ($c/a$) less than and larger than 1 respectively. One local minimum is $(a, c)=(5.97, 5.67)$~{\AA}, with tetragonality $c/a = 0.95$. The other local minimum with a lower energy is $(a, c)=(5.52, 6.66)$~{\AA}, with tetragonality $c/a = 1.21$. Both energy minima have total magnetic moments of 3~$\mu_{B}$ per f.u. and display half-metallicity. The energy difference between them is $0.051$~eV/f.u. In fact, CrTiAs displays half-metallicity in almost the entire blue region of Fig.~\ref{fig:crtias_energy_vs_c_a}. We found a few other compounds that behave similar to CrTiAs; these will be discussed in Sec.~\ref{ssec:half_metallic_ferromagnets}.     

\begin{table*}
\caption{Calculated total ($M_{tot}$) and $X$-, $Y$-, $Z$-site-projected partial spin magnetic moments ($m$) of CoMnAl, RhCrSn, and CrTiAs in different local minima of energy. All magnetic moment values listed are in units of $\mu_B$. (PG $=$ pseudogap, HM $=$ half-metal)}
\label{tab:magnetic_moments}
\begin{tabular}{llccccc}
\toprule
Compound & Magnetic state & Structure & $M_{tot}$ & $m(X)$ & $m(Y)$ & $m(Z)$ \\
\midrule
CoMnAl &  Ferrimagnetic (PG)        & $C1_{b}$      & 1.04  & $-0.25$   & 1.38      & $-0.10$\\
CoMnAl &  Ferromagnetic             & $C1_{b}$      & 3.60  & 0.46      & 3.20      & $-0.09$\\
RhCrSn &  Ferrimagnetic 1 (HM)      & $C1_{b}$      & 1.00  & $-0.22$   & $1.36$    & $-0.09$\\
RhCrSn &  Ferrimagnetic 2           & $C1_{b}$      & 3.25  & $-0.10$   & 3.29      & $-0.06$\\
CrTiAs &  Ferromagnetic (HM)        &  Tetragonal   & 3.00  & 2.35      & 0.48      & $-0.01$\\ 
CrTiAs &  Ferromagnetic (HM)        &  Tetragonal   & 3.00  & 2.46      & 0.38      & $-0.01$\\ 
\bottomrule
\end{tabular}%

\end{table*}%

We list the total and partial magnetic moments for CoMnAl, RhCrSn and CrTiAs at different local energy minima in Table \ref{tab:magnetic_moments} for comparison. These compounds might have interesting properties if they can be synthesized. CrTiAs is particularly interesting because it is unusual to find two metastable half-metallic phases so close in energy. Tetragonal half-metallic phases are also rare and might be interesting for applications that require uniaxial magnetic anisotropy.  

\subsection{Calculation of Energetic Quantities}
\label{ssec:energetic_quantities}

\subsubsection{Formation Energy}
\label{sssec:formation_energy}
The formation energy of a half-Heusler compound \textit{XYZ} is defined as 
\begin{equation}\label{eqn:formation_energy}
\Delta E_{f}\,(XYZ) = E\,(XYZ) - \frac{1}{3}\left(\mu_X + \mu_Y + \mu_Z\right)
\end{equation}
where $E(XYZ)$ is the total energy per atom of the half-Heusler compound, and $\mu_i$ is the reference chemical potential of element $i$, chosen to be consistent with those used in the OQMD (See Ref.~\cite{oqmd_npj_2015} for details). A negative value of $\Delta E_{f}$ indicates that at zero temperature, the half-Heusler compound is more stable than its constituent elements. It is a \textit{necessary but not sufficient} condition for ground state thermodynamic stability. It does not, for example, guarantee the stability of a half-Heusler phase over another competing phase or mixture of phases.

\subsubsection{Distance from the Convex Hull}
\label{sssec:hull_distance}
\begin{figure}
\centering
\includegraphics[width=0.9\columnwidth]{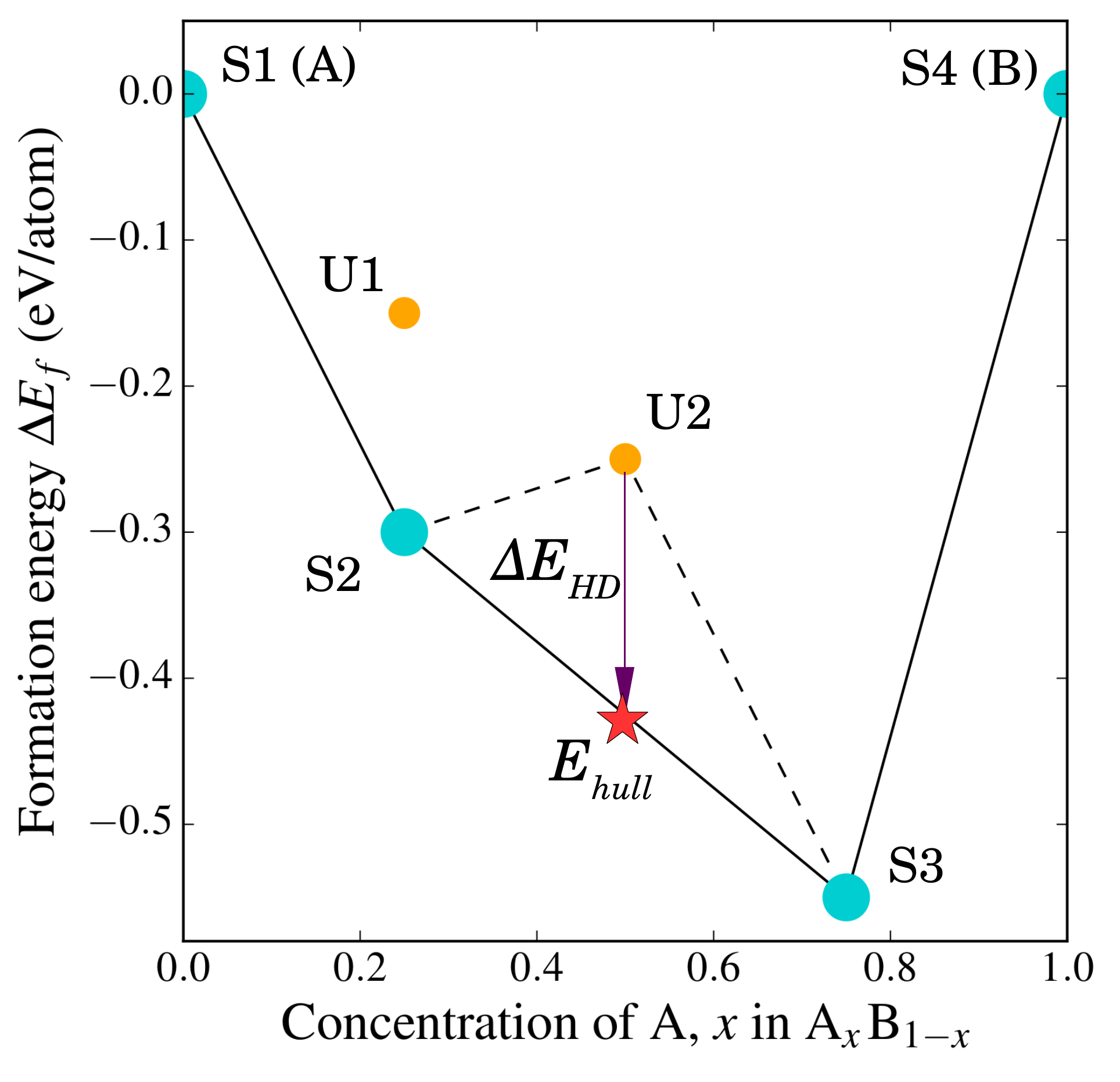}
\caption{A schematic convex hull in the A--B chemical space. Phases S$_i$ lie \textit{on} the convex hull and are thermodynamically stable, i.e., for each phase S$_i$, there is no other phase or combination of phases at its composition lower in energy. Phases U$_i$ are off the convex hull and thus unstable. For example, the formation energy of phase U2 is higher than that of a linear combination of phases S2 and S3. The distance from the convex hull ($\Delta E_{\rm HD}$) of phase U2 is given by the difference between its formation energy and the energy of the convex hull at its composition ($E_{hull}$, represented by the crimson star).}
\label{fig:schematic_convex_hull}
\end{figure}

A compound can be thermodynamically stable only if it lies \textit{on} the convex hull of formation energies of all phases in the respective chemical space. Every phase on the convex hull has a formation energy lower than any other phase or linear combination of phases in the chemical space at that composition. Thus, any phase on the convex hull is, by definition, thermodynamically stable (e.g., phases S1, S2, S3, and S4 in in Fig.~\ref{fig:schematic_convex_hull}). Conversely, any phase that does not lie on the convex hull is thermodynamically unstable -- there is another phase or combination of phases on the convex hull which is lower in energy. For example, in Fig.~\ref{fig:schematic_convex_hull}, phase U1 is unstable because there exists another phase (S2) at the composition that has a lower formation energy; similarly, phase U2 is unstable because a linear combination of phases S2 and S3 has a lower energy at that composition (``$E_{hull}$'').

A measure of thermodynamic stability of a phase is its distance from the convex hull. In other words, the farther away a phase is from the convex hull, higher is the thermodynamic driving force for it to transform or decompose into another phase or combination of phases. The distance from the convex hull $\Delta E_{\rm HD}$ for a phase with formation energy $\Delta E_{f}$ can be calculated as
\begin{equation}\label{eqn:hull_distance}
\Delta E_{\rm HD} = E_{hull} - \Delta E_{f}
\end{equation}
where $E_{hull}$ is the energy of the convex hull at the composition of the phase (see Fig.~\ref{fig:schematic_convex_hull} for an illustration). The energy of the convex hull at any composition is given by a linear combination of energies of stable phases. This is thus a linear composition-constrained energy minimization problem~\cite{akbarzadeh2007, kirklin2013}, and is available as a look-up feature called ``grand canonical linear programming'' (GCLP) on the OQMD website (\texttt{http://oqmd.org/analysis/gclp}). Obviously, the hull distance $E_{\rm HD}$ for a phase on the convex hull (hence thermodynamically stable) is 0, i.e., there is no other phase or linear combination of phases lower in energy than the phase at that composition. We note here that the distance from the convex hull of a phase depends on the completeness of the set of phases considered in the construction of the convex hull. Ideally, for calculating the convex hull of a system $X$--$Y$--$Z$, one would  investigate all possible compounds that can be formed from elements $X$, $Y$, and $Z$ (no matter how large or complex their structure), which is not feasible. A practical approach is to construct the convex hull using all the currently reported compounds in the $X$--$Y$--$Z$ phase space. Here, we have limited our universe of considered phases to those in the OQMD, which includes all of the binary and ternary phases that have been reported in the ICSD, and $\sim$350,000 hypothetical compounds based on common structural prototypes. Thus, the calculated formation energy of each \textit{XYZ} half-Heusler compound is compared against the calculated formation energies of all phases and all linear combinations of phases with total composition \textit{XYZ} in the OQMD database.

Further, as we will demonstrate in Sec.~\ref{ssec:formation_energy_hull_distance}, the distance of a phase from the convex hull (or simply ``hull distance'') $\Delta E_{\rm HD}$, apart from being a measure of its thermodynamic stability, is an indicator of the likelihood of its synthesis in experiments. We also note that since we use 0 K DFT energetics in our analysis, a phase that is above the convex hull may be either actually metastable or stabilized (i.e., moved on to the convex hull, and thus become experimentally accessible) due to (a) finite temperature contributions to the free energy such as phonons, magnons, configurational entropy, and/or (b) other external conditions such as pressure. Thus, while a phase that is above the convex hull may be experimentally realizable under carefully controlled conditions, we assert that the hull distance is still the best measure available of the \textit{likelihood} of its experimental synthesis (see Sec.~\ref{ssec:formation_energy_hull_distance} for further discussion).


\section{RESULTS AND DISCUSSION}
\label{sec:results_discussion}

\subsection{Energetics: Formation Energy and Distance from the Convex Hull}
\label{ssec:formation_energy_hull_distance}

In this section, we systematically investigate the energetics of  378 half-Heusler compounds in the $C1_b$ structure. For each \textit{XYZ} half-Heusler compound, we calculate its formation energy $\Delta E_{f}$ using Eq.~\ref{eqn:formation_energy} and distance from the convex hull $\Delta E_{\rm HD}$ using Eq.~\ref{eqn:hull_distance}. We explore the relationship between formation energy and hull distance of compounds at the compositions considered in this work (focusing on experimentally reported compounds and half-Heuslers, in particular), followed by an analysis of the trends in energetic quantities with composition.

\begin{figure}
\includegraphics[width=\columnwidth]{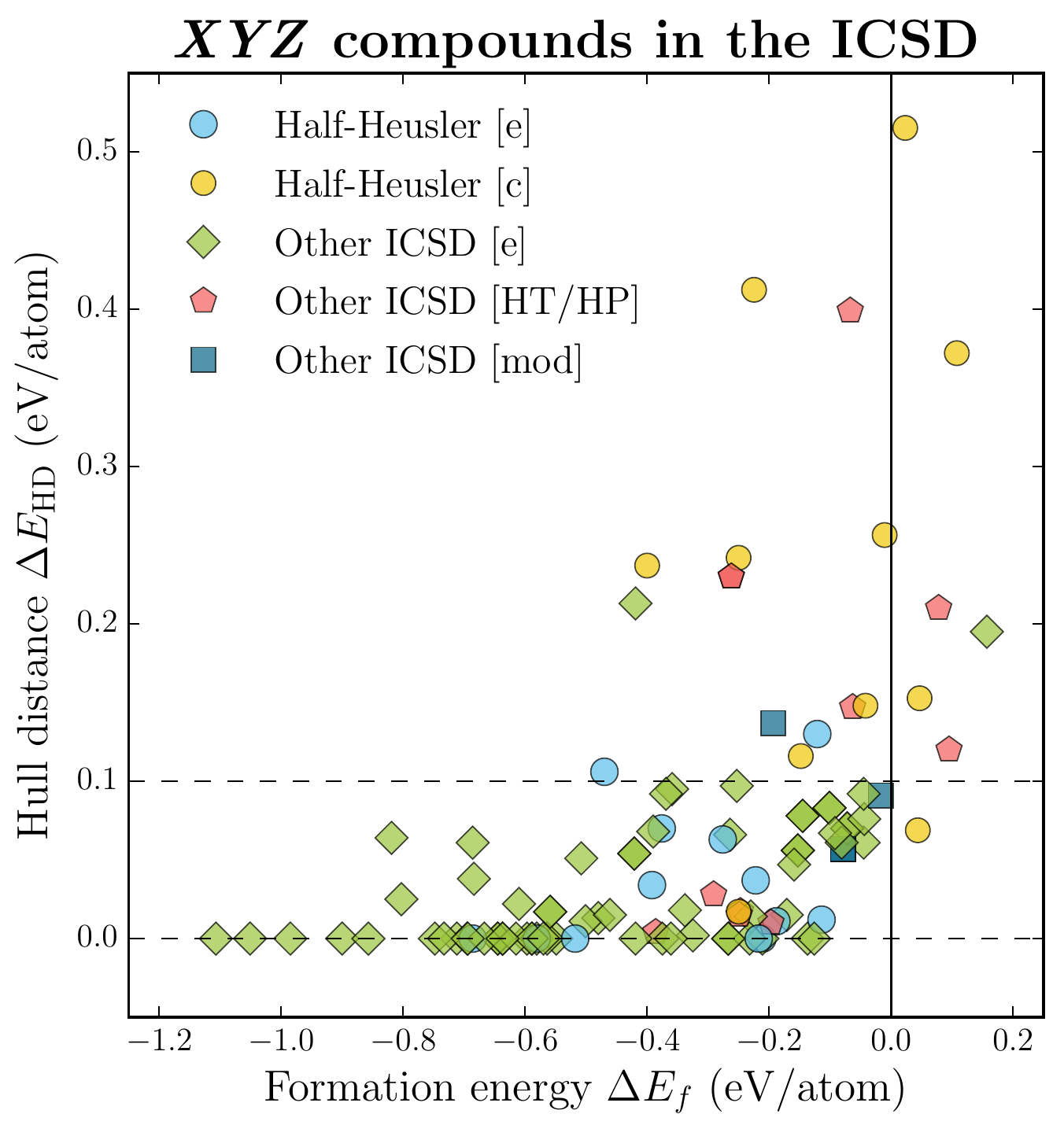}
\caption{DFT-calculated formation energy vs. hull distance of all compounds reported in the ICSD in the \textit{XYZ} compositions considered in this work. A hull distance $\Delta E_{\rm HD} = 0$ indicates a stable ground state compound on the convex hull. Blue circles indicate half-Heuslers in the ICSD\ that have been experimentally synthesized. Green diamonds indicate \textit{XYZ} phases other than C1$_b$ that have been experimentally synthesized. Red pentagons indicate \textit{XYZ} phases experimentally reported to be stable at high temperature or pressure. Blue-green squares indicate reported \textit{XYZ} phases  with  site occupations that differ from the OQMD\ calculation.  Yellow circles indicate  C1$_b$ phases sourced into ICSD from electronic structure calculations rather than from experiment.          }
\label{fig:icsd_formation_vs_hull_distance}
\end{figure}

For each composition \textit{XYZ} considered here, in an effort to identify all the compounds experimentally synthesized at the composition, we begin by compiling a list of all compounds reported (if any) in the Inorganic Crystal Structure Database (not limited to the half-Heusler $C1_b$ structure), and tabulate their formation energies and hull distances as calculated in the OQMD --- a total of 110 compounds (with 98 distinct compositions) and corresponding energies. The above sets of formation energies and hull distances are displayed in Fig.~\ref{fig:icsd_formation_vs_hull_distance}. The experimentally reported half-Heuslers are shown as blue circles.  Other experimentally reported \textit{XYZ} phases are shown as green diamonds.  (The yellow circles, red pentagons and blue-green squares will be discussed later.)   From Fig. \ref{fig:icsd_formation_vs_hull_distance}, it is clear that the vast majority of the reported compounds that have been experimentally synthesized (blue circles and green diamonds) lie on or close to the calculated convex hull --- 37 compounds are on the convex hull (i.e., a hull distance of 0~eV/atom) and an additional 52 lie relatively close to it (i.e., a hull distance less than about 0.1~eV/atom).  

The red pentagons represent \textit{XYZ} phases  that have been reported to exist at high pressure or high temperature. The blue-green squares represent experimentally reported \textit{XYZ} phases with partial site occupancies  e.g., RhFeAs is reported to have occupancies of (0.75 Fe, 0.25 Rh) on the $3f$, and (0.25 Fe, 0.75 Rh) on the $3g$ Wyckoff positions in the $P\overline{6}2m$ Fe$_2$P structure~\cite{guerin1979}, whereas the calculation in the OQMD corresponds to a structure in which $3f$ and $3g$ are respectively completely occupied by Fe and Rh).  Finally, the yellow circles represent phases that have not been experimentally synthesized, but have been sourced into the ICSD from previous first-principles calculations.     

The only two exceptions to the above observation that  experimentally reported \textit{XYZ} compounds (stoichiometric and ordered, at ambient conditions) have a hull distance less than about 0.1~eV/atom, are FeFeSn (at $\Delta E_f=0.156$ eV/atom) and MnTiAs (at $\Delta E_f=-0.408$ eV/atom). Both of these compounds have a calculated hull distance of about 0.2~eV/atom.  There appears to be some ambiguity about the exact composition of the former compound: FeFeSn in the $P63/mmc$ Ni$_2$In structure. The phase has been reported twice, once with the Fe$_2$Sn stoichiometry~\cite{djega1970}, and more recently with an Fe off-stoichiometry (Fe$_{1.68}$Sn)~\cite{vasilev1982}. MnTiAs has been reported in the $P\overline{6}2m$ Fe$_2$P structure, synthesized using a sealed-silica tube technique followed by annealing~\cite{johnson1973,RoyMontreuil1972813}. Since only its energy in a ferromagnetic configuration has been calculated in the OQMD, it is possible that other magnetic configurations may be energetically more favorable.

Overall, we find that most experimentally reported compounds in the \textit{XYZ} compositions considered here (90 of 99) have a hull distance less than about 0.1~eV/atom. Thus, even though a larger hull distance does not preclude the experimental realization of a compound, the likelihood of its synthesis and stability at ambient conditions is low.

We reiterate that while all the experimentally observed phases (blue circles and green diamonds in Fig.~\ref{fig:icsd_formation_vs_hull_distance}) might ideally be expected to lie on the calculated convex hull (i.e., with $\Delta E_{\rm HD}=0$), in practice, we find some as much as 0.1~eV/atom above it. As mentioned in Sec.~\ref{ssec:energetic_quantities}, the reasons for this include inaccuracies in DFT, actual metastability, and finite  temperature effects. We speculate that the latter may be very important because most of the phases in Fig.~\ref{fig:icsd_formation_vs_hull_distance} were synthesized at high temperatures, typically by arc melting or solid state diffusion followed by annealing. Even when properties are measured at low temperatures, if synthesis and processing are done at high temperatures, atomic positions and structures corresponding to the processing temperature may be ``frozen-in" for periods long compared to laboratory time scales.  Thus although the free energy at the processing temperature may be more relevant for determining relative phase stability we find empirically that the total energy as determined by DFT is a reasonable substitute with an uncertainty of about 0.1 eV/atom.    

Overall, it is clear from Fig.~\ref{fig:icsd_formation_vs_hull_distance} that the distance of a phase from the DFT-calculated zero temperature convex hull $\Delta E_{\rm HD}$ is a good indicator of the likelihood of its synthesis in experiments. This insight has important implications for the potential application of half-Heusler compounds. We expect that the further a compound lies from the convex hull, the less likely will be its successful synthesis, especially if the synthesis is limited to equilibrium processing.

\begin{figure}
\includegraphics[width=\columnwidth]{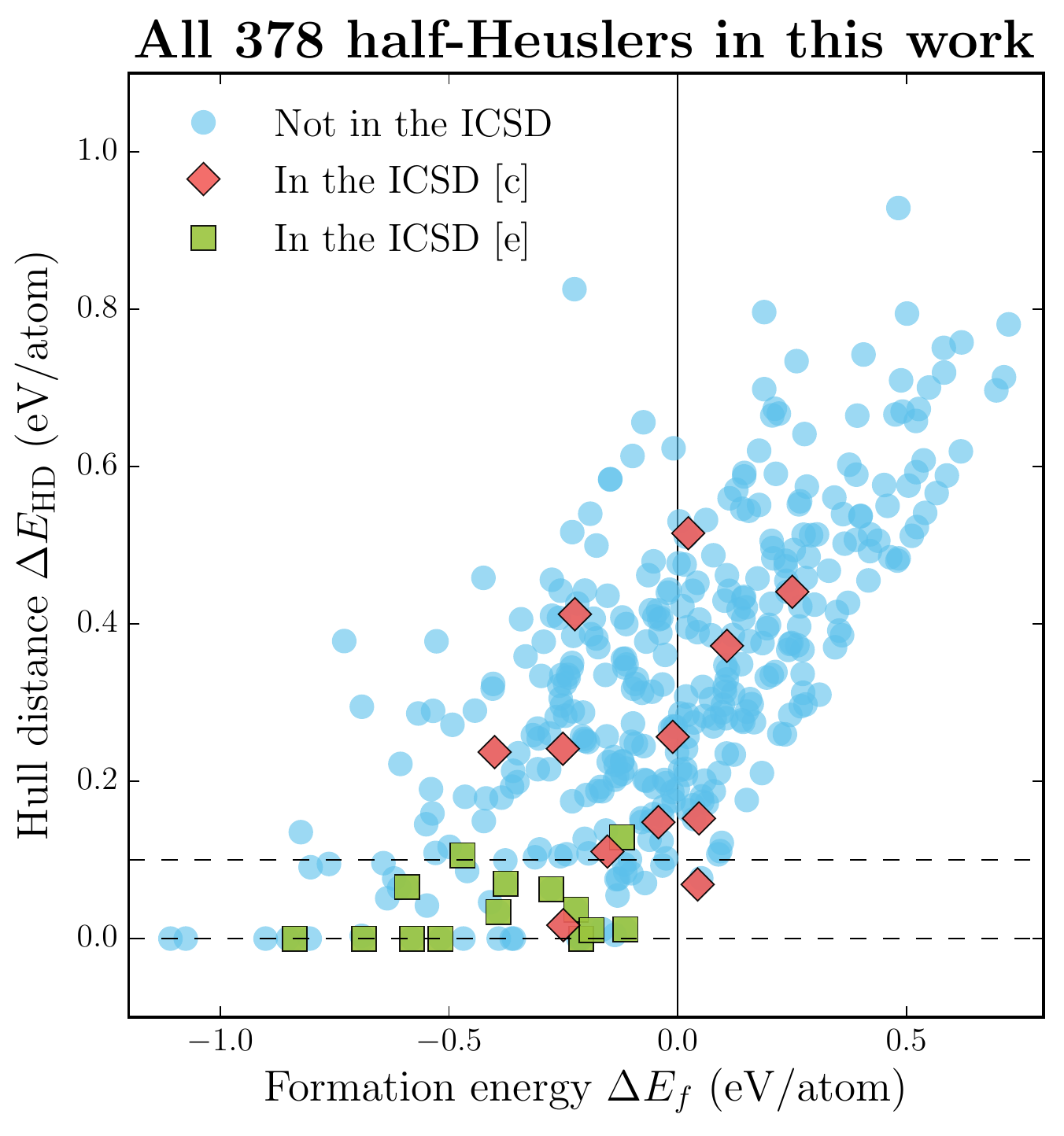}
\caption{DFT-calculated formation energy vs. hull distance of all the 378 \textit{XYZ} half-Heusler compounds considered in this work. A hull distance $\Delta E_{\rm HD} = 0$ indicates a stable ground state compound on the convex hull. Almost all the experimentally reported half-Heusler compounds (green squares, ``In ICSD [e]'') have a hull distance less than $\sim$0.1~eV/atom (the window represented by the two horizontal dashed lines); half-Heusler compounds sourced into the ICSD from previous computational work are represented by red diamonds (labeled ``In ICSD [c]'').}
\label{fig:hh_formation_vs_hull_distance}
\end{figure}

We extend the comparison of formation energies and hull distances to all the 378 half-Heusler compounds considered in this work (see Fig.~\ref{fig:hh_formation_vs_hull_distance}), and find that: (a) There is a large variation in formation energy of the half-Heusler compounds, ranging from $-1.1$~eV to 0.7 eV, with a large number (197) possessing a negative formation energy indicating stability against decomposition into constituent elements. (b) There is a relatively small number (24) of the half-Heusler compounds considered in this work that are reported in the ICSD. (c) As observed previously, of the half-Heusler compounds reported in the ICSD, almost all the experimentally synthesized ones (green squares, labeled ``In ICSD [e]'') lie on or close to the convex hull, with hull distances between 0.0 and about 0.1~eV/atom. (d) In contrast, most of the half-Heusler compounds in the ICSD sourced from previous calculations (red diamonds, labeled ``In ICSD [c]'') lie above the convex hull, with hull distances up to 0.5~eV/atom.  Overall, consistent with our previous observations for all compounds reported in the ICSD, we find that \textit{the distance of a half-Heusler compound from the convex hull is a good measure of the likelihood of its experimental synthesis}, and a hull distance of less than $\sim$0.1~eV/atom seems to be the corresponding approximate threshold. Our calculations predict about 50 (out of 378) half-Heusler compounds to be within the empirical threshold of a hull distance of $\sim$0.1~eV/atom, of which about 35 have not been previously reported. Further, we calculate 16 half-Heusler compounds considered in this work to lie on the convex hull ($E_{\rm HD} = 0$~eV/atom) of which (a) 6 have been reported in the $C1_b$ structure, (b) 6 have been reported in other structures ($Pnma$ and $P\overline{6}2m$ structures), and (c) 4 (RhTiP, RuVAs, CoVAs, CoTiAs) do not have any reported compounds at the composition in the ICSD. Thus, \textit{our calculations predict a number of new, hitherto unknown, half-Heusler compounds for further experimental investigation}. We discuss the properties of these predicted compounds in relevant later sections, Sec.~\ref{ssec:SP_semiconductors}--\ref{ssec:half_metallic_ferromagnets}.

\begin{figure*}[t]
\centering
\includegraphics[width=0.9\textwidth]{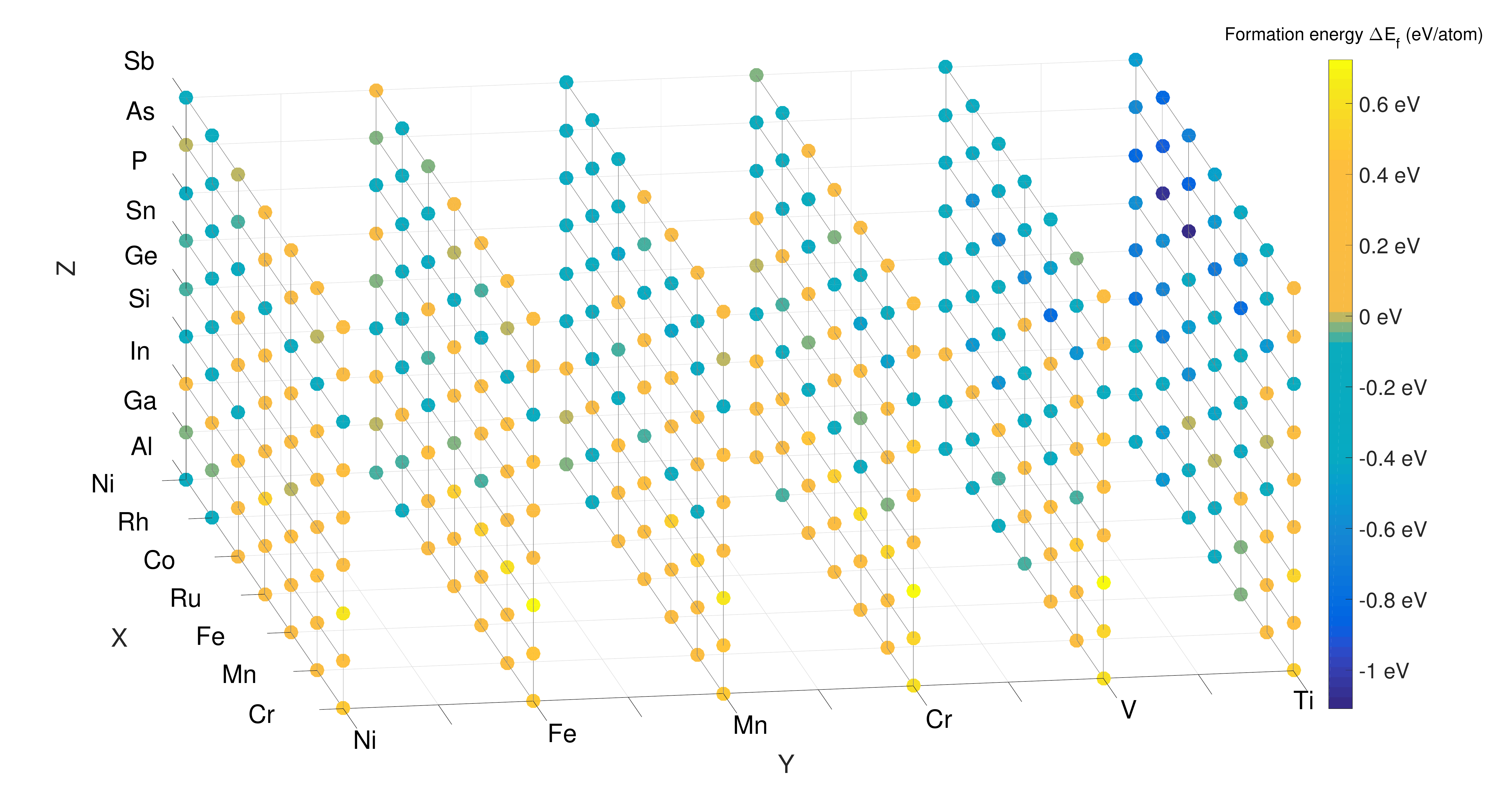}
\caption{DFT formation energies per atom for the 378 half-Heusler compounds considered in this work (colors represent the formation energy; blue and yellow = increasingly negative and positive formation energies respectively). The 3 coordinates represent the $X$, $Y$, and $Z$ species of the corresponding \textit{XYZ} compound.}
\centering
\label{fig:formation_energy_grid_xyz}
\end{figure*}

\begin{figure*}[t]
\centering
\includegraphics[width=0.9\textwidth]{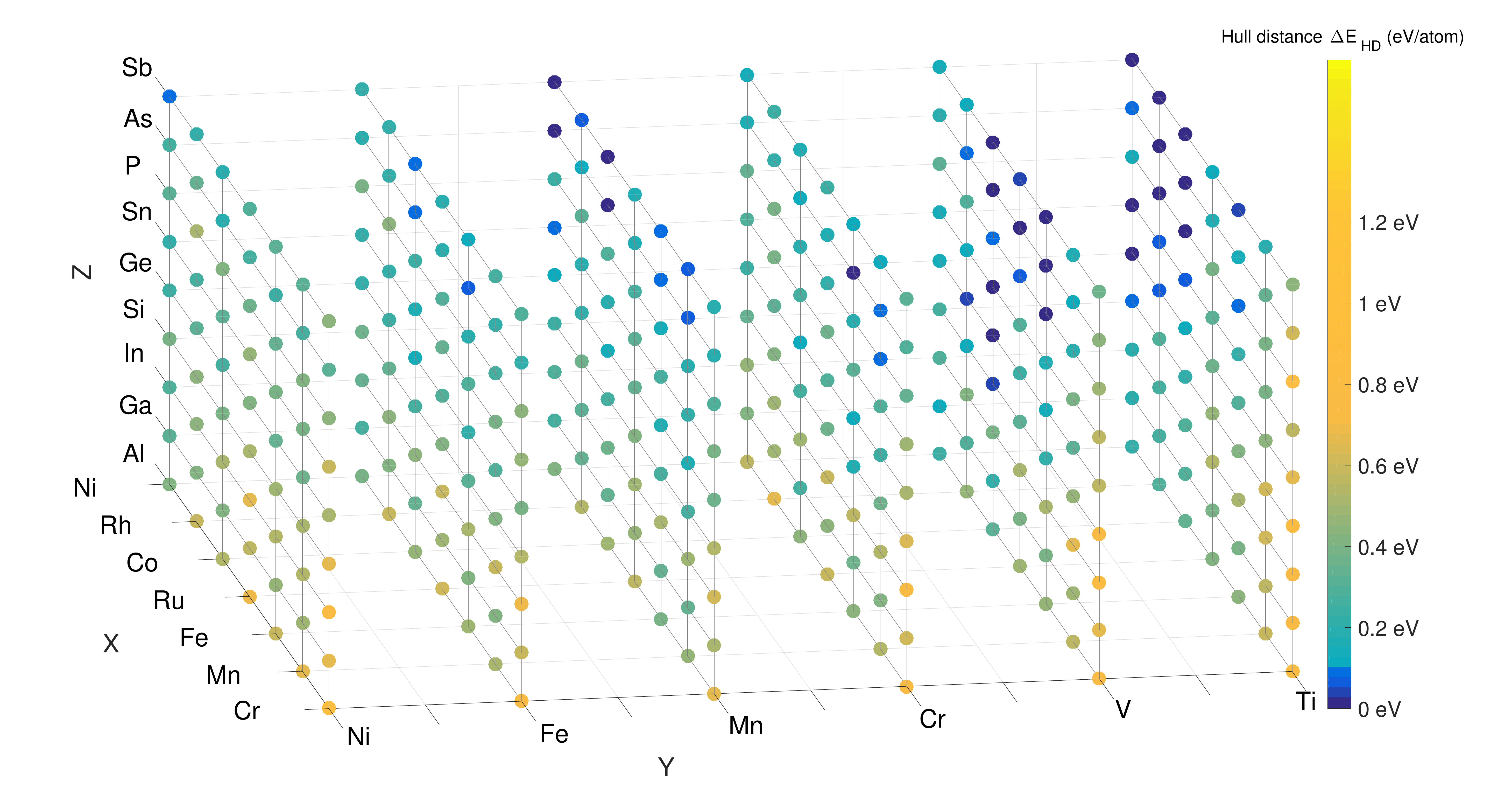}
\caption{DFT-calculated distances from the convex hull for the 378 half-Heusler compounds considered in this work (colors represent the hull distance; blue and yellow = increasingly close to and away from the convex hull respectively. All shades of blue represent hull distances $\Delta E_{\rm HD} \leq 0.1$~eV/atom). The 3 coordinates represent the $X$, $Y$, and $Z$ species of the corresponding \textit{XYZ} compound.}
\centering
\label{fig:hull_distance_grid_xyz}
\end{figure*}

\begin{figure}
\centering
\includegraphics[width=\columnwidth]{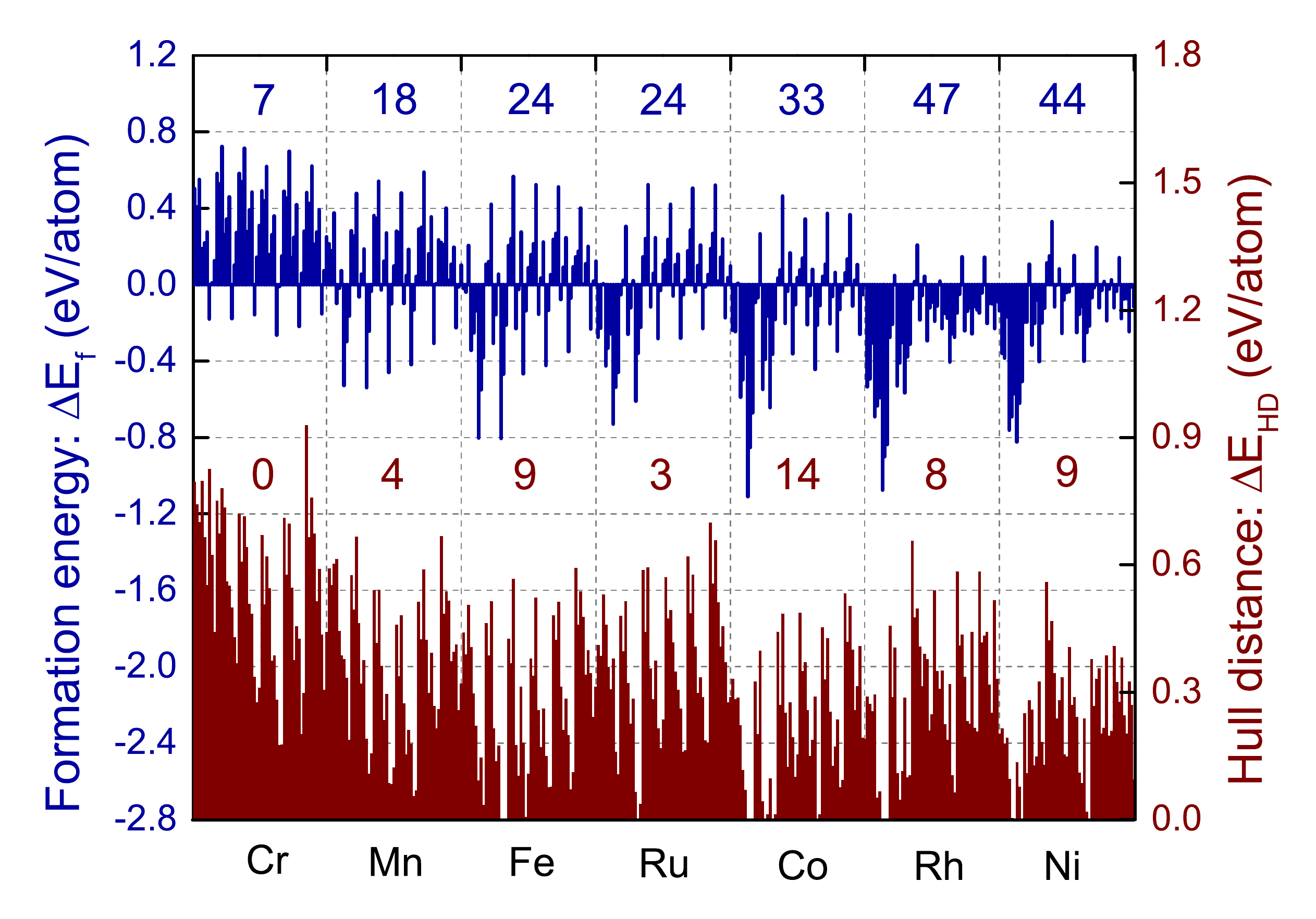}
\caption{DFT formation energies and hull distances for potential half-Heusler compounds grouped by the element on the $X$-site. The numbers near the top (in blue) and center (in brown) of each column denote the number of compounds with negative formation energy $\Delta E_f$ and hull distance $\Delta E_{\rm HD} \leq 0.1$~eV/atom, respectively, in the corresponding $Z$-element group. Within a given $X$-element column, the compounds are ordered first by the element on the $Y$-site (same order as in Fig.~\ref{fig:formation_energy_hull_distance_vs_y}) and then by the element on the $Z$-site (same order as in Fig.~\ref{fig:formation_energy_hull_distance_vs_z}), i.e., $Z$ varies more rapidly than $Y$.}
\label{fig:formation_energy_hull_distance_vs_x}
\end{figure}

\begin{figure}
\centering
\includegraphics[width=\columnwidth]{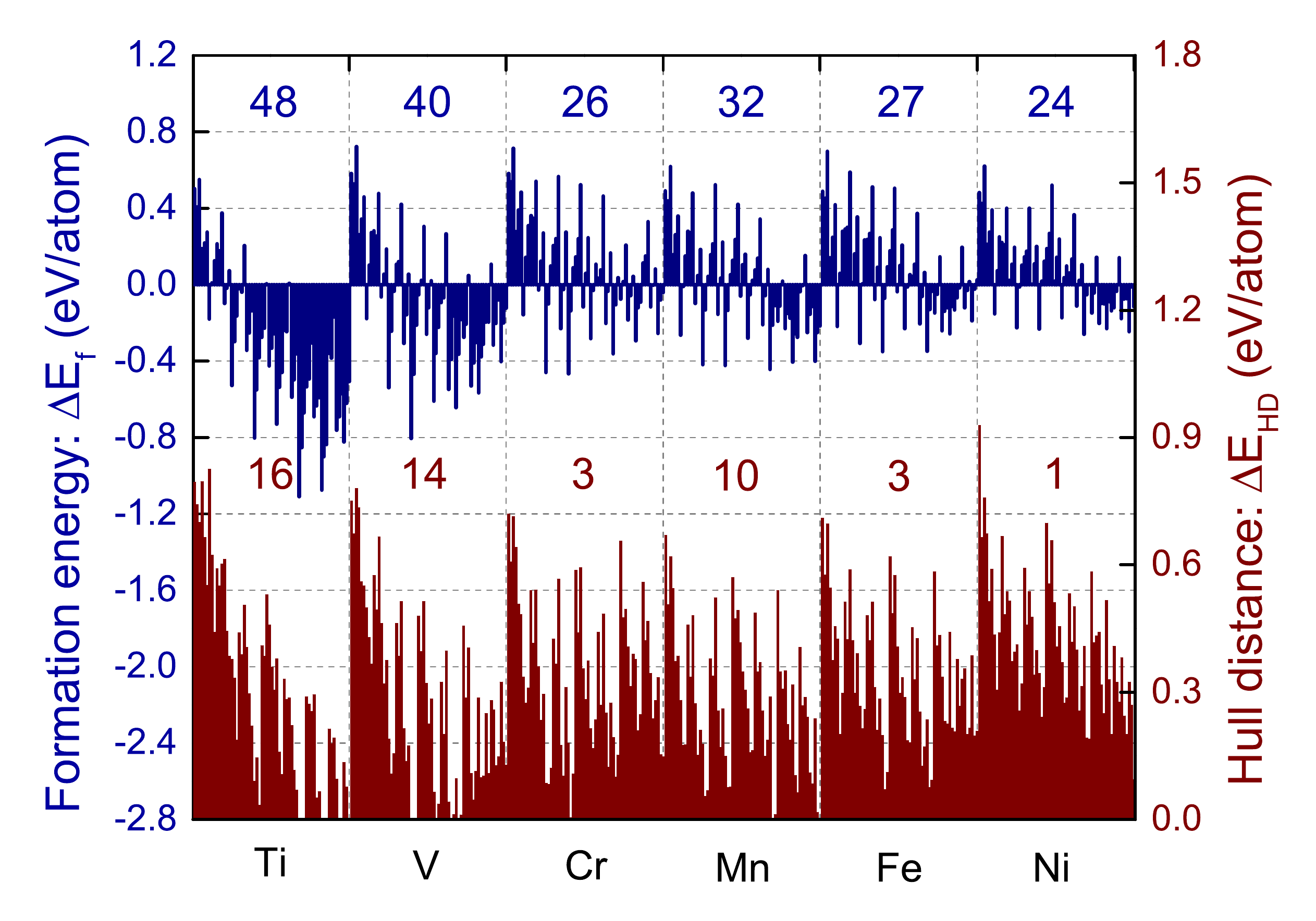}
\caption{DFT formation energies and hull distances for potential half-Heusler compounds grouped by the element on the $Y$-site. The numbers near the top (in blue) and center (in brown) of each column denote the number of compounds with negative formation energy $\Delta E_f$ and hull distance $\Delta E_{\rm HD} \leq 0.1$~eV/atom, respectively, in the corresponding $Y$-element group. Within a given $Y$-element column, the compounds are ordered first by the element on the $X$-site (same order as in Fig.~\ref{fig:formation_energy_hull_distance_vs_x}) and then by the element on the $Z$-site (same order as in Fig.~\ref{fig:formation_energy_hull_distance_vs_z}), i.e., $Z$ varies more rapidly than $X$.}
\label{fig:formation_energy_hull_distance_vs_y}
\end{figure}

\begin{figure}
\centering
\includegraphics[width=\columnwidth]{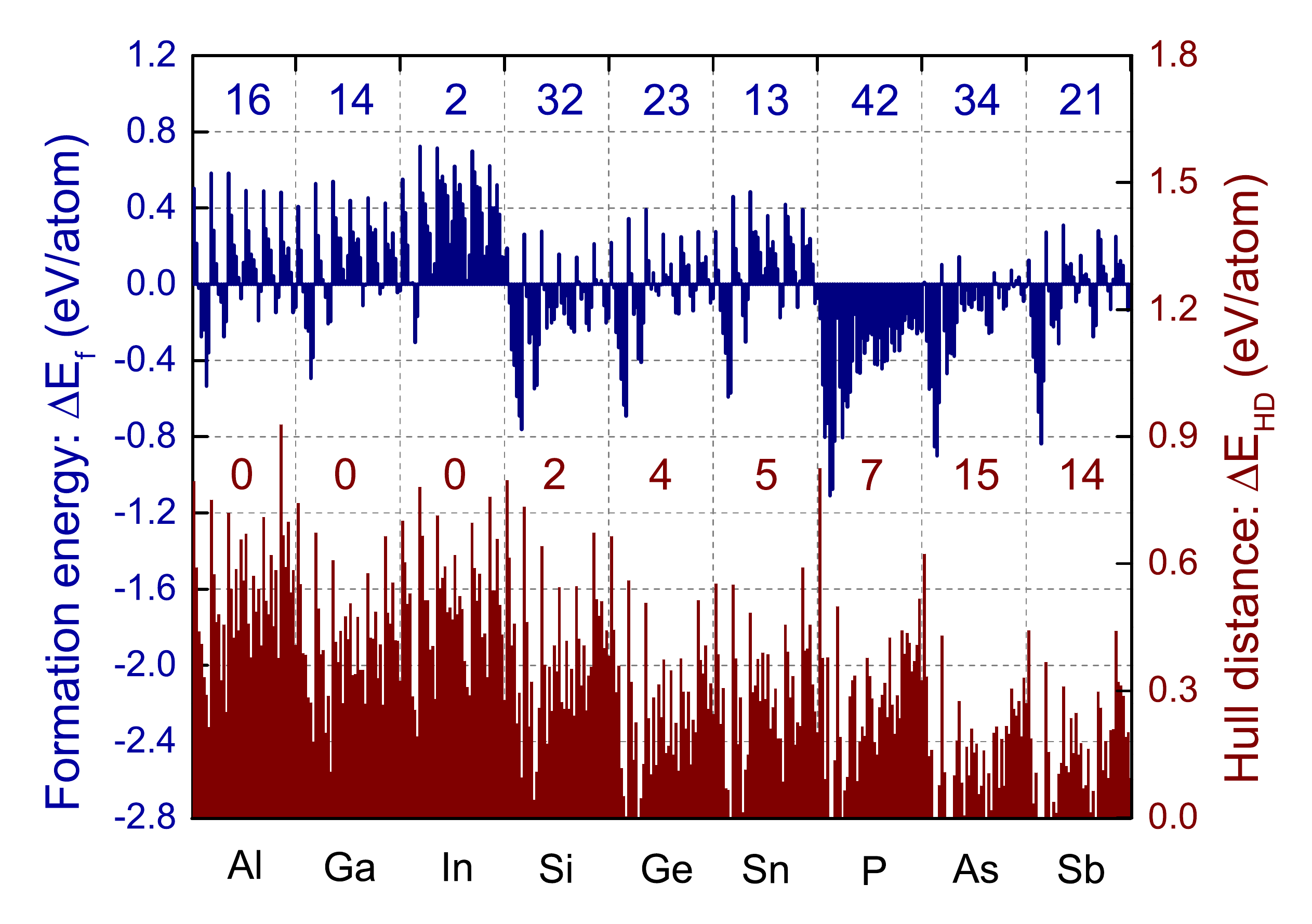}
\caption{DFT formation energies and hull distances for potential half-Heusler compounds grouped by the element on the $Z$-site. The numbers near the top (in blue) and center (in brown) of each column denote the number of compounds with negative formation energy $\Delta E_f$ and hull distance $\Delta E_{\rm HD} \leq 0.1$~eV/atom, respectively, in the corresponding $Z$-element group. Within a given $Z$-element column, the compounds are ordered first by the element on the $X$-site (same order as in Fig.~\ref{fig:formation_energy_hull_distance_vs_x}) and then by the element on the $Y$-site (same order as in Fig.~\ref{fig:formation_energy_hull_distance_vs_y}), i.e., $Y$ varies more rapidly than $X$.}
\label{fig:formation_energy_hull_distance_vs_z}
\end{figure}

We now analyze the variation in formation energies and hull distances with composition of all the 378 half-Heusler compounds in the $C1_b$ structure. All the formation energies are represented in a 3-dimensional plot in Fig.~\ref{fig:formation_energy_grid_xyz}, and the corresponding hull distances are plotted in Fig.~\ref{fig:hull_distance_grid_xyz}. It can be seen from the darker blue colors that compounds with $X=$ (Ni, Co, Rh), $Y=$ (V, Ti) and $Z=$ (group 5 elements P, As, Sb) tend to have lower formation energies and lie closer to or on the convex hull. This is largely consistent with known empirical rules for the stability of half-Heusler compounds, with the most electronegative (e.g., P, As, Sb) and the most electropositive elements (e.g., Ti) forming the NaCl-like sublattice, and the intermediate electronegative element (e.g., Co, Ni) occupying alternate tetrahedral sites~\cite{zeier2016}. To better illustrate the relation between structural stability and composition, we arranged the formation energy and hull distance data according to $X$, $Y$, and $Z$ element respectively in Figs.~\ref{fig:formation_energy_hull_distance_vs_x}--\ref{fig:formation_energy_hull_distance_vs_z}.  

From Fig.~\ref{fig:formation_energy_hull_distance_vs_x}, we see that Co, Rh, and Ni on the $X$-site form more compounds with negative formation energy (and more compounds with smaller hull distances) than other elements on the $X$-site. From Fig.~\ref{fig:formation_energy_hull_distance_vs_y}, it can be seen that Ti and V on the $Y$-site form more compounds with negative formation energies and smaller hull distances than other elements. Similarly, when the compounds are ordered by the element on the $Z$-site as shown in Fig.~\ref{fig:formation_energy_hull_distance_vs_z}, one can see that there is a trend for formation energy to decrease and stability to increase with group number, i.e., compounds with group 5 elements on the $Z$-site are in general more stable than those with group 4 elements which are more stable than those with group 3 elements on the $Z$-site. For a fixed number of valence electrons on the $Z$ atom, i.e. 3, 4 or 5, the formation energy is typically lower for the smaller atom, (i.e. Al, Si, P).  However, the trend is different for the  hull distance with the larger atoms, especially Sn and Sb leading to greater stability relative to other phases.  
The large number of compounds with very low formation energies with $Z=$ P is striking in Fig.~\ref{fig:formation_energy_hull_distance_vs_z}, but can also be observed in Figs.~\ref{fig:formation_energy_hull_distance_vs_x} and \ref{fig:formation_energy_hull_distance_vs_y} as the periodic pattern of low formation energy compounds, i.e., compounds 7, 16, 25, 34, etc. for fixed $X$ in Fig.~\ref{fig:formation_energy_hull_distance_vs_x} or fixed $Y$ in Fig.~\ref{fig:formation_energy_hull_distance_vs_y}. Although the  $Z=$ P compounds have lower formation energies, there are more compounds with $Z=$ (As, Sb) that lie on or closer to the convex hull (hull distance close to 0~eV/atom).  

We speculate that the smaller $Z$ atom allows for smaller interatomic distances  increasing the binding and reducing the total energy, however the relatively open structure of the $C1_b$ phase means that there are fewer nearest neighbor bonds, so that the C1$_b$ phase is at a disadvantage compared to more closely packed phases. Thus, while low formation energies sometimes correspond to low hull distances, a low formation energy is not sufficient to ensure the thermodynamic stability of a compound, which further depends on other competing phases in the chemical space.

We also observed that group 3 elements (Al, Ga, In) on the $Z$-site yield many compounds with distorted structures lower in energy than the cubic or tetragonal structures. For $Z=$ Al, Ga, and In, the corresponding numbers of compounds with a lower-energy distorted structure are 19, 18, and 22 respectively. Thus, most of the 72 compounds that relaxed to a distorted phase had a group 3 element on the $Z$-site.

Prompted by the trend of decreasing formation energy with decreasing the atomic number of the element on the $Y$-site in Fig.~\ref{fig:formation_energy_hull_distance_vs_y}, we investigated $Y=$ Sc and found three additional semiconductors (NiScP, NiScAs, and NiScSb) with a negative formation energy (listed in Table~\ref{tab:SP_semiconductors} but not included in Figs.~\ref{fig:formation_energy_hull_distance_vs_x}--\ref{fig:formation_energy_hull_distance_vs_z}). The NiSc(P,As,Sb) compounds have lower formation energies than the corresponding CoTi(P,As,Sb) compounds. We also found two additional $C1_{b}$ half-metals (CrScAs and CrScSb) with a negative formation energy. These are listed in Table~\ref{tab:half_metallic_ferromagnets} but not included in Figs.~\ref{fig:formation_energy_hull_distance_vs_x}--\ref{fig:formation_energy_hull_distance_vs_z}). 

Another interesting trend among the half-Heusler compounds is the preference for the transition metal with the larger atomic number to occupy the $X$-site ($\frac{1}{4}$,$\frac{1}{4}$,$\frac{1}{4}$) in an \textit{XYZ} Heusler compound. Our calculations included formation energies for CrMn$Z$, CrFe$Z$, CrNi$Z$, and MnFe$Z$, which can be compared directly with the formation energies calculated for MnCr$Z$, FeCr$Z$, NiCr$Z$, and FeMn$Z$ where $Z$ represents one of 9 non-transition metal atoms considered. Of these 36 pairs we found no exceptions to the rule that the energy is lower if the atomic number of the element on the $X$-site is larger than that of the element on the $Y$-site. Of course, this rule may be violated if structures other than $C1_b$ are considered. For example, CrNiAl and CrNiIn had lower formation energies than NiCrAl and NiCrIn, respectively, because the lowest energy relaxed structures for CrNiAl and CrNiIn were distorted triclinic cells. If comparisons are restricted to all compounds in the C1$_b$ structure, then the above observation of lower formation energy corresponding to the transition metal atom with the larger atomic number occupying the $X$-site is consistently true (at least for the 36 pairs we considered).                      

In addition, the formation of band gaps plays an important role in structural stability. From Figs.~\ref{fig:formation_energy_hull_distance_vs_x}--\ref{fig:formation_energy_hull_distance_vs_z}, it can be seen that there are five compounds (CoTiP, RhTiP, CoTiAs, RhTiAs, RhTiSb) with formation energies less than $-0.83$~eV/atom. All of them have Co or Rh on the $X$-site, Ti on the $Y$-site, and P, As or Sb on the $Z$-site, consistent with previous observations in this section, but another common characteristic of these five compounds is that they are all 18-electron Slater-Pauling semiconductors, with 3 electrons per atom in both spin channels. We speculate that a gap in one spin channel at the Fermi energy contributes to the stability of the compound, and that gaps in both spin channels contribute even more to stability, resulting in the compounds with the lowest formation energies in our database. 

The next 18 compounds in the order of increasing formation energy have Co, Rh, Ru, Fe or Ni on the $X$-site, Ti (13 out of 18) or V on the $Y$-site, and a group 5 element (P, As, Sb) or group 4 element (Si or Ge) on the $Z$-site. All except four (RuTiP, RhTiSi, RhVP, CoTiSi, which have competing lower-energy $Pnma$ or $P\overline{6}2m$ phases) lie on or close to the convex hull with hull distances lower than $\sim$0.1~eV/atom. Of the 18 compounds, 7 (FeVP, NiTiSi, NiTiGe, NiTiSn, CoTiSb, RuVP, CoVSi) are also 18-electron Slater-Pauling semiconductors and are listed in Table~\ref{tab:SP_semiconductors}.

\begin{figure}
\centering
\includegraphics[width=\columnwidth]{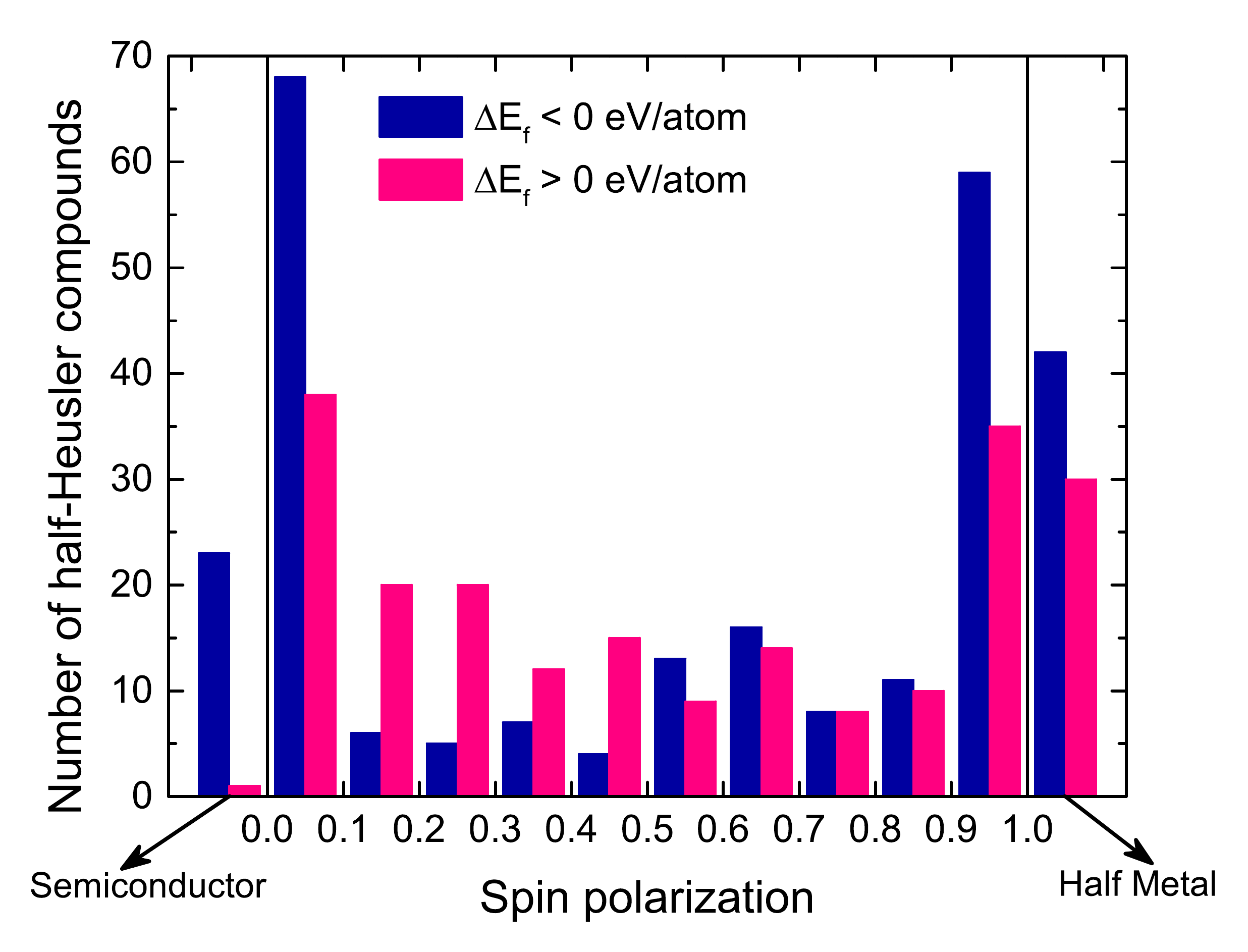}
\caption{The distribution of half-Heusler compounds with negative ($\Delta E_{f} < 0$~eV/atom) and positive ($\Delta E_{f} > 0$~eV/atom) formation energies as a function of spin polarization $\mathcal{P}(E_F)$ (given by Eq.~\ref{eqn:spin_polarization}). In the central region, we show the number of half-Heusler compounds grouped by 10 percentage points of spin polarization. In an additional region to the left, we show the 24 semiconductors, including 23 compounds with a negative formation energy and 1 with a positive formation energy. In the additional region on the right, we show 72 half-metals, including 42 and 30 with negative and positive formation energies respectively.}
\label{fig:formation_energy_vs_polarization}
\end{figure}

\begin{figure}
\includegraphics[width=\columnwidth]{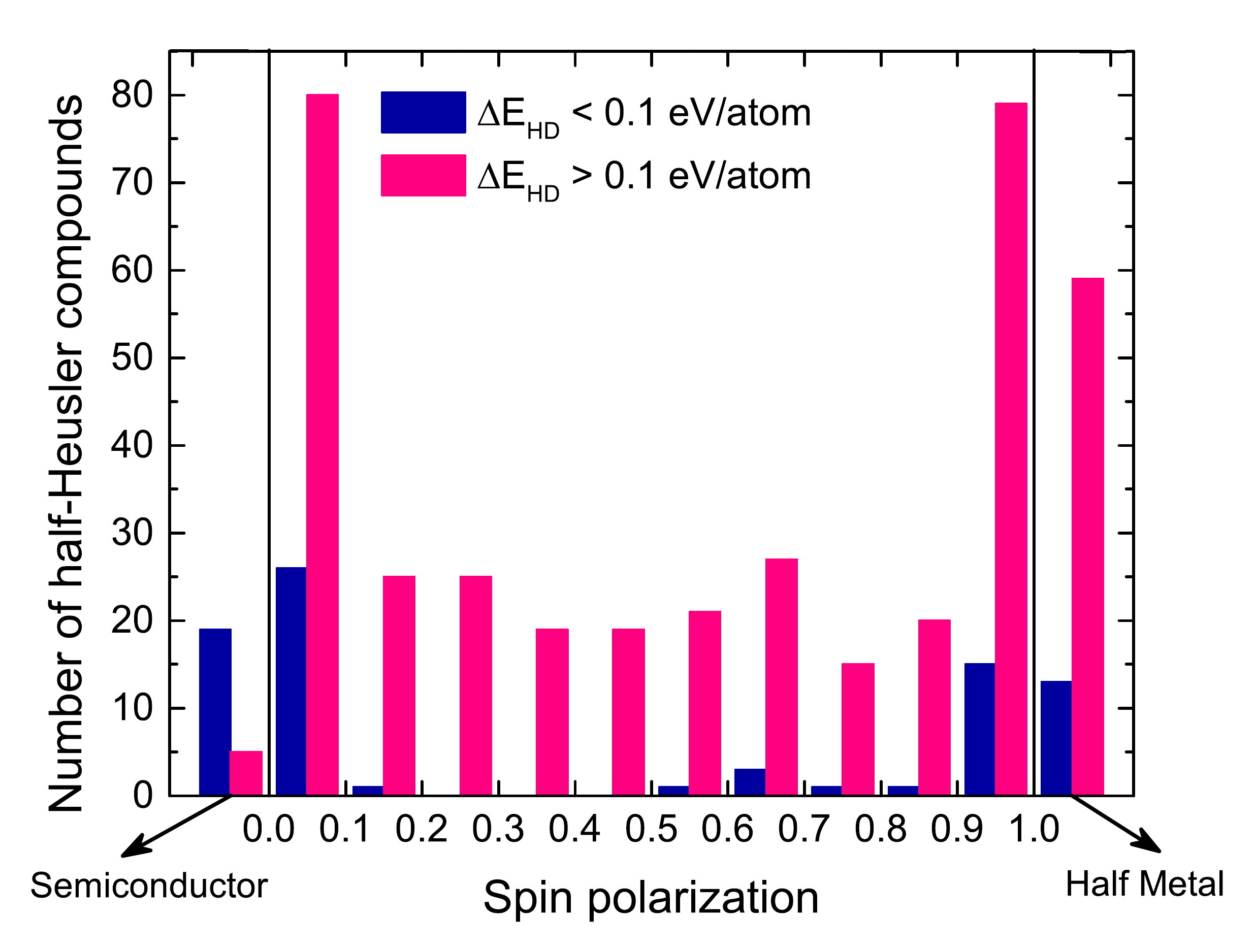}
\caption{The distribution of half-Heusler compounds that lie on or close to ($\Delta E_{\rm HD} < 0.1$~eV/atom) and far away ($\Delta E_{\rm HD} > 0.1$~eV/atom) from the convex hull, as a function of spin polarization $\mathcal{P}(E_F)$ (given by Eq.~\ref{eqn:spin_polarization}). In the central region, we show the number of half-Heusler compounds grouped by 10 percentage points of spin polarization. We show the 24 semiconductors (of which 19 compounds have $E_{\rm HD} < 0.1$~eV/atom) in the additional region to the left, and show the 72 half-metals (of which 15 compounds have $E_{\rm HD} > 0.1$~eV/atom). Clearly, the existence of a gap at the Fermi level in one or both spin channels contributes to the stability of a half-Heusler compound.}
\centering
\label{fig:hull_distance_vs_polarization}
\end{figure}

We calculated the electronic structure of each compound and obtained its spin polarization $\mathcal{P}$ at Fermi level $E_{F}$ using:
\begin{equation}\label{eqn:spin_polarization}
\mathcal{P}(E_{F})=\frac{N_{\uparrow}(E_{F})-N_{\downarrow}(E_{F})}{N_{\uparrow}(E_{F})+N_{\downarrow}(E_{F})}
\end{equation}
where $N_{\uparrow}$ and $N_{\downarrow}$ are the densities of states for majority (spin-up) and minority (spin-down) electrons, respectively. The distribution of spin polarization $\mathcal{P}(E_F)$ of the 378 half-Heusler compounds separated into those with positive and negative formation energies (and similarly, with hull distances greater than and less than $\sim$0.1~eV/atom) is shown in Figs.~\ref{fig:formation_energy_vs_polarization} and \ref{fig:hull_distance_vs_polarization}. A correlation between a negative formation energy and gaps at the Fermi energy is apparent. In particular, we have separated on the left in Fig.~\ref{fig:formation_energy_vs_polarization} 24 compounds with exactly zero polarization that are semiconductors. Only one of these has a positive formation energy. We have also separated on the right, 72 compounds that are fully spin polarized, i.e., they are half-metals. The majority (42) of these half-metals have negative formation energies. In fact, a majority of the near half-metals also have negative formation energies. The contribution of gaps at Fermi energy to stability of a compound is  even more striking in Fig.~\ref{fig:hull_distance_vs_polarization}. Almost all the half-Heusler compounds that are on or close to the convex hull (within $\sim$0.1~eV/atom of it) are either semiconductors (or near semiconductors with close to $\mathcal{P}(E_F) = 0$), or half-metals (or near half-metals with close to $\mathcal{P}(E_F) = 1$). In other words, having a gap (or even almost a gap) at the Fermi energy in one or both spin channels seems to contribute greatly to the stability of a compound, consistent with our previous observations. 

Among the half-Heusler compounds considered in this work, we identify a total of 27 18-electron semiconductors and 45 half-metals with negative formation energies. We discuss semiconductors in Sec.~\ref{ssec:SP_semiconductors} and half-metals in Secs.~\ref{ssec:zero_moment_half_metals} and \ref{ssec:half_metallic_ferromagnets}. 

\subsection{Slater-Pauling Semiconductors}
\label{ssec:SP_semiconductors}

Table \ref{tab:all_18_sp_compounds} lists the 60 half-Heuslers in our database with 18 valence electrons per formula unit. The row labels give the $X$ atoms and their number of valence electrons. The column labels give the $Y$ atoms and their number of valence electrons. Each entry in the table represents 3 compounds. Thus CoV4 represents CoVSi, CoVGe, or CoVSn. Note that the final number (e.g., the 4 in CoV4) is actually redundant since the total number of valence electrons is 18.

\begin{table}
\caption{List of the 60 18-electron half-Heusler compounds considered in this work. Row and column labels indicate the atom on the $X$- and $Y$-sites respectively. Compounds in bold (and underlined) are Slater-Pauling semiconductors with no moment on any atom.  \textit{Z} = 5 means \textit{Z} = P, As, or Sb; \textit{Z} = 4 means \textit{Z} = Si, Ge, or Sn; \textit{Z} = 3 means \textit{Z} = Al, Ga, or In.}
\label{tab:all_18_sp_compounds}
\begin{tabular}{ccccccc}
\toprule
$X$/$Y$& 3Sc & 4Ti & 5V & 6Cr & 7Mn & 8Fe \\
\midrule
10Ni & \textbf{\underline{NiSc5}} & \textbf{\underline{NiTi4}} & \textbf{\underline{NiV3}} & 	   &  	   &  \\
 9Co &  						  & \textbf{\underline{CoTi5}} & \textbf{\underline{CoV4}} & CoCr3 &  	   &  \\
 9Rh &  						  & \textbf{\underline{RhTi5}} & \textbf{\underline{RhV4}} & RhCr3 &  	   &  \\
 8Fe &  						  &  						   & \textbf{\underline{FeV5}} & FeCr4 & FeMn3 &  \\
 8Ru &  						  &  						   & \textbf{\underline{RuV5}} & RuCr4 & RuMn3 &  \\
 7Mn &  						  &  						   &  						   & MnCr5 & MnMn4 & MnFe3 \\
 6Cr &  						  &  						   &  						   &  	   & CrMn5 & CrFe4 \\
\bottomrule
\end{tabular}

\end{table}

For the 27 18-electron half-Heusler compounds that are Slater-Pauling semiconductors, we tabulate the DFT-calculated properties such as lattice constant, band gap, gap type, formation energy, and hull distance in Table~\ref{tab:SP_semiconductors}. Remarkably all of the 27 systems with Sc, Ti, or V as on the $Y$-site have similar electronic structure in that they are in Slater-Pauling states with zero magnetic moment on all atoms and are therefore semiconductors (band gaps are given in Table~\ref{tab:SP_semiconductors}). It is also remarkable that the decrease in energy due to the creation of the gaps in both spin channels associated with the Slater-Pauling state is sufficient to eliminate the magnetic moment on all atoms including those that are usually found to be magnetic (e.g., Ni, Co, Fe).

\afterpage{
\begin{table*}
\caption{For each of the 27 18-electron \textit{XYZ} half-Heusler compounds that are Slater-Pauling semiconductors, we list the calculated lattice constant $a$, band gap $E_g$ within DFT, the type of gap, the formation energy of the compound in the $C1_b$ and $Pnma$ structures, distance from the convex hull for the $C1_b$ phase $\Delta E_{\rm HD}^{C1_b}$, previous experimental reports, whether or not a full-Heusler $X_2YZ$ phase (in the $L2_1$ structure) has been observed, and the formation energy of the observed full-Heusler $L2_1$ phase where applicable. ([Legend] Gap type: \orange{D} = direct, \blue{I} = indirect band gap.)}
\label{tab:SP_semiconductors}
\begin{tabular}{|l|c|c|c|c|c|c|c|c|c|}
\toprule
\textit{XYZ} & $a$ & $E_g$ & Gap & $\Delta E_f^{C1_b}$ & $\Delta E_f^{Pnma} $ & $\Delta E_{\rm HD}^{C1_b}$ & Experimental & $X_2YZ$ & $E_f^{L2_1}$ \\
 & ({\AA}) & (eV) & type & \multicolumn{3}{c|}{(eV/atom)} & reports & reports & (eV/atom) \\
\midrule
NiScP   & 5.67  & 0.57 &  \orange{D}  & $-1.139$ & $-1.308$ & 0.169 & $Pnma$\cite{Kleinke1998218} & & \\ 
NiScAs  & 5.82  & 0.46 &  \orange{D}  & $-0.994$ & $-1.029$ & 0.035 & & & \\ 
NiScSb  & 6.10  & 0.25 &  \orange{D}  & $-0.911$ & $-0.773$ & 0 & $F\overline{4}3m$\cite{pecharskii1983crystal,Dwight_Haschke_Eick_1974} & & \\
& & & & & & & & & \\
NiTiSi  & 5.56  & 0.74 & \blue{I} & $-0.762$ & $-0.846$ & 0.084 & $Pnma$\cite{landrum1998tinisi,Shoemaker:a04664} & & \\
NiTiGe  & 5.65  & 0.62 & \blue{I} & $-0.691$ & $-0.684$ & 0     & $Pnma$\cite{PSSA:PSSA2210660104} & & \\ 
NiTiSn  & 5.93  & 0.44 & \blue{I} & $-0.571$ & $-0.392$ & 0     & $F\overline{4}3m$\cite{Stadnyk20103023,jeitschko1970transition,Pierre199495} & \cite{VANENGEN1983374,Pierre199495} & $-0.473$ \\ 
& & & & & & & & & \\
CoTiP   & 5.43  & 1.39 & \blue{I} & $-1.109$ & $-1.245$ & 0.136     & $Pnma$\cite{lomnitskaya1981interaction} & & \\ 
CoTiAs  & 5.61  & 1.29 & \blue{I} & $-0.852$ & $-0.804$ & 0     & $Pnma$\cite{Rundqvist1967} & & \\ 
CoTiSb  & 5.88  & 1.06 & \blue{I} & $-0.670$ & $-0.415$ & 0     & $F\overline{4}3m$\cite{C0DT00742K,szytula1973crystal} & &\\ 
& & & & & & & & & \\
RhTiP   & 5.74  & 0.87 & \blue{I} & $-1.075$ & $-1.258$ & 0.183     & & &\\ 
RhTiAs  & 5.87  & 0.85 & \blue{I} & $-0.901$ & $-0.890$ & 0     & $Pnma$\cite{roymontreuil1979influence,deyris1974structural} & & \\ 
RhTiSb  & 6.12  & 0.75 & \blue{I} & $-0.837$ & $-0.621$ & 0     & $F\overline{4}3m$\cite{Dwight1974279} & &\\ 
& & & & & & & & & \\
NiVAl   & 5.57  & 0.10 &  \orange{D}  & $-0.196$ & $-0.164$ & 0.230 & & \cite{Marazza1975341} & $-0.389$ \\ 
NiVGa   & 5.55  & 0.30 &  \orange{D}  & $-0.195$ & $-0.270$ & 0.108 & & \cite{Ni2VGaMarkiv} & $-0.287$ \\ 
NiVIn   & 5.84  & 0.26 &  \orange{D}  & 0.106    & 0.078    & 0.280 & & &\\ 
& & & & & & & & & \\
CoVSi   & 5.41  & 0.55 & \blue{I} & $-0.548$ & $-0.584$ & 0.036 & $Pnma$\cite{conrad2005refinement,landrum1998tinisi} & \cite{Co2Vsi} & $-0.424$ \\ 
CoVGe   & 5.50  & 0.68 & \blue{I} & $-0.391$ & $-0.300$ & 0     & $Pnma$\cite{Jeitschko1969} & & \\ 
CoVSn   & 5.79  & 0.65 & \blue{I} & $-0.164$ & 0.046    & 0.012 & & \cite{Carbonari1996313,VANENGEN1983374} & $-0.092$ \\  
& & & & & & & & & \\
RhVSi   & 5.69  & 0.31 & \blue{I} & $-0.529$ & $-0.657$ & 0.109 & & & \\ 
RhVGe   & 5.77  & 0.43 & \blue{I} & $-0.410$ & $-0.391$ & 0.046 & & & \\ 
RhVSn   & 6.04  & 0.39 & \blue{I} & $-0.302$ & $-0.244$ & 0.113 & & \cite{Suits1976423} & $-0.349$ \\ 
& & & & & & & & & \\
FeVP    & 5.31  & 0.32 & \blue{I} & $-0.804$ & $-0.907$ & 0.103     & $Pnma$\cite{RoyMontreuil1972813} & &\\ 
FeVAs   & 5.49  & 0.37 & \blue{I} & $-0.468$ & $-0.357$ & 0     & $P\overline{6}2m$\cite{RoyMontreuil1972813} & &\\ 
FeVSb   & 5.78  & 0.38 & \blue{I} & $-0.211$ & 0.083    & 0     & $F\overline{4}3m$\cite{kripyakevich1963crystal,Stadnyk200530} & &\\
& & & & & & & $P6_{3}/mmc$\cite{Kong20113003,Evers199793} & & \\ 
RuVP    & 5.62  & 0.19 & \blue{I} & $-0.609$ & $-0.767$ & 0.158 & & & \\ 
RuVAs   & 5.76  & 0.24 & \blue{I} & $-0.358$ & $-0.295$ & 0     & & & \\ 
RuVSb   & 6.02  & 0.20 & \blue{I} & $-0.222$ & $-0.031$ & 0.037 & $F\overline{4}3m$\cite{Evers199793} && \\    
\bottomrule
\end{tabular}

\end{table*}
}

We suggest that the absence of moments in these materials results from the atoms on the $Y$-site (Sc, Ti, and V) being difficult to magnetically polarize. Since the total moment in the Slater-Pauling state for an 18-electron half-Heusler must be zero, zero moment on the atoms on the $Y$-site (due to the broad, high-lying $d-$states of Sc, Ti, and V) and the $Z$-site (which is even more difficult to polarize than the atom on the $Y$-site) implies zero moment on the atom on the $X$-site. Another important factor in their stability may be that the large difference in the number of $d$-electrons in the atoms on the $X$- and $Y$-sites leads to a large difference in the on-site energies of the $d$-states, which in turn contributes to relatively large energy gaps. Slater and Koster showed in 1954 that for $X$-$Y$ compounds with only nearest-neighbor interactions involving $d$-states, there will be no states between the $d$-onsite energies of the atoms on the $X$- and $Y$-sites~\cite{PhysRev.94.1498}.

If our entire dataset of 384 $C1_b$ half-Heusler compounds is ordered by calculated formation energies (see Table~\ref{tab:SP_semiconductors}), the ``lowest-formation energy'' list is dominated by 18-electron semiconductors. A low formation energy does not, however, guarantee stability against other phases with still lower energy. The experimental literature suggests that the most common competing phase for the 18-electron half-Heusler semiconductors is the orthorhombic MgSrSi-type $Pnma$ phase with 4 formula units per cell. We calculate the formation energies for these competing $Pnma$ phases, list them in Table \ref{tab:SP_semiconductors} for comparison with the formation energy of the corresponding $C1_b$ phases, and include them in the construction of the respective convex hulls. Interesting patterns can be observed in the formation energies of the $C1_b$ semiconductors and those of the competing $Pnma$ compounds: As one proceeds from left to right and top to bottom in Table \ref{tab:all_18_sp_compounds} (or alternatively as the difference in the number of valence electrons of the atoms on the $X$- and $Y$- sites decreases) the formation energies tend to increase (stability against decomposition into elements decreases) for both $C1_b$ and $Pnma$ phases. In addition, for a given $X$ and $Y$, the formation energies increase as the size of the atom on the $Z$-site increases, thus the formation energy increases from \textit{XY}P to \textit{XY}As to \textit{XY}Sb, and from \textit{XY}Si to \textit{XY}Ge to \textit{XY}Sn. This increase in formation energy is faster for the $Pnma$ compounds than for the $C1_b$ semiconductors so that in most cases, for a given $X$ and $Y$, the $Pnma$ compound has the lower energy for $Z=$ P or Si. When $Z=$ As or Ge, the formation energies are similar, and when $Z=$ Sb or Sn, the $C1_b$ semiconductor has the lower energy. 

The NiV(Al,Ga,In) sequence is an exception to the above pattern due to the relatively high formation energies of both phases ($C1_b$ and $Pnma$) and the low formation energy of the Ga phases relative to the Al phases. Our calculated formation energies indicate that neither of the $C1_b$ or $Pnma$ compounds would be an equilibrium phase at low temperature due to the very low formation energy of competing binary phases -- NiAl ($B2$), Ni$_2$Ga$_3$ and Ni$_2$In$_3$ (Al$_3$Ni$_2$ structure type), respectively. In fact, we predict NiVIn to have a positive formation energy in both the $Pnma$ and semiconducting $C1_b$ phases. NiVIn is the only one of the 27 semiconducting $C1_b$ compounds that we find to have a positive formation energy.

The trend of the formation energy of the $Pnma$ phase being lower than that of the $C1_b$ phase for the smaller $Z$ atoms, and vice versa for the larger $Z$ atoms, may be explained by the atomic size of the $Z$ atom and its effect on the lattice constant. For the $C1_b$ phase, the $X$ atom has 4 nearest neighbors that are $Z$ atoms and 4 that are $Y$ atoms, all at the same distance. The $X$--$Z$ distances in the $Pnma$ structure are smaller than in the $C1_b$ structure for small $Z$ atoms, but about the same for the larger ones. However, for the large $Z$ atoms, the $X$--$Y$ distances in the $C1_b$ structure are significantly smaller than in the $Pnma$ structure. Thus, in the case of large $Z$ atoms, the smaller $X$--$Y$ distances in the $C1_b$ structure seem to result in stronger interatomic binding when compared to the corresponding $Pnma$ phase, leading to lower formation energies.

The hull distance is determined by the energy differences of the C1$_b$ and \textit{Pnma} phases for most of these systems.  Exeptions are the NiV3 and RhV4 systems for which OQMD predicts that both the C1$_b$ and Pnma phases are undercut by a mixture of binaries (NiAl+V for the case of NiVAl).  For CoVSn and RuVSb, the C1$_b$ phase is almost degenerate with a mixture of binaries.

Overall, there is good agreement between theory and experiment displayed in Table \ref{tab:SP_semiconductors}. First, all the six half-Heusler compounds that have been experimentally synthesized (NiScSb, NiTiSn, CoTiSb, RhTiSb, FeVSb, and RuVSb) are predicted to lie on or close to the convex hull, with the energy of the $C1_b$ structure correctly predicted to be lower than that of the $Pnma$ structure in all cases. Of the six compounds, all except RuVSb are predicted to lie \textit{on} the convex hull. RuVSb is predicted to have a small hull distance $E_{\rm HD} = 0.037$~eV/atom with a linear combination of binaries (RuV$_3$--RuV--RuSb$_2$) in the OQMD predicted to be lower in energy. Second, of the systems for which the calculated formation energy of the $Pnma$ phase is lower than that of the $C1_b$ phase, most (five, i.e., NiScP, NiTiSi, CoTiP, CoVSi, and FeVP) are experimentally observed in the $Pnma$ structure. In two cases (NiTiGe, CoTiAs) the difference between the energies of the two structures is very small ($\Delta E^{Pnma - C1_b} = 0.007$ and $0.011$~eV/atom, respectively), and both have been experimentally observed in the $Pnma$ structure. The two cases for which the calculated formation energy of the $C1_b$ phase is considerably lower than the $Pnma$ phase while experimental reports of the $Pnma$ phase exist are CoTiAs and CoVGe, with $\Delta E^{Pnma - C1_b} = 0.048$ and $0.091$~eV/atom, respectively. The source of these discrepancies is not clear, though errors in DFT, unusual magnetic ordering, and finite temperature contributions to the free energy are the usual suspects (see Sec.~\ref{ssec:energetic_quantities}). 

In all the above cases discussed, the lower energy phase ($C1_b$ or $Pnma$) is predicted to lie on the convex hull. In particular, according to our calculations, the formation energy of CoVGe in the $C1_b$ structure is considerably lower than that of the experimentally reported $Pnma$ structure by 0.091~eV/atom. Similarly, FeVAs has been experimentally observed in the Fe$_2$P $P\overline{6}2m$ structure but our calculations indicate that the $C1_b$ structure is lower in energy than the $P\overline{6}2m$ structure by 0.143~eV/atom. In both the above cases, the calculated difference in formation energies is sufficiently large that efforts to fabricate the corresponding $C1_b$ phases are justified.

The 18-electron semiconductors in Table~\ref{tab:SP_semiconductors} for which we did not find experimental reports are NiScAs, RhTiP, NiV(Al,Ga,In), CoVSn, RhV(Si,Ge,Sn), RuVP, and RuVAs. Based on our calculations and the phases in the OQMD for each of those systems, we predict the following compounds to be thermodynamically stable in the corresponding structures: NiScAs ($Pnma$), RhTiP ($Pnma$), RuVP ($Pnma$) and RuVAs ($C1_b$). CoVSn in the $C1_b$ structure is predicted to be only just above the convex hull with $\Delta E_{\rm HD} = 0.012$~eV/atom. In all other cases, we find a linear combination of other phases in the OQMD, usually binaries, to have a lower energy than both the \textit{XYZ} $C1_b$ and $Pnma$ phases, with $\Delta E_{\rm HD}$ ranging from 0.037~eV/atom (RuVSb) to 0.280~eV/atom (NiVIn). As discussed in Sec.~\ref{ssec:formation_energy_hull_distance}, we expect that the further a compound lies from the convex hull, the less likely will be its successful synthesis.

For several of the \textit{XYZ} 18-electron $C1_b$ semiconductors in Table~\ref{tab:SP_semiconductors}, there are reports of a corresponding $L2_1$ (full-Heusler) phase at the $X_2YZ$ composition (namely, Ni$_2$TiSn, Ni$_2$VAl, Ni$_2$VGa, Co$_2$VSi, Co$_2$VSn, and Rh$_2$VSn). We include these reports in the table (see the rightmost columns in Table~\ref{tab:SP_semiconductors} for references to the reports and corresponding calculated formation energies) for completeness, and because it is sometimes difficult to distinguish between the $L2_1$ and $C1_b$ phases during experimental characterization. However, our calculated formation energies for these reported $L2_1$ phases predict all of them to lie above the convex hull, with a linear combination of other phases to having a lower energy in each case: (a) Ni$_2$TiSn: NiTiSn--Ni$_3$Ti--Ni$_3$Sn$_2$ (lower in energy by $\Delta E_{\rm HD} = 0.029$~eV/atom), (b) Ni$_2$VAl: Ni$_2$V--NiAl--NiV$_3$ ($\Delta E_{\rm HD} = 0.050$~eV/atom), (c) Ni$_2$VGa: Ni$_2$V--NiV$_3$--Ni$_{13}$Ga$_9$ ($\Delta E_{\rm HD} = 0.025$~eV/atom), (d) Co$_2$VSi: CoVSi--Co$_3$V--Co$_2$Si ($\Delta E_{\rm HD} = 0.031$~eV/atom), (e) Co$_2$VSn: CoSn--Co$_3$V--CoV$_3$ ($\Delta E_{\rm HD} = 0.079$~eV/atom), and (f) Rh$_2$VSn: RhSn--RhV ($\Delta E_{\rm HD} = 0.016$~eV/atom). The calculated hull distances $\Delta E_{\rm HD}$ of all the $L2_1$ phases are small -- all except Co$_2$VSn are within 0.050~eV/atom (possible reasons for experimentally observed phases being predicted to lie above the convex hull are discussed in Sec.~\ref{ssec:formation_energy_hull_distance}). We note that the Rh$_2$VSn phase was observed to occur in a tetragonally-distorted structure (space group $P4_2/ncm$, $c/a = 1.27$) in slowly cooled samples, and in a two-phase mixture of $L2_1$ and tetragonal phases in quenched samples~\cite{Suits1976423}. Thus, the $L2_1$ structure seems to be stable only at high temperatures. Similarly, Ni$_2$VAl was experimentally observed in a two-phase mixture, with XRD data insufficient to distinguish between the $C1_b$ and CsCl structure types~\cite{Marazza1975341}. Further, the experimentally reported lattice parameter $a = 6.33$~{\AA} is not only $\sim$12\% larger than the DFT-calculated value but also $\sim$8\% larger than that of Ni$_2$VGa. Since a Ga atom is larger than an Al atom, one would expect the lattice constant of Ni$_2$VGa to be similar or larger than that of Ni$_2$VAl. Thus, we conclude that the reported lattice constant of Ni$_2$VAl is unreasonable and call for careful recharacterization of the phase.

\begin{figure}
\includegraphics[width=\columnwidth]{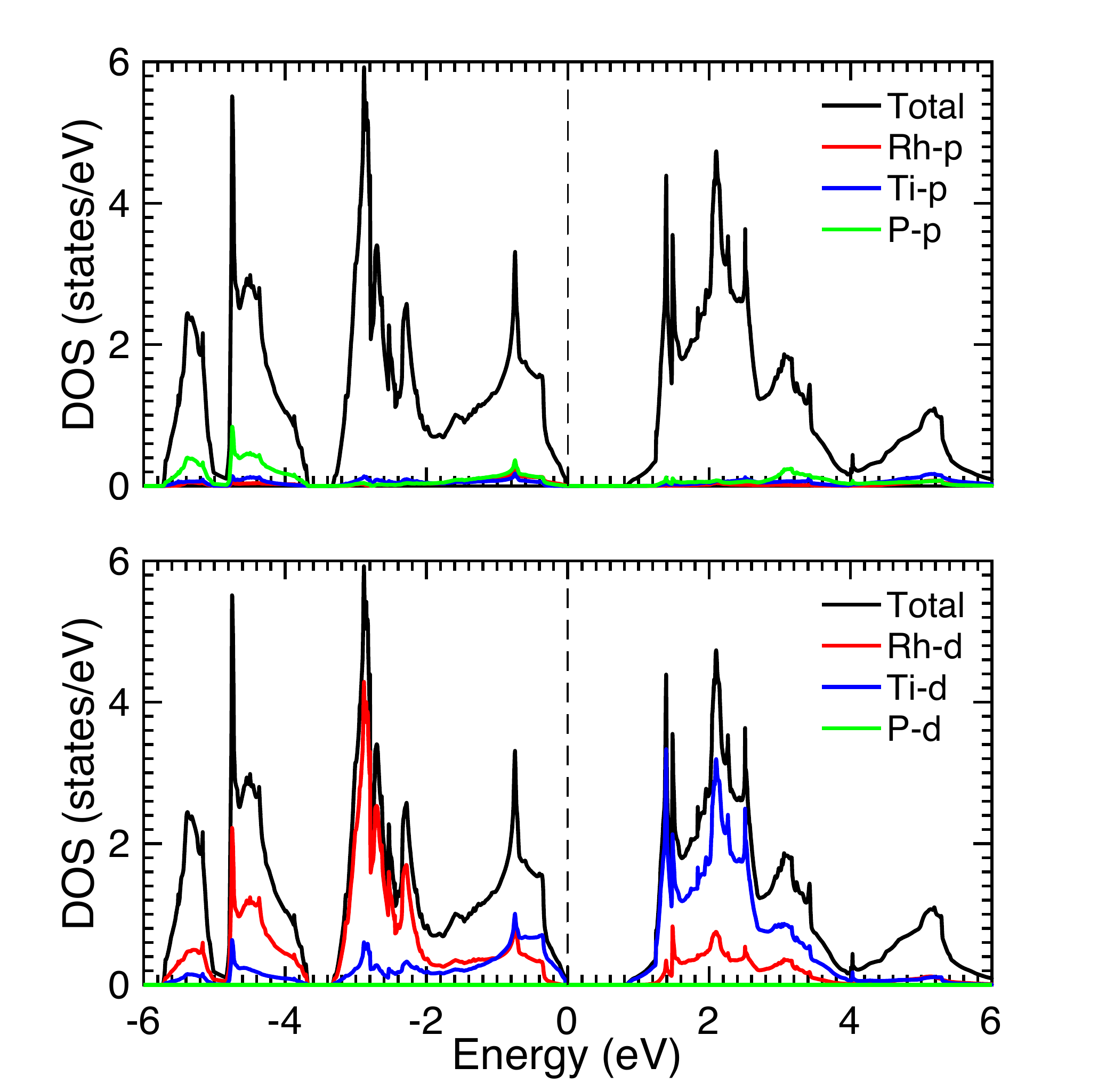}
\caption{Density of electronic states (DOS) of the 18-electron half-Heusler semiconductor RhTiP. The upper panel presents the total DOS (black) and the DOS projected on the $p$-orbitals of Rh (red), Ti (blue) and P (green).The lower panel presents the contribution from the corresponding $d$-orbitals of each atom. Zero energy corresponds to the Fermi level.}
\centering
\label{fig:rhtip_dos}
\end{figure}

We now briefly discuss the bonding and electronic structure of these 18-electron Slater-Pauling semiconductors. Several authors (see for example Ref.~\cite{kandpal2006covalent} and references therein) have suggested that these 18-electron semiconductors can be viewed as covalently bonded \textit{XZ} negative ions forming a zincblende lattice ``stuffed" with positive $Y$ ions. In this picture, for instance, NiSc5 ($5=$ P, As, Sb) compounds would be viewed as a covalently bonded (Ni5)$^{3-}$ zincblende lattice stuffed with Sc$^{3+}$ ions. We investigated the electronic structure of CoTiP, RhTiP, CoTiSb, RhTiSb, FeVSb and RuVSb in more detail to test these ideas. The atom-projected density of states (DOS) of RhTiP is presented in Fig.~\ref{fig:rhtip_dos}. The 27 semiconducting 18-electron half-Heuslers have a similar electronic structure. The nearest neighbor interactions are between the $X$ and $Z$ atoms (Rh and P in this case) and between the $X$ and $Y$ atoms (Rh and Ti). As mentioned previously, a lattice having only the $X$-$Z$ atoms or only the $X$-$Y$ atoms would have the zincblende crystal structure.  

\begin{figure}
\includegraphics[width=\columnwidth]{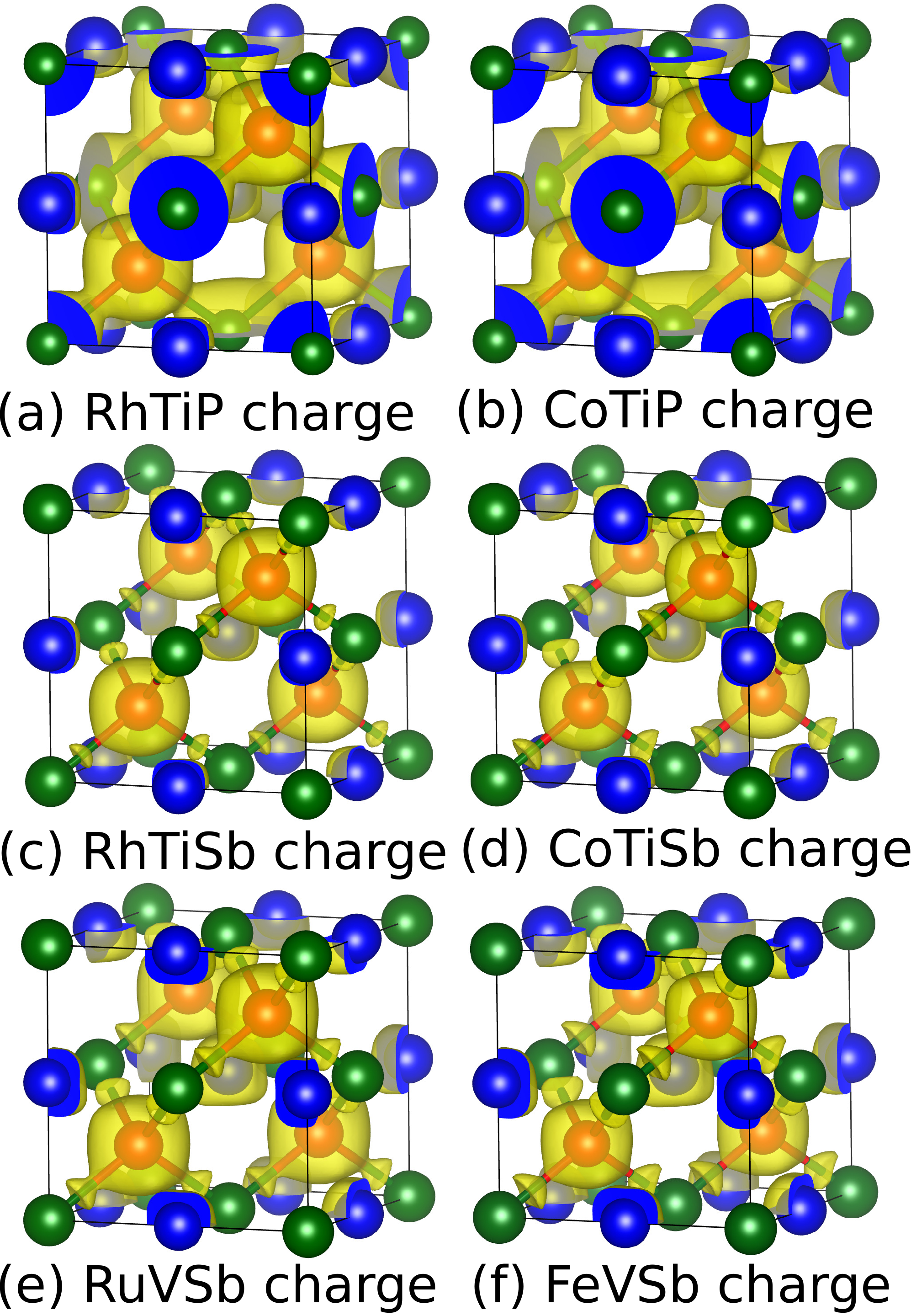}
\caption{Charge densities for (a) RhTiP, (b) CoTiP, (c) RhTiSb, (d) CoTiSb, (e) RuVSb, and (f) FeVSb, for an isovalue of 0.405~$e\cdot\text{\AA}^{-3}$, respectively. The figures were generated using Visualization for Electronic and Structural Analysis (VESTA)~\cite{Momma:db5098}.}
\centering
\label{fig:chargedensity}
\end{figure}

As can be seen from Fig.~\ref{fig:rhtip_dos}, the energy ordering of the atomic orbitals is $Z$-$s$ (in this case forming a narrow band more than 10~eV below the Fermi level and not shown in the figure), followed by $Z$-$p$, followed by $X$-$d$ and finally $Y$-$d$.  The $Z$-$p$--$Y$-$d$ interaction generates a hybridization gap well below the Fermi energy, while the $Y$-$d$--$X$-$d$ interaction generates a hybridization gap (the Slater-Pauling gap) at the Fermi energy. The electronic structure of the other 18 electron semiconductors is similar except that the $Z$-$s$ states are somewhat higher ($\approx -7$ to $-8$~eV) for group 4 $Z$ elements and higher still ($\approx-5$ to $-7$~eV) for group 3 $Z$ elements. Additionally, the hybridization gap between the $X$-$d$ and the $Z$-$p$ states is not fully formed in the systems with group 3 and 4 $Z$ elements. The DOS can be interpreted in terms of a more covalent bond between the $X$ and $Z$ atoms and a more ionic bond between the $X$ and $Y$ atoms. This picture is supported by plots of the charge density shown in Fig.~\ref{fig:chargedensity}. These show a much larger charge density between the $X$ and $Z$ atoms than between the $X$ and $Y$ atoms. We also calculate the net charge within spheres of radius 1.45~{\AA}. These are difficult to interpret in terms of ``ionic charges" because of well-known ambiguities in how one partitions space among atoms in a solid. Nevertheless, all of the spheres are calculated to have net positive charges, with the positive charges on the $Y$ atoms larger than those on the $X$ or $Z$ atoms.

\subsection{Zero-Moment Half-Metals}
\label{ssec:zero_moment_half_metals}

Even more interesting than the non-magnetic 18-electron half-Heusler semiconductors discussed in the previous section are those 18-electron half-Heuslers that may be ferrimagnetic zero-moment Slater-Pauling half-metals. In contrast to an \textit{XYZ} half-Heusler with $Y=$ Ti or V, one with $Y=$ Cr, Mn, or Fe (see columns 6Cr, 7Mn and 8Fe of Table~\ref{tab:all_18_sp_compounds}) may have a ground state with a moment on the $Y$ atom. If such a system is in a Slater-Pauling state, we expect it to have zero net moment per f.u. so any moment on the $Y$ atom should be approximately balanced by an equal and opposite moment on the $X$ atom. The word, ``approximately", is needed in the last sentence because the magnetization density cannot be unambiguously partitioned among the atoms, and because there may be a small moment on the $Z$ atom. This type of system should not be confused with an antiferromagnet because, even if the total magnetic moment is zero, it will have, unlike the typical antiferromagnet, different electronic structures for the two spin channels. 

A zero-moment half-metal would be interesting since it would not respond to magnetic fields (assuming the internal exchange fields are sufficiently strong) but its transport currents would be nominally fully spin polarized. Another interesting feature would be that any magnetic anisotropy might lead to a potentially infinite magnetic anisotropy field, $H_K$, since $H_K=2K/\mu_0 M_s$ where $K$ is the magnetic anisotropy energy density, and $M_s$ is the saturation magnetization. This anisotropy field would be associated with an extremely high (nominally infinite) ferromagnetic resonance (FMR) frequency. Robust materials of this type, if they can be fabricated, might offer the potential for magnetoelectronics that is competitive in terms of switching speed with traditional semiconductor electronics~\cite{PerpSwitchIEEE}.          

\afterpage{
\begin{table*}
\caption{33 18-electron half-Heuslers with $Y=$ Cr, Mn, or Fe. Successive columns present: calculated lattice constant $a$, formation energy $\Delta E_f$, electronic ground state, total spin moment $M_{tot}$, local moments for atoms on $X$, $Y$, and $Z$-sites: $m(X)$, $m(Y)$, and $m(Z)$, gap type, and magnetic lattice constant $a_{\rm mag}$. ([Legend] Electronic ground state: \pblue{NMM} = non-magnetic metal, \orange{HM} = half-metal, \green{FiM} = ferrimagnetic metal, MM = magnetic metal. Gap type: \orange{P} = pseudo gap, \pblue{M}/m = gap in the major/minor spin channel.)}
\label{tab:zero_moment_half_metals}
\begin{tabular}{|l|c|c|c|c|c|c|c|c|c|}
\toprule
\textit{XYZ} & $a$ & $\Delta E_f$ & Electronic & $M_{tot}$ & $m(X)$ & $m(Y)$ & $m(Z)$ & Gap type & $a_{\rm mag}$ \\
 & ({\AA}) & (eV/atom) & ground state & \multicolumn{4}{c|}{($\mu_B$/f.u.)} & & ({\AA}) \\
\midrule
CoCrAl & 5.45 & 0.033 & \pblue{NMM} & 0.000 & 0.000 & 0.000 & 0.000 & \orange{P} & 5.90 \\
CoCrGa & 5.45 & 0.078 & \pblue{NMM} & 0.000 & 0.000 & 0.000 & 0.000 & \orange{P} & 5.80 \\
CoCrIn & 5.76 & 0.465 & \pblue{NMM} & 0.000 & 0.000 & 0.000 & 0.000 & \orange{P} & 5.85 \\
      &       &       &       &       &       &       &       &       &  \\
RhCrAl & 5.73 & $-0.075$ & \pblue{NMM} & 0.000 & 0.000 & 0.000 & 0.000 & \orange{P} & 6.35 \\
RhCrGa & 5.74 & 0.065 & \pblue{NMM} & 0.000 & 0.000 & 0.000 & 0.000 & \orange{P} & 6.25 \\
RhCrIn & 6.19 & 0.207 & MM & $-3.704$ & 0.020 & $-3.461$ & $-0.009$ & m (1.2~eV) & 6.25 \\
      &       &       &       &       &       &       &       &       &  \\
FeCrSi & 5.33 & $-0.231$ & \pblue{NMM} & 0.000 & 0.000 & 0.000 & 0.000 & \orange{P} & 5.65 \\
FeCrGe & 5.45 & $-0.025$ & \pblue{NMM} & 0.000 & 0.000 & 0.000 & 0.000 & \orange{P} & 5.60 \\
FeCrSn & 5.85 & 0.274 & \orange{HM} & 0.000 & 1.827 & $-1.861$ & 0.066 & \pblue{M} & 5.80 \\
      &       &       &       &       &       &       &       &       &  \\
RuCrSi & 5.61 & $-0.115$ & \pblue{NMM} & 0.000 & 0.000 & 0.000 & 0.000 & \orange{P} & 6.40 \\
RuCrGe & 5.71 & 0.059 & \pblue{NMM} & 0.000 & 0.000 & 0.000 & 0.000 & \orange{P} & 6.30 \\
RuCrSn & 5.99 & 0.246 & \pblue{NMM} & 0.000 & 0.000 & 0.000 & 0.000 & \orange{P} & 6.40 \\
      &       &       &       &       &       &       &       &       &  \\
MnCrP & 5.30 & $-0.460$ & \pblue{NMM} & 0.000 & 0.000 & 0.000 & 0.000 & \orange{P} & 5.40 \\
MnCrAs & 5.51 & $-0.101$ & \orange{HM} & 0.000 & 1.487 & $-1.478$ & 0.033 & \pblue{M} & 5.45 \\
MnCrSb & 5.95 & 0.097 & near \orange{HM} & $-0.014$ & 2.709 & $-2.711$ & 0.046 & \pblue{M} (0.07~eV) & 5.55 \\
      &       &       &       &       &       &       &       &       &  \\
FeMnAl & 5.42 & 0.157 & \pblue{NMM} & 0.000 & 0.000 & 0.000 & 0.000 & \orange{P} & 5.50 \\
FeMnGa & 5.49 & 0.214 & near \orange{HM} & $-0.012$ & 1.414 & $-1.482$ & 0.078 & \pblue{M} (0.08~eV) & 5.40 \\
FeMnIn & 5.95 & 0.523 & near \orange{HM} & $-0.546$ & 2.451 & $-3.116$ & 0.078 & \pblue{M} (0.26~eV) & 5.50 \\
      &       &       &       &       &       &       &       &       &  \\
RuMnAl & 5.67 & 0.128 & \pblue{NMM} & 0.000 & 0.000 & 0.000 & 0.000 & \orange{P} & 5.80 \\
RuMnGa & 5.69 & 0.236 & \green{FiM} & 0.110 & 0.105 & $-0.236$ & 0.200 & \orange{P} & 5.70 \\
RuMnIn  & 6.15 & 0.421 & non-SP \orange{HM} & 4.000 & $-0.114$ & $-3.741$ & $-0.013$ & \pblue{M} & 5.70 \\
      &       &       &       &       &       &       &       &       &  \\
MnMnSi & 5.37 & $-0.098$ & weakly MM & 0.007 & 1.315 & $-1.329$ & 0.064 & \orange{P} & 5.20 \\
MnMnGe & 5.58 & 0.047 & \orange{HM} & 0.000 & 2.273 & $-2.317$ & 0.068 & \pblue{M} & 5.20 \\
MnMnSn & 6.05 & 0.184 & near \orange{HM} & 0.174 & 3.111 & $-3.348$ & 0.036 & \pblue{M} (0.17~eV) & 5.20 \\
      &       &       &       &       &       &       &       &       &  \\
CrMnP & 5.42 & $-0.264$ & near \orange{HM} & 0.000 & 1.725 & $-1.758$ & 0.066 & \pblue{M} (0.01~eV) & 5.20 \\
CrMnAs & 5.79 & $-0.009$ & \orange{HM} & 0.000 & 2.620 & $-2.676$ & 0.038 & \pblue{M} & 5.20 \\
CrMnSb & 6.10 & 0.151 & near \orange{HM} & 0.194 & 3.086 & $-3.371$ & $-0.007$ & \pblue{M} (0.18~eV) & 5.15 \\
      &       &       &       &       &       &       &       &       &  \\
MnFeAl & 5.52 & 0.292 & \green{FiM} & 0.001 & 1.534 & $-1.556$ & 0.053 & \orange{P} & 5.30 \\
MnFeGa & 5.56 & 0.299 & \green{FiM} & 0.007 & 1.999 & $-2.032$ & 0.041 & \orange{P} & 5.20 \\
MnFeIn & 5.96 & 0.589 & \orange{HM} & 0.000 & 2.676 & $-2.708$ & 0.003 & \pblue{M}   & 5.30 \\
      &       &       &       &       &       &       &       &       &  \\
CrFeSi & 5.45 & 0.141 & \green{FiM}    & 0.050 & 1.300 & $-1.385$ & 0.056 & \orange{P} & 5.40 \\
CrFeGe & 5.65 & 0.249 & \orange{HM}    & 0.000 & 2.129 & $-2.164$ & 0.021 & \pblue{M} & 5.35 \\
CrFeSn & 6.01 & 0.417 & \orange{HM}    & 0.000 & 2.596 & $-2.664$ & $-0.014$ & \pblue{M} & 5.50 \\
\bottomrule
\end{tabular}

\end{table*}
}

We investigate the electronic structure and formation energies of the 18-electron half-Heuslers in columns 6Cr, 7Mn and 8Fe of Table~\ref{tab:all_18_sp_compounds}. The calculated properties for these systems -- lattice constants, formation energies, magnetic moments, gap types -- are summarized in Table~\ref{tab:zero_moment_half_metals}. Many of these compounds have positive formation energies but the earlier observed trends of formation energies of the \textit{XYZ} half-Heusler compounds decreasing with the group number of the $Z$ element, and increasing with the size of the $Z$-element are still evident.

Many of these compounds are predicted to be nonmagnetic at the equilibrium lattice constant. The total moment per f.u. and the moments within spheres of radius 1.45~{\AA} surrounding the atoms are listed in Table~\ref{tab:zero_moment_half_metals}. All of the 33 compounds develop magnetic moments when the lattice is artificially expanded. The approximate lattice constant associated with the onset of magnetic moments is listed in the rightmost column of the table. If the 33 18-electron systems are sorted by ``groups'' having a common $X$ and $Y$ but different $Z$ element, we find that the onset of magnetism occurs at approximately the same lattice constant for the three members of the same ``group''. 

The lattice constant associated with the onset of magnetism varies between groups roughly according to our notion of the tendency of the various atoms to magnetically polarize. Thus, groups with Mn and Cr as the $X$ and $Y$ elements tend to become magnetic at smaller lattice constants. One interesting feature is that groups for which the atomic number of $Y$ exceeds that of $X$ polarize at a smaller lattice constant. That is, CrMn5, MnFe3, and CrFe4 polarize at smaller lattice constants than MnCr5, FeMn3, and FeCr4 respectively. Another interesting feature is that the moment on the $Y$ atom (within the 1.45~{\AA} sphere) always exceeds that on the $X$ atom. This is true even when $X$ and $Y$ are interchanged (e.g., FeMn4 and MnFe4). The reason for this may be that $X$ has a full complement of 8 nearest neighbors at a distance of $a\frac{\sqrt{3}}{4}$, while $Y$ only has 4 at this distance, and thus has, in a sense, ``more space". This notion of more space for the $Y$ atom  also helps explain the onset of magnetism at smaller lattice constants for CrFe4 compared to FeCr4, for MnFe3 compared to FeMn3, and CrMn5 compared to MnCr5, especially if the transition to the Slater-Pauling state is determined by the magnetic polarization of the atom on the $X$-site, i.e., the one with less space. In other words, Mn is easier to polarize than Cr which is easier to polarize than Fe. Remember that a magnetic Slater-Pauling state for an 18-electron half-Heusler requires that the $X$ atom have a moment that is approximately equal in magnitude and opposite in sign to the moment on the $Y$ atom. 

For the CoCr3, RhCr3 and RuCr4 groups, the magnetic states form with a much larger moment on the Cr than on the Co, Rh or Ru atoms. Although these magnetic states are ferrimagnetic in the sense that the small Co, Rh, or Ru moments align oppositely to the larger Cr moments, their different magnitudes lead to a non-zero moment per formula unit precluding the Slater-Pauling state which would have zero moment. We speculate that the reason for this behaviour is that the Cr atom forms a large moment (especially as a $Y$ atom with only four nearest neighbors) more easily when compared to Co, Rh, and Ru atoms. The density of electronic states (DOS) for systems with $Y=$ Cr typically shows pseudogaps rather than gaps at the Fermi energy for the equilibrium lattice constant~\cite{Heusl89:online}. We speculate that the reason for this is that there is insufficient contrast between the $d-$onsite energies of the Co, Rh or Ru atoms and the Cr atom to support a gap. On (artificial) expansion of the lattice, a large magnetic moment forms on the Cr atoms, the minority $d-$onsite energy of the Cr atoms shifts upward significantly while the minority Co, Rh or Ru $d-$onsite energy shifts down slightly, creating a large difference between the minority $d-$onsite energies and a Slater-Pauling gap. However, this gap is not at the Fermi energy because the $X$ and $Y$ moments are \textit{not} approximately equal and opposite, so the system is not a half-metal even with an expanded lattice. 

The RuMn3 compounds behave similarly to the three groups described in the previous two paragraphs, {with the interesting exception that at a lattice constant of approximately 6.1~{\AA}, these three compounds form a non-Slater-Pauling (non-SP) half-metallic state with a moment of 4~$\mu_B$}. In this state, the moment is largely on the Mn site with a small parallel moment on the Ru site. This non-SP half-metal actually seems to be the equilibrium state for $C1_b$ RuMnIn; however, the gap is very small and the formation energy is significantly greater than zero. The reason these compounds do not form in the Slater-Pauling state even for an expanded lattice is the same as that for the CoCr3, RhCr3, and RuCr4 groups of compounds: the $X$ element cannot match the moment of the more easily polarizable $Y$ element. 

The 7 other groups in Table \ref{tab:zero_moment_half_metals}, show an interesting competition between the Slater-Pauling state at larger lattice constants and a nonmagnetic state with a pseudogap near the Fermi energy for smaller lattice constants. As the lattice is expanded, these compounds undergo a transition into a Slater-Pauling state with zero total moment as opposite and approximately equal moments form on the $X$ and $Y$ atoms. If the moment on the $X$ atom is taken to be positive, the gap is in the majority channel. Surprisingly, this result seems to be independent of whether the $X$ or the $Y$ atom has the larger number of valence electrons. Thus for MnCr5, as one expands the lattice, the Mn and Cr atoms acquire moments around $a=5.5$~{\AA}. In this case, the moment enhances the contrast between the atomic potentials in one channel and decreases it in the other.  The gap forms in the channel with the increased contrast, i.e., since Mn has more electrons than Cr, a positive moment on the Mn and a negative moment on the Cr will increase the contrast in the majority channel and lead to a gap in the majority channel. More explicitly, neglecting charge transfer, Mn without a moment has 3.5 valence electrons/atom in each spin channel. Similarly, Cr without a magnetic moment has 3 valence electrons/atom in each spin channel. If Mn atoms gain a moment of 1.5~$\mu_B$ and Cr atoms gain a moment of -1.5~$\mu_B$, then one spin channel will contain 4.25 electrons on Mn and 2.25 electrons on Cr, whereas the other spin channel will have 2.75 electrons/atom on Mn and 3.75 electrons/atom on Cr atoms. Thus, the contrast between the atoms in the two spin channels \textit{increases} in one spin channel from 0.5 electrons/atom to 2 electrons/atom and in the other from 0.5 electrons/atom to 1 electron/atom. It is not surprising that the gap is in the channel with the larger contrast, i.e., the majority channel if Mn is assumed to have a positive moment. It \emph{is} surprising however that for CrMn5, CrFe4, and MnFe3, the gap is also in the majority channel if the sign of the moment on the $X$ atom is taken to be positive.  For these compounds, the moments increase rapidly, even discontinuously, as the lattice is expanded. The moments are generally larger, sufficiently large in fact to cause large contrast in the majority channel and support a gap in that channel. Of course, the contrast in the number of electrons per atom is even larger for the minority channel, without inducing a gap. It is clear that, at least in this case, the contrast in the number of electrons/atom/spin channel is not the only factor controlling the origin of the gap. 

According to our calculations, many of these compounds that are non-magnetic with pseudogaps could be converted to half-metals if the lattice could be expanded. One way to expand the lattice is to insert a larger atom on the $Z$-site. Unfortunately, from the point of view of fabricating a zero-moment half-metal, as one substitutes larger $Z$ elements to increase the lattice constant, the formation energy also appears to increase. It is also possible to make the lattice constant too large for the Slater-Pauling zero net-moment half-metallic state. In this case the gap continues to be large as the lattice expands and the moments increase in magnitude, but the Fermi energy moves below the Slater-Pauling gap. An example of this effect is MnCrSb which has a gap, but it lies above the Fermi level, whereas MnCrAs with a smaller lattice constant and smaller moments is a half-metal. 

Although a number of zero net-moment half-Heusler half-metals are listed in Table~\ref{tab:zero_moment_half_metals}, only two (MnCrAs and CrMnAs) have negative formation energies. However, both are predicted to lie above the convex hull with hull distances $\Delta E_{\rm HD} = 0.083$ and $0.175$~eV/atom respectively, due to a low-energy competing binary phase MnAs (MnP structure type, space group $Pnma$). Non-equilibrium processing techniques such as epitaxial growth would likely be needed to synthesize the $C1_b$ phases.                        
        
\subsection{Half-metallic ferromagnets}
\label{ssec:half_metallic_ferromagnets}

\afterpage{
\begin{table*}
\caption{DFT-calculated properties of 45 half-metallic \textit{XYZ} half-Heusler compounds with negative formation energy. Successive columns present: number of valence electrons per formula unit $N_{V}$, calculated lattice constant $a$, total spin moment $M_{tot}$ per f.u., local moments for atoms on the $X$-, $Y$-, and $Z$-sites: $m(X)$, $m(Y)$, and $m(Z)$, formation energy $\Delta E_f$, distance from the convex hull $\Delta E_{\rm HD}$, band gap $E_g$, experimental reports of compounds with composition \textit{XYZ}, and experimental reports of corresponding $X_2YZ$ full-Heusler compounds, if any. All half-Heusler compounds listed exhibit an indirect band gap, with the exception of CrScAs and CrTiAs, both of which exhibit a direct gap.}
\label{tab:half_metallic_ferromagnets}
\begin{tabular}{|l|c|c|c|c|c|c|c|c|c|c|c|}
\toprule
\textit{XYZ} & $N_{V}$ & $a$ & $M_{tot}$ & $m(X)$ & $m(Y)$ & $m(Z)$ & $\Delta E_f$ & $\Delta E_{\rm HD}$ & $E_g$  & Experimental & $X_{2}YZ$ \\
 & & ({\AA}) & \multicolumn{4}{c|}{($\mu_{B}$)} & \multicolumn{2}{c|}{(eV/atom)} & (eV) & reports & reports \\
\midrule
CrScAs & 14 & 6.11(6.13) & $-4$ & $-3.276$ & $-0.445$ &  0.044   & $-0.128$ &  0.790 & 0.73 & & \\
CrScSb & 14 & 6.43       & $-4$ & $-3.330$ & $-0.335$ &  0.059   & $-0.106$ &  0.522 & 0.99 & & \\
CrTiAs & 15 & 5.52(6.66) & $-3$ & $-2.458$ & $-0.376$ &  0.011   & $-0.009$ &  0.623 & 0.69 & $P\overline{6}2m$\cite{Johnson19731067} & \\
MnVAs  & 17 & 5.59       & $-1$ & $-1.574$ &  0.625   & $-0.039$ & $-0.243$ &  0.107 & 0.87 & $P4/nmm$\cite{RoyMontreuil1972813} & \\
MnVSb  & 17 & 5.92       & $-1$ & $-2.157$ &  1.180   & $-0.027$ & $-0.034$ &  0.156 & 0.94 & $P6_{3}/mmc$\cite{noda1984high} & \\
CrMnAs & 18 & 5.71       &  0   & $-2.618$ &  2.671   & $-0.036$ & $-0.009$ &  0.175 & 0.98 & $P4/nmm$\cite{Nylund1972115} & \\
MnCrAs & 18 & 5.51       &  0   & $-1.474$ &  1.464   & $-0.031$ & $-0.101$ &  0.083 & 0.73 & $P4/nmm$\cite{Nylund1972115} &  \\
MnMnAs & 19 & 5.63       &  1   & $-2.068$ &  2.995   & $ 0.023$ & $-0.131$ &  0.055 & 1.11 & $P\overline{6}2m$\cite{Jeitschko:a09138}, $P4/nmm$\cite{yuzuri1960magnetic, pearson1985cu2sb} & \\
FeCrAs & 19 & 5.48       &  1   & $-0.652$ &  1.640   & $-0.036$ & $-0.137$ &  0.005 & 0.96 & $P\overline{6}2m$\cite{Hollan1966, Nylund1972115} & \\
RuCrAs & 19 & 5.74       &  1   & $-0.316$ &  1.357   & $-0.064$ & $-0.030$ &  0.168 & 0.58 & $Pnma$\cite{deyris1974structural, kanomata1991magnetic} &\\
CoCrGe & 19 & 5.47       &  1   & $-0.339$ &  1.391   & $-0.078$ & $-0.035$ &  0.124 & 0.96 & $P6_{3}/mmc$\cite{wirringa2000crystal}, $Cmcm$\cite{wirringa2000crystal} & \\
CoVAs  & 19 & 5.53       &  1   & $-0.123$ &  1.092   & $-0.028$ & $-0.363$ &    0   & 1.22 & $Pnma$\cite{Johnson19731067} & \\
CoVSb  & 19 & 5.81       &  1   & $-0.246$ &  1.217   & $-0.041$ & $-0.182$ &  0.011 & 0.90 & $F\overline{4}3m$\cite{terada1970magnetic, Evers199793}, $P6_{3}/mmc$\cite{noda1979synthesis} & \\
RhCrSi & 19 & 5.65       &  1   & $-0.148$ &  1.230   & $-0.093$ & $-0.183$ &  0.407 & 0.57 & & \\
RhCrGe & 19 & 5.75       &  1   & $-0.191$ &  1.293   & $-0.098$ & $-0.056$ &  0.314 & 0.67 & & \\
RhVAs  & 19 & 5.81       &  1   & $-0.156$ &  1.165   & $-0.061$ & $-0.377$ &  0.099 & 0.88 & $Pnma$\cite{deyris1974structural, roy1984analyse} & \\
RhVSb  & 19 & 6.06       &  1   & $-0.166$ &  1.189   & $-0.062$ & $-0.312$ &  0.103 & 0.80 & & \\
NiVSn  & 19 & 5.87       &  1   &  0.004   &  1.017   & $-0.056$ & $-0.079$ &  0.148 & 0.48 & & \cite{Ni2VSn, Ni2VSn2} \\
FeMnP  & 20 & 5.32       &  2   & $-0.401$ &  2.380   & $-0.040$ & $-0.424$ &  0.149 & 0.83 & $Pnma$\cite{Nylund1971}, $P\overline{6}2m$\cite{Chenevier198757} & \\
FeMnAs & 20 & 5.51       &  2   & $-0.708$ &  2.690   & $-0.041$ & $-0.134$ &  0.075 & 1.08 & $P4/nmm$\cite{Hollan1966}, $P\overline{6}2m$\cite{Tobola2001274} & \\ 
RuMnP  & 20 & 5.59       &  2   & $-0.313$ &  2.350   & $-0.072$ & $-0.280$ &  0.261 & 0.65 & & \\
RuMnAs & 20 & 5.76       &  2   & $-0.410$ &  2.489   & $-0.087$ & $-0.053$ &  0.158 & 0.78 & $P\overline{6}2m$\cite{roymontreuil1979influence, deyris1974structural, Deyris1975603} & \\
CoCrP  & 20 & 5.32       &  2   & $-0.036$ &  2.054   & $-0.077$ & $-0.362$ &  0.193 & 1.34 & $Pnma$\cite{Nylund1972115} & \\
CoCrAs & 20 & 5.52       &  2   & $-0.276$ &  2.305   & $-0.090$ & $-0.104$ &  0.101 & 1.07 & $P\overline{6}2m$\cite{Nylund1972115} & \\
CoMnSi & 20 & 5.36       &  2   & $-0.037$ &  2.169   & $-0.162$ & $-0.209$ &  0.257 & 0.78 & $Pnma$\cite{PSSA:PSSA2210350161}, $P6_{3}/mmc$\cite{doi101021/ic50147a032} & \cite{Co2Vsi, Co2MnSi2, Co2MnSi3} \\
CoMnGe & 20 & 5.49       &  2   & $-0.254$ &  2.394   & $-0.152$ & $-0.057$ &  0.150 & 0.99 & $Pnma$\cite{Nizio1982281}, $P6_{3}/mmc$\cite{doi101021/ic50147a032} & \cite{Co2MnGe1, Co2MnGe2} \\
RhCrP  & 20 & 5.65       &  2   & $-0.234$ &  2.297   & $-0.121$ & $-0.293$ &  0.377 & 0.91 & & \\
RhCrAs & 20 & 5.81       &  2   & $-0.291$ &  2.384   & $-0.135$ & $-0.121$ &  0.209 & 0.95 & $P\overline{6}2m$\cite{Michel1984, Ohta1995157, kanomata1991magnetic} & \\
NiCrSi & 20 & 5.44       &  2   & $ 0.067$ &  2.040   & $-0.158$ & $-0.116$ &  0.345 & 0.85 & $Pnma$\cite{landrum1998tinisi} & \\
NiCrGe & 20 & 5.54       &  2   & $-0.022$ &  2.150   & $-0.163$ & $-0.001$ &  0.237 & 0.64 & & \\
NiVSb  & 20 & 5.89       &  2   &  0.070   &  1.878   & $-0.078$ & $-0.122$ &  0.129 & 0.41 & $F\overline{4}3m$\cite{kripyakevich1963crystal} & \\
CoMnP  & 21 & 5.34       &  3   &  0.078   &  2.947   & $-0.108$ & $-0.443$ &  0.290 & 1.29 & $Pnma$\cite{Fruchart:a19268, PSSA:PSSA2210570225} & \\
CoMnAs & 21 & 5.53       &  3   & $-0.092$ &  3.130   & $-0.109$ & $-0.211$ &  0     & 1.16 & $Pnma$\cite{Johnson19731067, Nylund1971} & \\
CoMnSb & 21 & 5.82       &  3   & $-0.178$ &  3.262   & $-0.111$ & $-0.108$ &  0.012 & 0.89 & $F\overline{4}3m$\cite{Buschow19831, Otto198733}, $Fd\overline{3}m$\cite{Senateur1972226} & \cite{Buschow19831,Co2MnSb1} \\
 & & & & & & & & & & $Fm\overline{3}m$\cite{PSSA:PSSA2210090109, PhysRevB.74.134426} & \\
CoFeGe & 21 & 5.50       &  3   &  0.507   &  2.601   & $-0.131$ & $-0.061$ &  0.126 & 0.44 & $P6_{3}/mmc$\cite{lecocq1963etude} & \\
RhMnP  & 21 & 5.67       &  3   & $-0.136$ &  3.229   & $-0.153$ & $-0.404$ &  0.318 & 0.89 & $P\overline{6}2m$\cite{Suzuki199877, kanomata1991magnetic} & \\
RhFeGe & 21 & 5.78       &  3   &  0.201   &  2.908   & $-0.102$ & $-0.139$ &  0.231 & 0.49 & & \\
RhFeSn & 21 & 6.05       &  3   &  0.192   &  2.978   & $-0.094$ & $-0.121$ &  0.225 & 0.48 & & \\
NiCrP  & 21 & 5.42       &  3   &  0.152   &  2.854   & $-0.121$ & $-0.254$ &  0.335 & 0.77 & $Pnma$\cite{orishchin1984cr, Nylund1972115}, $P\overline{6}2m$\cite{orishchin1984cr} & \\
NiCrAs & 21 & 5.62       &  3   &  0.037   &  2.991   & $-0.146$ & $-0.079$ &  0.153 & 0.56 & $P\overline{6}2m$\cite{Ohta1995157, Nylund1972115} & \\
NiMnSi & 21 & 5.45       &  3   &  0.120   &  3.028   & $-0.207$ & $-0.251$ &  0.242 & 0.85 & $P6_{3}/mmc$\cite{doi101021/ic50147a032}, $Pnma$\cite{PSSA:PSSA2210200133, landrum1998tinisi} & \\
NiFeGa & 21 & 5.56       &  3   &  0.332   &  2.792   & $-0.132$ & $-0.017$ &  0.266 & 0.59 & & \cite{Ni2FeGa1} \\
NiMnP  & 22 & 5.46       &  4   &  0.333   &  3.607   & $-0.062$ & $-0.400$ &  0.237 & 0.87 & $P\overline{6}2m$\cite{fruchart1969crystallographic, Nylund1971}, $Pnma$\cite{Nylund1972115} & \\
NiMnAs & 22 & 5.64       &  4   &  0.278   &  3.688   & $-0.082$ & $-0.250$ &  0.017 & 0.69 & $P\overline{6}2m$\cite{NiMnAs1}, $Pnma$\cite{Nylund1972115} & \\
NiMnSb & 22 & 5.91       &  4   &  0.222   &  3.764   & $-0.081$ & $-0.217$ &  0     & 0.48 & $F\overline{4}3m$\cite{Buschow19831, PSSA:PSSA2210090109, vanEngelen1994247} & \cite{Buschow19831, Kanomata1987286, Ni2MnSb2} \\
\bottomrule
\end{tabular}

\end{table*}
}

Considering our entire database of 384 $C1_b$ systems rather than just the 18-electron systems, we find 75 half-metals of which 45 are calculated to have negative formation energy. In addition we find 34 half-Heusler compounds that are \textit{near} half-metals with negative formation energy. In this section, we will focus on these compounds which are listed with their properties -- number of valence electrons, spin magnetic moments, formation energy, hull distance, band gap, spin polarization -- in Tables~\ref{tab:half_metallic_ferromagnets} and \ref{tab:near_half_metallic_ferromagnets}, respectively. 

Although we restrict our attention to the systems with negative values of the calculated formation energy, our results to not conclusively exclude the existence of $C1_b$ systems with a calculated positive formation energy. Apart from limitations of DFT, there may be significant contributions to the entropy and free energy from several types of thermal disorder. In particular, the open structure of the half-Heuslers may be conducive to soft-phonon modes which may reduce its free energy relative to competing phases. This is a complex phenomenon because the magnetic and vibrational excitations may be coupled. There is also the possibility of configurational entropy arising from substitutional disorder, especially due to off-stoichiometry, i.e., excess $X$ in the vacant sublattice, vacancies in the $Y$, $Z$ sublattices, etc., but this is beyond the scope of our current work.

When we tested for stability against tetragonal distortions, only one of the 45 half-metallic half-Heusler compounds in Table~\ref{tab:half_metallic_ferromagnets} was calculated to have a lower energy in a tetragonal structure,  CrTiAs.  The total energy landscape for CrTiAs as a function of lattice constants $a$ and $c$ was discussed in Sec.~\ref{ssec:lowest_energy_structure}, where it was shown to have two local minima, one with $c/a>1$ and another at a slightly higher energy with $c/a<1$. CrScP, CrScAs, CrScSb, CrTiP, and CrTiSb behave similarly to CrTiAs with the exception that the $Z=$ Sb compounds have a single global energy minimum. The $Z=$ (P, As) compounds have two energy minima that lie along a line describing volume conserving distortions. The calculated energy difference between these minima is extremely small, especially for the $Y=$ Sc compounds. If any of these phases can be fabricated, they would be expected to have anomalous properties.        

Table~\ref{tab:CrScTi5_minima} shows the lattice constants corresponding to the local energy minima for CrSc(P,As,Sb) and CrTi(P,As,Sb), the calculated magnetic spin moments at each minimum and the energy difference between the minima. CrScP appears in Table~\ref{tab:near_half_metallic_ferromagnets} and not in Table~\ref{tab:half_metallic_ferromagnets} because it is a near half-metal at its global minimum, but there is a nearly cubic local minimum that is only 2~meV higher in energy which is calculated to be half-metallic. For CrTiP, neither of the solutions is half-metallic although the one for $a>c$ comes close. Its energy, however is significantly higher than the phase with $c>a$. CrTiP is not included in Table~\ref{tab:near_half_metallic_ferromagnets} because its Fermi energy falls rather far from the gap ($M_{tot}=$ 2.53$~\mu_B$ rather than 3.00~$\mu_B$ per f.u.). CrScSb and CrTiSb are both predicted to be half-metallic and have only small tetragonal distortions. Although CrTiSb is predicted to be a half-metal in its ground state, it is omitted from Table~\ref{tab:half_metallic_ferromagnets} because its formation energy is calculated to be positive. 

It may be important to note that these anomalous energy landscapes would imply soft long-wavelength phonons which would impact several physical properties, possibly including enhanced stability of these phases because of the associated contributions to the entropy and free energy. Observation of these anomalous phases may, unfortunately, be difficult because of the very low formation energy of the competing $B1$ Sc(P,As,Sb) and orthorhombic $P6_3/mmc$ Ti(P,As, Sb) binary phases.

\begin{table}
\caption{Lattice constants and total magnetic moments per f.u. corresponding to the different energy minima, and the energy difference between the minima, for CrSc5 and CrTi5 compounds. The compounds with integer moments are predicted to be half-metals.}
\label{tab:CrScTi5_minima} 
\begin{tabular}{lccccc}
\toprule
\textit{XYZ} & Min.~(1) & $M^{(1)}_{tot}$ & Min.~(2) & $M_{tot}^{(2)}$ & $\Delta E_{(1)-(2)}$ \\ 
 & ($a_1$, $c_1$) & $(\mu_B)$ & ($a_2$, $c_2$) & ($\mu_B$) & (eV/f.u.) \\
\midrule
CrScP   & (5.64, 6.65) & $-3.9999$ & (5.96, 5.90) & $-4$     & $-0.002$  \\
CrScAs  & (5.95, 6.48) & $-4$      & (6.11, 6.13) & $-4$     &  0.002    \\
CrScSb  & ---          & ---       & (6.43, 6.42) & $-4$     & ---       \\
CrTiP   & (5.24, 6.67) & $-2.5301$ & (5.78, 5.50) & $-2.983$ & $-0.094$  \\
CrTiAs  & (5.52, 6.66) & $-3$      & (5.97, 5.67) & $-3$     & $-0.052$  \\
CrTiSb  & ---          & ---       & (6.19, 6.15) & $-3$     & ---       \\
\bottomrule
\end{tabular}

\end{table} 

\begin{figure*}
\includegraphics[width=0.9\textwidth]{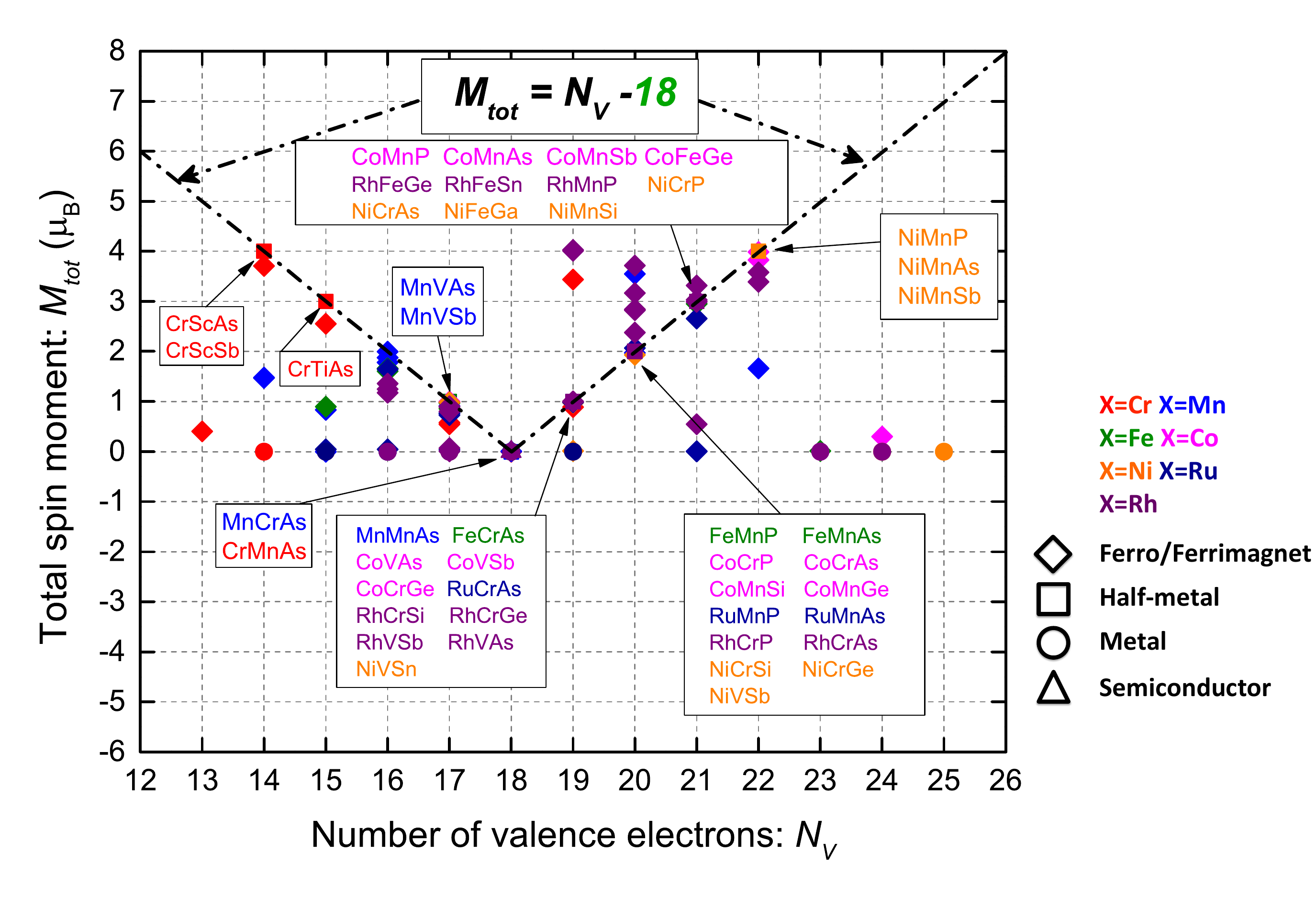}
\caption{Calculated total magnetic moment $M_{tot}$ as a function of the total number of valence electrons $N_V$ per f.u. for the 203 half-Heusler compounds with negative formation energies. The dash-dot line represents the Slater-Pauling rule $M_{tot}=N_{V}-18$, and all the 45 half-metals listed in the boxes follow this rule precisely. Different colors indicate different sets of half-Heusler compounds based on the element on the $X$-site. Diamond, square, circle, and triangle symbols indicate ferro/ferrimagnets, half-metals, metals, and semiconductors, respectively. To avoid confusion about the signs of magnetic moments, we uniformly use the absolute values of magnetic moments in this diagram.}
\label{fig:spin_moment_vs_valence_electrons}
\end{figure*}

By comparing the number of valence electrons in each system to the corresponding total magnetic moment per f.u., we can see that all of the half-metals follow the Slater-Pauling rule:
\begin{equation}\label{eqn:SP_rule}
M_{tot}=N_{V}-18
\end{equation}
where $M_{tot}$ is the total magnetic moment and $N_{V}$ is the total number of valence electrons per \textit{XYZ} f.u. In Fig.~\ref{fig:spin_moment_vs_valence_electrons}, we summarize the calculated total magnetic moments as a function of the total number of valence electrons for all the investigated half-Heusler compounds (including the 6 additional compounds with $Y=$ Sc) with negative formation energy (203 of the 384 compounds considered in this work). In the figure we use different colors and geometric symbols to distinguish their properties, i.e., semiconductor, metal, half-metal, and ferro/ferrimagnets. The dash-dot line represents the Slater-Pauling expression (from Eq.~\ref{eqn:SP_rule}). The 45 half-metallic half-Heusler compounds with negative formation energy are listed in eight boxes classified by their total magnetic moments per f.u.

The five half-metals CrScAs, CrScSb, CrTiAs, MnVAs, and MnVSb have band gaps in the majority-spin channel since $N_{V}<18$, while the band gaps of the half-metals with $N_{V}>18$ are in the minority-spin channel. Two half-metals, MnCrAs and CrMnAs, have $N_{V}=18$, and in this case the choice of the majority/minority spin channel is arbitrary. However, the channels are different and in both cases the gap occurs in the channel for which the atom on the $X$-site has more electrons. In other words, if Mn has a positive moment in MnCrAs then the gap is in the majority channel. If Cr has a positive moment in CrMnAs, then the gap is in the majority channel.    

In order to analyze the magnetic configurations of the half-metallic half-Heusler compounds, we also list in Table~\ref{tab:half_metallic_ferromagnets} the local magnetic moments within spheres of radius 1.45~{\AA} centered at the $X$-, $Y$-, and $Z$-sites. We find that the magnetic configurations can be divided into several categories by the number of valence electrons $N_{V}$. The three half-metals with $N_{V}<17$ (CrScAs, CrScSb, CrTiAs) have relatively large moments on the $X$ sublattice with much smaller ferromagnetically aligned moments on the $Y$ sublattice. The small moments on $Y$ are due to the difficulty in magnetically polarizing Sc and Ti atoms. Perhaps the unusual (for a Slater-Pauling half-metal) tetragonal distortions result in additional space for the magnetic atom on the $X$-site.

For half-Heusler compounds with $N_{V}=17$, the half-metals are ferrimagnetic with larg moments on $X$- and smaller antiparallel moments on $Y$-sites. For compounds with $N_{V}=18$, the net magnetic moment is zero: the $X$ and $Y$ sublattices have approximately equal but antiparallel spin moments. For half-Heuslers with $19~\leq~N_V~\leq~20$, most of the spin moment is on the $Y$ sublattice, while the $X$ sublattice has a moment that is small and usually opposite to that of the $Y$ sublattice. For compounds with $21~\leq~N_V~\leq~22$, the half-metals tend to be ferromagnetic with large localized moments on the $Y$ sublattice, and small spin moments on the $Z$ sublattice.  

We found no $C1_b$ half-metals with $M_{tot} > 4$. This limit can be understood if one makes the approximation that the local moment on the non-transition metal atom (the $Z$ sublattice) is zero and that the number of majority spin electrons on either of the transition metal atoms is less than 5.5. This limit arises from the fact that there are only 5 $d$-states per spin channel per transition metal atom. Transition metal atoms have approximately one $s-$electron more or less degenerate with the $d-$states, shared between majority and minority. This leads to the $s-d$ bands holding 5.5 or fewer electrons per transition metal atom per spin channel. The requirement that $N_{V,X}^{\uparrow} + N_{V,Y}^{\uparrow} < 11$, together with the requirements that $N_{V,Z}^{\uparrow} = N_{V,Z}^{tot}/2$ and $M_{tot} = N_{V,X}^{tot} + N_{V,Y}^{tot} + N_{V,Z}^{tot} - 18$, leads to the limit $M_{tot} < 2 + N_{V,Z}^{tot}/2$. Since the largest value of $N_{V,Z}^{tot}$ that we considered was 5 and since $M_{tot}$ must be an integer for half-metals, we obtain $M_{tot} \leq 4$. Thus, although one might imagine obtaining a large moment half-metal by choosing a system with large $N_V$, e.g. NiNiP ($N_V=25$), a Slater-Pauling state with $M_{tot}=7$ cannot be obtained because achieving large moments on Ni is not possible. Note that this limit does not apply for very small values of $N_V$, e.g., $C1_b$ CrTiIn ($N_V=13$) is predicted to be a Slater-Pauling half-metal with $M_{tot}=5$. Unfortunately, from the point of view of synthesizing high-moment half-Heusler half-metals, it is also predicted to have a large positive formation energy.    

\begin{figure*}[t]
\includegraphics[width=0.9\textwidth]{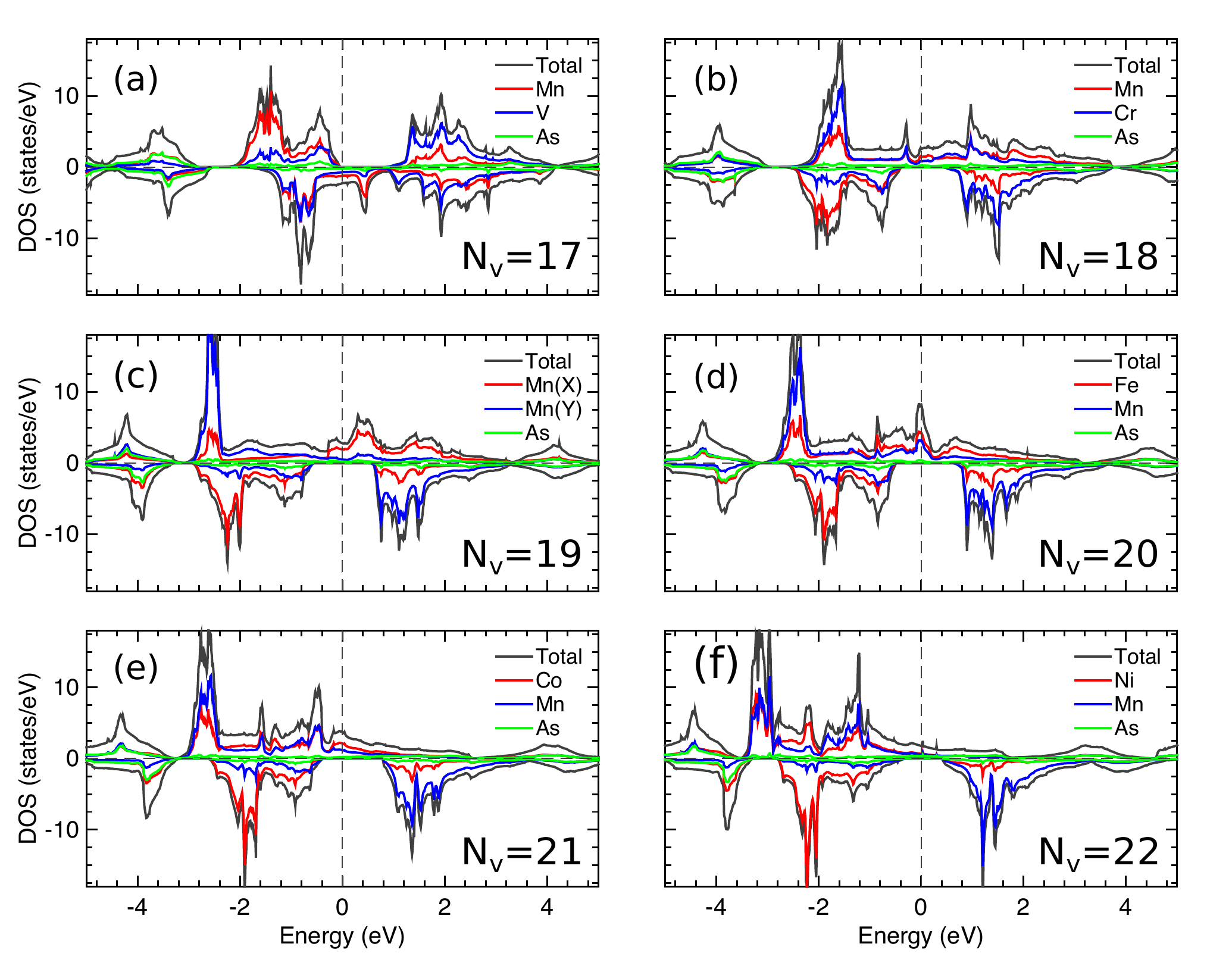}
\caption{Calculated total and atom-resolved densities of electronic states for MnVAs, MnCrAs, MnMnAs, FeMnAs, CoMnAs and NiMnAs. In each subplot, the upper (lower) panel shows the majority (minority) spin channel. The number of valence electrons per f.u. $N_{V}$ is also indicated for each system. Zero energy corresponds to the Fermi level.}
\label{fig:mn_half_heusler_dos}
\end{figure*} 

We analyzed the electronic structure -- density of electronic states (DOS) and band dispersion -- for each half-Heusler compound considered in this work. The size of the bandgap for each half-metal is listed in Table~\ref{tab:half_metallic_ferromagnets}. All C1$_b$ half-metals with a negative formation energy have indirect band gaps. The valence band maximum is at $\Gamma$ or $L$ or occasionally at $W$ in the Brillouin zone. The conduction band minimum is uniformly at $X$. Two half-metals that are tetragonally distorted, CrTiAs and CrScAs, have direct gaps at $\Gamma$. It should be noted that in MnVSb and RuMnAs (both counted as half-metals and listed in Table~\ref{tab:half_metallic_ferromagnets}), and in CoCrSi (listed as a near half-metal in Table~\ref{tab:near_half_metallic_ferromagnets}) the Fermi level just touches the band edge but their total magnetic moment still follows the Slater-Pauling rule. The precise location of the Fermi energy in these compounds might be further refined by future calculations. 

Fig.~\ref{fig:mn_half_heusler_dos} shows the total and atom-resolved DOS for six half-Heusler half-metals with $X$ or $Y=$ Mn, all with negative formation energy, ordered by the number of valence electrons. In accordance with the Slater-Pauling rule (see Eq.~\ref{eqn:SP_rule}), the states in one spin channel are filled to the Fermi level located in the band gap separating the filled and unfilled states. There are precisely 3 electrons per atom in this spin channel, and it can be seen that the gapped channel appears very similar for all the 6 compounds. There are 9 total electrons in the gapped channel for all the 6 compounds, and the remaining electrons accumulate in the other spin channel. Compounds with different number of total electrons will have different number of electrons in the metallic channel, implying changes in the energy levels of the transition metal orbitals. It can be seen in Fig.~\ref{fig:mn_half_heusler_dos}(a)--(f) that the energy levels and DOS shift downwards in energy in the metallic channel as the number of valence electrons per f.u. increases.

From Table~\ref{tab:half_metallic_ferromagnets}, we see that the formation energy of the half-metallic half-Heuslers ranges from $-0.44$~eV/atom to barely negative. The 30 half-Heusler half-metals with positive formation energy are not shown in the table, but the calculated data is available online at \texttt{heusleralloys.mint.ua.edu}~\cite{Heusl89:online}. Although there is a wide range of formation energies for both, the half-Heusler half-metals are typically less stable than the half-Heusler semiconductors. This difference may be due to the former having a gap in only one spin channel while the latter have a gap in both spin channel (see Figs.~\ref{fig:formation_energy_vs_polarization}--\ref{fig:hull_distance_vs_polarization}, and the corresponding discussion in Sec.~\ref{ssec:formation_energy_hull_distance}).

Using our calculated $C1_b$ formation energies, and the formation energies of all the other phases in the OQMD database, we calculate the hull distance $\Delta E_{\rm HD}$ for all the half-metallic half-Heusler compounds (listed in Table~\ref{tab:half_metallic_ferromagnets}. Of the 45 $C1_b$ half-metals with negative formation energy, our calculations predict 3 (CoVAs, CoMnAs, and NiMnSb) to lie on the convex hull of the respective systems. However, the $Pnma$ phase has been observed experimentally for CoVAs and CoMnAs. Since the two $Pnma$ phases are not in the OQMD, we calculated their formation energies and found them to be indeed lower than that of the respective $C1_b$ phase by 0.479 and 0.073~eV/atom for CoVAs and CoMnAs, respectively. We also verified that the $C1_b$ phase of NiMnSb is more stable than the $Pnma$ phase by 0.172~eV/atom. On the other hand, we found experimental reports indicating that three systems in Table~\ref{tab:half_metallic_ferromagnets} (CoVSb, NiVSb, and CoMnSb) have been observed in the $C1_b$ structure, yet all three compounds are predicted to lie above the convex hull, with a linear combination of other phases predicted to be lower in energy: \\

(a) CoVSb: is predicted to lie near the convex hull with a mixture of phases CoSb$_3$--VCo$_3$--V$_3$Co lower in energy by $\Delta E_{\rm HD} = 0.011$~eV/atom. In fact, CoVSb has been synthesized and studied extensively~\cite{Evers199793,CoVSb2,terada1970magnetic,CoVSbMag,CoVSbNeutron,CoVSBweakitin,CoVSb3,noda1979synthesis}, and the $C1_b$ phase seems to be well established, but the compound is a weak itinerant ferromagnet rather than a half-metal. It is possible that the spin fluctuations associated with this type of magnetism help to stabilize the phase. \\

(b) NiVSb: a linear combination of binary phases NiSb--V$_3$Sb--Ni$_2$V is predicted to be lower in energy than the ternary $C1_b$ phase by $\Delta E_{\rm HD} = 0.130$~eV/atom. NiVSb was reported in the $C1_b$ structure in 1963~\cite{NiVSb63}, but more recent studies of the Ni--V--Sb system, not only report failure in synthesizing the $C1_b$ NiVSb phase but also find a mixture of three binary phases (NiSb, V$_3$Sb, and VSb) coexisting at the equiatomic composition~\cite{NiVSb}, in qualitative agreement with our calculations. \\

\begin{figure}
\includegraphics[width=\columnwidth]{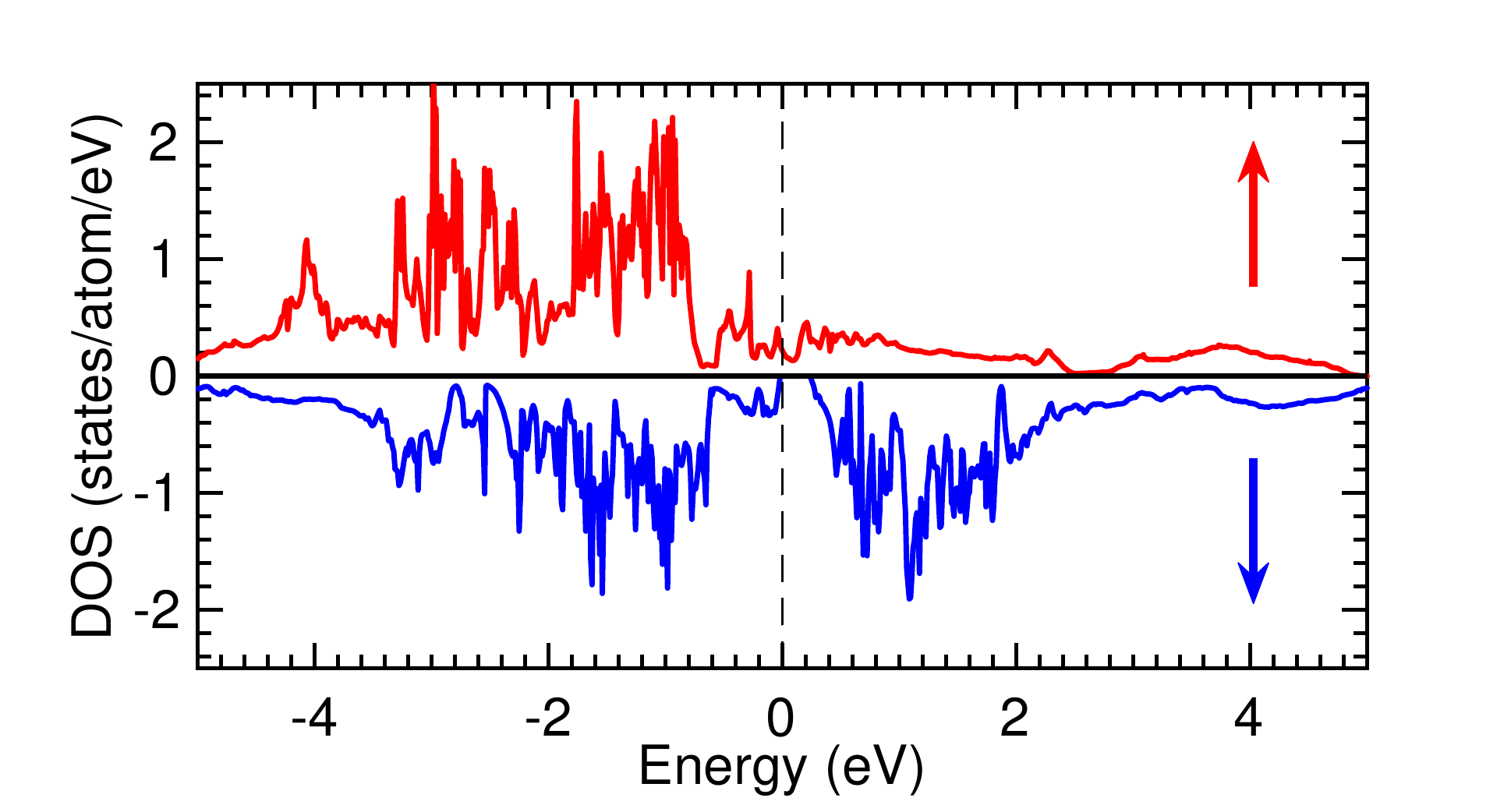}
\caption{Calculated density of electronic states (DOS) of CoMnSb in the Co$_2$MnSb-MnSb superstructure reported in Ref.~\cite{PhysRevB.74.134426}. Zero energy corresponds to the Fermi level.}
\label{fig:CoMnSb_DOS}
\end{figure}

(c) CoMnSb: has been reported in the $C1_b$ structure~\cite{Buschow19831,Otto198733}, but recent work~\cite{PhysRevB.74.134426} has shown its structure to be more complicated -- a superstructure consisting of alternating layers of Co$_2$MnSb and MnSb. Its observed magnetic moment is substantially larger than the Slater-Pauling value of 3~$\mu_B$/f.u. Our calculations confirm the lower energy of the Co$_2$MnSb-MnSb superstructure (by 0.012~eV/atom), but in contrast to those in Ref.~\cite{PhysRevB.74.134426}, they indicate that that the superstructure is a non-Slater-Pauling half-metal with a moment of 3.75~$\mu_B$/f.u. (30~$\mu_B$ for a 24-atom supercell). The minority channel is predicted to have 2.875 rather than 3 electrons/atom. Our calculated density of electronic states for the Co$_2$MnSb-MnSb superstructure is shown in Fig.~\ref{fig:CoMnSb_DOS}.

Further, we find that 8 $C1_b$ half-Heusler compounds in Table~\ref{tab:half_metallic_ferromagnets} lie close to the convex hull with $\Delta E_{\rm HD} \leq \,\sim$0.1~eV/atom. In almost all such cases, we find experimental reports of other non-$C1_b$ compounds at the composition (space group of the structure(s) experimentally reported, and hull distance $\Delta E_{\rm HD}$ of the $C1_b$ half-Heusler compound in eV/atom): FeCrAs ($P\overline{6}2m$, 0.005), NiMnAs ($P\overline{6}2m$, $Pnma$, 0.017), MnMnAs ($P\overline{6}2m$, $P4/nmm$, 0.055), MnCrAs ($P4/nmm$, 0.083), RhVAs ($Pnma$, 0.099), CoCrAs ($P\overline{6}2m$, 0.101), RhVSb (none, 0.103), MnVAs ($P4/nmm$, 0.107). Attempts to synthesize some of the above compounds in the half-Heusler structure using non-equilibrium processing techniques may be merited. In addition, we found no experimental reports of phases for a dozen of the potential half-metallic half-Heuslers in Table~\ref{tab:half_metallic_ferromagnets}. However, our calculated formation energies for these 12 compounds indicate that they all lie above the convex hull of the respective system, with $\Delta E_{\rm HD} >0.1$~eV/atom. 

We list in Table~\ref{tab:near_half_metallic_ferromagnets} the 34 \textit{XYZ} half-Heusler phases with negative formation energy which our calculations predict to be ``near half-metals'', i.e., they have a gap in one of the spin channels at 3 electrons/atom and the Fermi energy falls just above or just below the gap. We also tabulate  the calculated properties for the above phases -- lattice constant, magnetic moments, formation energy, hull distance, spin polarization at Fermi energy, compounds reported experimentally at each composition, and reports of corresponding $X_2YZ$ full-Heusler phases. These near half-metallic half-Heusler systems may be of interest for spintronic applications, especially if the position of the Fermi energy can be adjusted, e.g., by alloying or by applied electrical bias. It can be seen from Table~\ref{tab:near_half_metallic_ferromagnets} that there is one tetragonal near-half-metal -- CrScP (see Table~\ref{tab:CrScTi5_minima} and relevant discussion). All other compounds prefer the cubic $C1_{b}$ cell to a tetragonal distortion of that cell.

\afterpage{
\begin{table*}
\caption{DFT-calculated properties of 34 near half-metallic \textit{XYZ} half-Heusler compounds with negative formation energy. Successive columns present: number of valence electrons per formula unit $N_{V}$, calculated lattice constant $a$, total spin moment $M_{tot}$ per f.u., local moments for atoms on the $X$-, $Y$-, and $Z$-sites: $m(X)$, $m(Y)$, and $m(Z)$, formation energy $\Delta E_f$, distance from the convex hull $\Delta E_{\rm HD}$, spin polarization at Fermi energy $\mathcal{P}(E_F)$, experimental reports of compounds with composition \textit{XYZ}, and experimental reports of corresponding $X_2YZ$ full-Heusler compounds.}
\label{tab:near_half_metallic_ferromagnets}
\begin{tabular}{|l|c|c|c|c|c|c|c|c|c|c|c|}
\toprule
\textit{XYZ} & $N_{V}$ & $a$ & $M_{tot}$ & $m(X)$ & $m(Y)$ & $m(Z)$ & $\Delta E_f$ & $\Delta E_{\rm HD}$ & $\mathcal{P}(E_F)$ & Experimental & $X_{2}YZ$ \\
 & & ({\AA}) & \multicolumn{4}{c|}{($\mu_{B}$)} & \multicolumn{2}{c|}{(eV/atom)} & & reports & reports \\
\midrule
CrScP  & 14 & 5.64 & $-3.9999$ & $-3.271$ & $-0.464$ & 0.021    & $-0.241$ & 0.838 & 0.9900 & & \\
 & & (6.65) & & & & & & & & & \\
MnTiAs & 16 & 5.74 & $-1.9946$ & $-1.848$ & $-0.027$ & $-0.036$ & $-0.298$ & 0.334 & 0.8902 & $P\overline{6}2m$\cite{Johnson19731067} & \\ 
FeTiP  & 17 & 5.45 & $-0.8837$ & $-0.621$ & $-0.180$ & $-0.055$ & $-0.802$ & 0.091 & 0.6074 & $Pnma$\cite{rundqvist1966crystal,ivanov2000florenskyite} & \\
FeTiAs & 17 & 5.65 & $-0.9896$ & $-1.080$ &    0.141 & $-0.042$ & $-0.550$ & 0.145 & 0.8895 & $P\overline{6}2m$\cite{Johnson19731067} & \\
FeTiSb & 17 & 5.95 & $-0.9550$ & $-1.301$ &    0.356 & $-0.023$ & $-0.382$ & 0.034 & 0.6670 & $F\overline{4}3m$\cite{kripyakevich1963crystal} & \\ 
FeVGe  & 17 & 5.56 & $-1     $ & $-1.136$ &    0.202 & $-0.052$ & $-0.157$ & 0.137 & 0.6487 & & \\
CoTiSi & 17 & 5.58 & $-0.9998$ & $-0.407$ & $-0.430$ & $-0.102$ & $-0.590$ & 0.222 & 0.7874 & $Pnma$\cite{PSSA:PSSA2210820124,Szytu198399,Jeitschko1969}, $P\overline{6}2m$\cite{Jeitschko:a07462}  & \cite{markiv1966phase,Carbonari1996313,Co2TiGe2} \\
CoTiGe & 17 & 5.64 & $-0.9734$ & $-0.415$ & $-0.401$ & $-0.078$ & $-0.498$ & 0.117 & 0.5117 & $P\overline{6}2m$\cite{Jeitschko:a07462} & \cite{Carbonari1996313,Co2TiGe2} \\
CoTiSn & 17 & 5.93 & $-0.9647$ & $-0.383$ & $-0.416$ & $-0.048$ & $-0.360$ & 0.070 & 0.7131 & $F\overline{4}3m$\cite{Pierre199393,Pierre199495} & \cite{Co2TiSn1,VANENGEN1983374,Co2TiSn2} \\
NiTiIn & 17 & 5.99 & $-0.9996$ & $-0.078$ & $-0.668$ & $-0.068$ & $-0.168$ & 0.280 & 0.9664 & & \\
MnVP   & 17 & 5.40 & $-0.8574$ & $-1.057$ &    0.268 & $-0.047$ & $-0.539$ & 0.190 & 0.2099 & & \\
CrMnP  & 18 & 5.42 &  0.0016   & $-1.715$ &    1.744 & $-0.064$ & $-0.264$ & 0.282 & 0.9959 & & \\
MnMnP  & 19 & 5.33 &  1        & $-0.964$ &    1.916 &  0.002   & $-0.419$ & 0.178 & 0.9914 & $P\overline{6}2m$\cite{rundqvist1961x,Hggstrm1986171}, $P321$\cite{Arstad1937} & \\
FeCrP  & 19 & 5.29 &  0.9978   & $-0.346$ &    1.298 & $-0.023$ & $-0.465$ & 0.180 & 0.8754 & $Pnma$\cite{kumar2004magnetization,zolensky2008andreyivanovite,Gurin1977381} & \\
FeMnSi & 19 & 5.32 &  0.9718   & $-0.306$ &    1.295 & $-0.056$ & $-0.155$ & 0.257 & 0.9228 & & \\
RuCrP  & 19 & 5.58 &  0.9958   & $-0.240$ &    1.249 & $-0.052$ & $-0.281$ & 0.215 & 0.9019 & & \\
CoVP   & 19 & 5.36 &  0.9949   & $-0.022$ &    0.974 & $-0.010$ & $-0.643$ & 0.096 & 0.9265 & & \\
CoCrSi & 19 & 5.36 &  1        & $-0.214$ &    1.254 & $-0.069$ & $-0.203$ & 0.251 & 0.9960 & $Pnma$\cite{landrum1998tinisi} & \\
RhVP   & 19 & 5.66 &  0.9998   & $-0.124$ &    1.116 & $-0.048$ & $-0.567$ & 0.286 & 0.9587 & & \\
NiVSi  & 19 & 5.47 &  0.9582   &   0.097  &    0.841 & $-0.040$ & $-0.316$ & 0.259 & 0.9097 & $Pnma$\cite{landrum1998tinisi,Jeitschko1969} & \\
NiVGe  & 19 & 5.58 &  0.9917   &   0.042  &    0.944 & $-0.052$ & $-0.203$ & 0.127 & 0.9655 & $Pnma$\cite{Jeitschko1969} & \\
NiVP   & 20 & 5.45 &  1.9054   &   0.168  &    1.656 & $-0.035$ & $-0.403$ & 0.324 & 0.7368 & $Pnma$\cite{palfij1979vnip} & \\
NiVAs  & 20 & 5.62 &  1.9956   &   0.127  &    1.802 & $-0.063$ & $-0.200$ & 0.182 & 0.5758 & $Pnma$\cite{RoyMontreuil1972813,Johnson19731067} & \\
FeFeP  & 21 & 5.31 &  2.9346   &   0.469  &    2.447 & $-0.047$ & $-0.350$ & 0.199 & 0.5125 & $P\overline{6}2m$\cite{fujii1979polarized,Catalano1973262}, $Pnma$\cite{britvin2002allabogdanite} & \\
 & & & & & & & & & & $P321$\cite{hendricks1930xxxv}, $Imm2$\cite{PhysRevB.82.085103} & \\
FeFeAs & 21 & 5.49 &  2.9999   &   0.459  &    2.547 & $-0.053$ & $-0.071$ & 0.071 & 0.9376 & $P4/nmm$\cite{elander1935crystal,Hggstrm1986171} & \\
RuFeAs & 21 & 5.76 &  2.8648   &   0.131  &    2.723 & $-0.028$ & $-0.012$ & 0.186 & 0.0665 & & \\
CoFeSi & 21 & 5.36 &  2.9966   &   0.573  &    2.514 & $-0.143$ & $-0.204$ & 0.254 & 0.8636 & $Pnma$\cite{landrum1998tinisi} & \cite{Co2FeSi1,Buschow19831,Co2FeSi2} \\
RhMnAs & 21 & 5.83 &  3.0289   & $-0.182$ &    3.346 & $-0.160$ & $-0.256$ & 0.105 & 0.6134 & $P\overline{6}2m$\cite{PSSA:PSSA2210840125,Kanomata1987286} & \\
RhFeSi & 21 & 5.68 &  2.9903   &   0.220  &    2.860 & $-0.104$ & $-0.240$ & 0.335 & 0.7374 & & \\
NiMnGe & 21 & 5.57 &  3.0076   &   0.032  &    3.147 & $-0.203$ & $-0.153$ & 0.111 & 0.6038 & $P6_{3}/mmc$\cite{doi101021/ic50147a032} & \cite{Ni2MnGe1} \\
 & & & & & & & & & & $Pnma$\cite{PSSA:PSSA2210380235,PSSA:PSSA2210880243,Fjellv1985291}, $Cmcm$\cite{PSSA:PSSA2210880243} & \\
NiFeAl & 21 & 5.56 &  2.9978   &   0.361  &    2.760 & $-0.137$ & $-0.069$ & 0.377 & 0.8543 & & \\
CoFeP  & 22 & 5.35 &  3.8481   &   0.920  &    2.854 & $-0.021$ & $-0.348$ & 0.236 & 0.1231 & $Pnma$\cite{Nylund1971,fruchart1969crystallographic} & \\
CoFeAs & 22 & 5.53 &  3.9795   &   0.982  &    2.944 & $-0.024$ & $-0.130$ & 0.076 & 0.6560 & $P\overline{6}2m$\cite{Nylund1972115} & \\
CoFeSb & 22 & 5.81 &  3.9793   &   0.995  &    2.982 & $-0.033$ & $-0.033$ & 0.093 & 0.5237 & & \\
\bottomrule
\end{tabular}

\end{table*}
}

The total magnetic moments/f.u.~calculated for all the 34 near half-metallic half-Heusler compounds in Table~\ref{tab:near_half_metallic_ferromagnets} are approximately integers. The difference between the total moment and an integer can be used to estimate how far the Fermi energy falls from the band gap in the gapped channel. For example, MnTiAs has 16 valence electrons and is calculated to have a moment of -1.9946~$\mu_B$/f.u. Thus $N^\uparrow + N^\downarrow=16$ and $N^\uparrow -N^\downarrow=1.9946$ which implies that $N^\uparrow=8.9973$ and $N^\downarrow=7.0027$. There are therefore 0.0027 unfilled states below the gap in the majority channel which occurs at the Slater-Pauling value of 9 electrons/f.u.  

Since Ti and Sc are hard to magnetically polarize, for most of the near-half-metals with $N_{V}\leq17$, their magnetic moments are mainly localized on $X$, and there are smaller parallel magnetic moments on $Y$, resulting in ferromagnetic states. Interestingly, three CoTi$Z$ ($Z=$ Si, Ge, and Sn) compounds behave differently with approximately equal spin moments on Co and Ti. We also found an 18-electron ferrimagnetic near-half-metal, CrMnP, with small total magnetic moment of 0.0016~$\mu_{B}$/f.u. As the number of valence electrons increases to 19, most of the near-half-metals are ferrimagnets whose spin moments are mainly localized on $Y$, while three NiV$Z$ ($Z=$ Si, Ge, and Sn) compounds prefer to be ferromagnets. The near-half-metals with $N_{V}\geq20$, tend to be ferromagnets with larg spin moments remaining on the atom in the $Y$-site. 

Of the 34 compounds in Table~\ref{tab:near_half_metallic_ferromagnets}, we find reports of experimental observation of the $C1_b$ phase for only two systems, FeTiSb and CoTiSn. Our calculations predict both the $C1_b$ phases to lie close to the convex hull with hull distances of $\Delta E_{\rm HD} = 0.034$ and 0.070~eV/atom, respectively. Both of these systems merit further discussion:\\
(a) FeTiSb: has been reported to exist in the $C1_b$ phase, however, recent experimental and theoretical studies~\cite{FeTiSbpaper} indicate that the composition of this phase is actually intermediate between the half- and full-Heusler compositions, close to Fe$_{1.5}$TiSb. DFT calculations showed that several layered systems consisting of equal amounts of FeTiSb and Fe$_2$TiSb would generate semiconducting Fe$_{1.5}$TiSb compounds with formation energy per atom lower than a linear combination of FeTiSb $+$ Fe$_2$TiSb. The system with the lowest calculated formation energy was shown to be a non-magnetic semiconductor Fe$_{1.5}$TiSb phase with primitive unit cells of FeTiSb and Fe$_2$TiSb alternatingly layered in the [111] direction. \\
(b) CoTiSn: both the $C1_b$ \textit{XYZ} and the $L2_1$ $X_2YZ$ phases have been reported in the system. However, more recent experimental studies~\cite{CoTiSn} have shown its composition also to be closer to Co$_{1.5}$TiSn. Similar to the case of FeTiSb, a system consisting of alternating layers of $C1_b$ CoTiSn and $L2_1$ Co$_2$TiSn primitive cells layered along [111] was calculated to have a formation energy per atom lower than a combination of $C1_b$ CoTiSn + $L2_1$ Co$_2$TiSn~\cite{VinayCoTiSn}, and was predicted to be a Slater-Pauling half-metal. 

Further, we find 6 of the near half-metallic half-Heusler compounds with negative formation energy to lie close to the convex hull, i.e., with hull distances $\Delta E_{\rm HD} \leq \,\sim$0.1~eV/atom. In most of the cases, we find experimental reports of other non-$C1_b$ compounds at the composition (space group of the structure(s) experimentally reported, and hull distance $\Delta E_{\rm HD}$ of the $C1_b$ half-Heusler compound in eV/atom): FeFeAs ($P4/nmm$, 0.071), CoFeAs ($P\overline{6}2m$, 0.076), FeTiP ($Pnma$, 0.091), CoFeSb (none, 0.093), CoVP (none, 0.096), RhMnAs ($P\overline{6}2m$, 0.105). In particular, we did not find any compounds experimentally reported at the composition for CoVP and CoFeSb, which merit experimental synthesis efforts. In the case of all the other \textit{XYZ} compositions for which we did not find any experimental reports of compounds (11 additional \textit{XYZ} compositions), the calculated formation energies of the $C1_b$ phases predict them to lie farther away ($\Delta E_{\rm HD} > 0.1$~eV/atom) from the convex hull.

\section{Summary and Conclusion}
\label{sec:summary_conclusion}
In this work, using density functional calculations, we studied the thermodynamic and structural stability, electronic structure, and magnetism of 378 \textit{XYZ} half-Heusler compounds ($X=$ Cr, Mn, Fe, Co, Ni, Ru, Rh; $Y=$ Ti, V, Cr, Mn, Fe, Ni; $Z=$ Al, Ga, In, Si, Ge, Sn, P, As, Sb), and an additional 6 compounds with $X=$ Ni, Cr, $Y=$ Sc, and $Z=$ P, As, Sb. We find that almost all of the 384 systems we studied exhibit a Slater-Pauling gap or a pseudogap in at least one of the spin channels, for some value of the lattice parameter. We find that having a gap at the Fermi energy in one or both spin channels seems to contribute significantly to the stability of a half-Heusler compound relative to other possible phases.

We calculated the formation energy of all 384 \textit{XYZ} compounds, and systematically compared their formation energy against all other phases or  linear combination of phases at that composition in the Open Quantum Materials Database (OQMD). We represent the phase stability of each compound using its distance from the convex hull -- the farther away a compound is from the convex hull, the less thermodynamically stable it is -- and assert that the calculated hull distance of the compound is a good measure of the likelihood of its experimental synthesis. We find low formation energies and (mostly) correspondingly low hull distances for compounds with $X =$ Co, Rh or Ni, $Y=$ Ti or V, and $Z=$ P, As, Sb, or Si. 

Of the 384 half-Heuslers considered, we find 26 18-electron Slater-Pauling semiconductors with negative formation energy. In these systems a gap exists at the Fermi energy in both spin channels. Overall the agreement between theory and experiment was found to be good, i.e., most of the 18-electron \textit{XYZ} compounds were correctly predicted to be in the $C1_b$ or the competing $Pnma$ structures, and all the experimentally reported compounds were found to lie on or close to (i.e., within $\sim$0.1~eV/atom) the convex hull. Our calculations predict CoVGe and FeVAs in the $C1_b$ structure to be sufficiently lower in energy than the experimentally reported $Pnma$ structure that efforts to fabricate the $C1_b$ are justified. Further, our calculations predict semiconducting RuVAs phase (in the $C1_b$ structure), NiScAs, RuVP, RhTiP phases (all in the $Pnma$ structure) to lie on the convex hull (i.e., thermodynamically stable), and CoVSn, RhVGe phases (in the $C1_b$ structure) to lie close to the convex hull of phases. We found no experimental reports of any compounds at these compositions, and thus these compounds present opportunities for experimental exploration.

We find two particularly interesting compounds, CrMnAs and MnCrAs, with 18 electrons/f.u. that are predicted to be zero-moment half-metals rather than semiconductors, and have negative formation energy. Both are calculated to lie above the convex hull, however, and non-equilibrium processing techniques may be necessary to synthesize them. 

Further, we find 45 half-Heusler half-metals with negative fromation energy. In these systems the Fermi energy falls in the Slater-Pauling gap for only one of the spin channels. We also find 34 half-Heusler near half-metals with negative formation energy. In these systems, there is a Slater-Pauling gap, but the Fermi energy falls very near but not quite in the gap. Our calculations predict a half-metal RhVSb, and two near half-metals, CoFeSb and CoVP, to lie within $\sim$0.1~eV/atom of the convex hull. The lack of experimental reports of any compound at the three compositions merits efforts to synthesize them.

Overall, our calculations correctly predict a large number of experimentally-reported half-Heusler compounds to be thermodynamically stable. In addition, they also predict a number of semiconducting, half-metallic, and near half-metallic half-Heusler compounds to lie above but close to the convex hull. Such compounds may be experimentally realized using suitable (non-equilibrium) synthesis conditions.

\begin{acknowledgments}
The authors acknowledge support from the National Science Foundation through grants DMR-1235230, DMR-1235396, and DMR-1309957 (VH and CW). The authors also acknowledge Advanced Research Computing Services at the University of Virginia and High Performance Computing staff from the Center for Materials for Information Technology at the University of Alabama for providing technical support that has contributed to the results in this paper. The computational work was done using High Performance Computing Cluster at the Center for Materials for Information Technology, University of Alabama, resources of the National Energy Research Scientific Computing Center (NERSC), a DOE Office of Science User Facility supported by the Office of Science of the U.S. Department of Energy under Contract No. DE-AC02-05CH11231 and Rivanna high-performance cluster at the University of Virginia. 
\end{acknowledgments}

\bibliography{halfheusler}

\end{document}